\documentclass{article}

\usepackage[left = 20 mm, right = 20 mm, top = 30 mm, bottom = 20 mm]{geometry}
\usepackage{amsmath, amssymb, verbatim, graphicx, subfig, caption, slashed, makecell, empheq, cite, jheppub}
\usepackage[toc,page]{appendix}

\newcommand{\minus}{\scalebox{0.75}[1.0]{$-$}}

\allowdisplaybreaks

\begin{document}

\title{Generation of quasiparticles by particle-antiparticle mixing and limitations of quantum mechanics}

\author{Chang-Hun Lee}
\affiliation{National Center for Theoretical Sciences, \\
101, Section 2, Kuang-Fu Road, Hsinchu, \\
Republic of China (Taiwan)}
\emailAdd{changhun@terpmail.umd.edu}

\begin{abstract}
{The mixing of a neutral unstable particle and its antiparticle has been usually discussed in the framework of quantum mechanics in the literature. In this paper, the particle-antiparticle mixing is discussed fully in the quantum field theory, and it is shown that quantum mechanics is not the proper non-relativistic limit of the quantum field theory in the presence of such mixing. Moreover, it is also shown that a discrepancy indeed exists between the results from quantum mechanics and the quantum field theory, which can sometimes make big differences in theoretical predictions of observables such as decay widths. This result has an implication for the mixing of neutral mesons and the CP violation in their decays.}
\end{abstract}

\maketitle

\section{Introduction}
In a neutral meson system, the meson and antimeson mix with each other, which causes CP violation in their decays. For decades, such CP violation effects have been experimentally observed in various decay channels of various mesons. The standard way to calculate the decay widths of neutral mesons and the CP asymmetries therein is finding the effective Hamiltonian of the neutral meson system and solving the associated effective Schr\"odinger equation. The self-energy matrix of the neutral mesons as unstable particles has an absorptive part, from which the non-Hermitian effective Hamiltonian is obtained. In other words, the mixing of neutral mesons and their time evolution have been studied in the framework of quantum mechanics.

The physics of mixing has also been studied in the quantum field theory in many works in the literature, and they are listed and briefly summarized in reference \cite{OscNeutMesonQFT}. Most works, however, discussed it in the context of flavor mixing of light neutrinos which are stable particles, and the difference between stable and unstable particles were thought to be relatively minor. Only a small number of works specifically focused on mixing of unstable particles. In reference \cite{IntNeuK}, the neutral kaons were studied using their dressed propagator. Since each physical unstable particle of kaons, which propagates like a free particle until it decays, is associated with a complex-valued pole of the dressed propagator, the mixing matrix that diagonalizes the dressed propagator will relate the flavor eigenstates and physical particles of kaons. The dynamics of neutral kaons such as decays was studied from it. In other words, the properties of unstable particles were studied from physical processes in which those particles appear as intermediate states. Note that the physical particles are usually called mass eigenstates in the literature, which is actually inappropriate as we will see. In reference \cite{ResCPVMixing}, a similar approach was adopted while focusing on gauge invariant renormalization of the fields of unstable particles, because physical observables in gauge theories of unstable particles sometimes seem to depend on renormalization schemes. In particular, the diagonal components of the self-energy of unstable particles were renormalized in the on-shell scheme. In reference \cite{FlavMixGaugeInvWFR}, another issue was raised about renormalization in theories of unstable particles with flavor mixing. It was claimed that the fields of in- and out-states of unstable particles of a physical process require different field-strength renormalization matrices when there exists flavor mixing, because of the absorptive part of the self-energy matrix of those unstable particles. In reference \cite{FOQFT}, the analysis of the light neutrino oscillation in the quantum field theory was applied to the oscillation of neutral mesons. To study the oscillation of unstable particles, the decay factor $e^{-\gamma t}$ was inserted by hand into the flavor oscillation formula which had been obtained for the canonically quantized free fields of mass eigenstates of mesons. An implicit assumption in that approach is that the fields of mass eigenstates are renormalized in the on-shell scheme.

However, the physics of mixing in the case of unstable particles is more subtle. In reference \cite{MixingRen}, it was shown that a consistent treatment of bare as well as renormalized theories of unstable particles with flavor mixing requires that the physical unstable particles should be interpreted as quasiparticles generated by various interacting fields. Such particles cannot be regarded as external states of a physical process, and should be treated only as intermediate states in the same vein as the approach discussed in reference \cite{IntNeuK}. Moreover, it was also shown that each of physical particles cannot be related to a single renormalized field by the same reason that led to the conclusion of reference \cite{FlavMixGaugeInvWFR}, and the fields of physical unstable particles and their canonically quantized fields should be distinguished. In other words, physical renormalization schemes such as the on-shell or complex-mass scheme cannot be applied to them. In addition, it is impossible to observe each of physical particles separately because of the interference between multiple physical particles, as opposed to a popular belief that each should be observable as a resonance peak in scattering mediated by unstable particles or as a longer-living one in the decays of unstable particles with much different lifetimes as in the neutral kaon system. Hence, physical unstable particles should not be simply called mass eigenstates, not only because the formers are defined by particles with specific decay widths as well as masses, but also because the phenomenology of physical unstable particles is different from that of mass eigenstates of stable particles or quantum mechanics.

This has an important implication for what the correct understanding of the time evolution of an unstable particle is. It is generally thought that a decay of an unstable particle in the plane-wave limit is a phenomenon in which its single one-particle state evolves in time and decays into other particles. However, each of physical unstable particles as a quasiparticle cannot be associated with such a single one-particle state in the Hilbert space. This means that the quantum mechanical approach based on the Schr\"odinger equation should fail especially in the case of multiple flavors, because every particle in the plane-wave limit in quantum mechanics is a one-particle state with definite mass, \textit{i.e.}, mass eigenstate, which evolves in time as a plane wave while keeping its identity as a specific linear combination of basis states of the Hilbert space. Even in the frameworks of quantum mechanics, it is easy to understand that such a description is flawed, since the left- and right-eigenstates of a non-Hermitian effective Hamiltonian are different. Hence, we should conclude that quantum mechanics is not a proper non-relativistic limit of the quantum field theory in theories of unstable particles with flavor mixing.

In this paper, we discuss those issues in detail using a toy model which imitates effective theories of neutral mesons. A real meson system is too complicated for that purpose, not only because many different kinds of fields contribute even to the simplest processes, but also because the non-perturbative quantum chromodynamics is involved. This often obscures understanding of its dynamics, and a simple model has an advantage over a real meson system to reveal the physics of meson mixing in a transparent way. In a sense, the purpose of this paper is to analyze the particle-antiparticle mixing by using the approach discussed in reference \cite{MixingRen} which studied the flavor mixing of unstable particles. Specifically, we will discuss in more details how the properties of physical unstable particles such as decay widths can be read from scattering mediated by those particles. Various different methods will be discussed, and it will be shown that they are mutually consistent although there exist some differences. We will also see that this result supports the interpretation that physical particles are quasiparticles which cannot be separately observable. Furthermore, it will be shown that the decay widths calculated in the quantum field theory are different from those obtained in quantum mechanics and the differences can sometimes be very large, in contrast to the popular belief that they are approximately identical. The issue of gauge invariance will not be studied in this paper.

We will work in the plane-wave limit where each external particle of a physical process is described by a momentum eigenstate with specific mass whose time evolution is a plane wave. Even though a particle is better described by a wave packet, it is beyond the scope of this paper. It is sometimes argued that oscillation cannot occur in the plane-wave limit \cite{OscNeutMesonQFT}. The energy of the decay product can be exactly specified in that limit, and thus we should be able to identify the unstable particle from which the decay product is generated. Hence, any oscillation between two different physical particles cannot happen in such a limit, and the external particles should be described by wave packets to allow oscillation. However, this reasoning is false, and it originates from mixing up two different analyses in time and energy domains which are related to each other by the Fourier transform. The oscillation is a concept in the time domain, and it is generated by the interference between two particles which have dominant contributions at different energies. In the energy domain, the interference occurs between two propagators of unstable particles, and it also contributes to each resonance peak when one of them is on-shell and the other is off-shell. Exactly measuring the total energy of the decay product excludes neither of two propagators of physical particles. We will discuss this issue in more details while deriving their decay widths. Moreover, the plane-wave limit does not necessarily assumes that the energy-momentum of each particle is exactly measured. A perturbation theory always assumes specific precision of calculation, and the plane-wave limit can be a perfect approximation as long as the precision of experimental measurements is better than the precision of calculation. 

It is also worth mentioning a difference between physical particles of this paper and mass eigenstates in the literature with regards to renomalization. Each of physical unstable particles in this paper will be defined as the on-shell contribution of the associated fields that corresponds to the canonical form of a propagator with a complex-valued pole. In other words, it is the particle that propagates like a free particle until it decays in the full interacting theory. On the other hand, a mass eigenstate in the literature is usually defined as a particle of a canonically quantized free field, \textit{i.e.}, a field in the interaction picture, whose mass is the pole mass of its propagator. Such a definition assumes that the field is renormalized in a physical scheme, and the mass eigenstate can also be regarded as a particle of an interacting field, \textit{i.e.}, a field in the Heisenberg picture. However, such an assumption is allowed only for stable particles as discussed in reference \cite{MixingRen}, and physical schemes are inapplicable to unstable particles especially when there exists flavor or particle-antiparticle mixing. As mentioned above, the canonically quantized fields and the fields of physical particles must be distinguished in the case of unstable particles with mixing, and the theory should be renormalized in a non-physical scheme.

This paper is organized as follows: in section \ref{sec:RenLag}, a renormalizable toy model that will imitate a neutral meson system is introduced, and the expression of the renormalized self-energy is derived; in section \ref{sec:DiagProp}, we discuss how to diagonalize the dressed propagator matrix of unstable particles; in section \ref{sec:Quasi}, the properties of physical particles and the limitations of quantum mechanics are discussed; in section \ref{sec:DerivDW}, the decay widths of unstable physical particles are derived in multiple ways in the quantum field theory; in section \ref{sec:QM}, the results are compared with those obtained in quantum mechanics, and it is shown that a discrepancy exists; in section \ref{sec:Example}, two numerical examples that imitate real meson systems are provided to test the formalism and also to understand the physics behind it.

\section{Toy model and renormalization of its Lagrangian density}		\label{sec:RenLag}
A real meson system is very complicated, not only because many particles contribute to even the simplest type of interactions, but also because the non-perturbative interactions of quantum chromodynamics are involved in many processes. However, the physics of particle-antiparticle mixing can be understood in a much simpler way, since what is really important is the dynamics at the level of an effective field theory of mesons. In this section, a toy model that will imitate such an effective field theory will be developed. Appropriately choosing model parameters, we can indeed make the model mimic real neutral meson systems quite closely, as will be shown in section \ref{sec:Example}. \\

The toy model consists of four different kinds of fields, $\Phi$, $\chi_i$, $\xi$, $\psi_i$, which interact with each other by
\begin{align}
	\boxed{\mathcal{L}_\text{int} = \minus \sum_i f_i \overline{\psi_i} \psi_i \Phi - \sum_i h_i \overline{\chi_i} \xi \Phi + \text{H.c}.}
\end{align}
The roles of the fields are as follows:
\begin{enumerate}
	\item $\Phi$ : a neutral complex scalar field corresponding to a spin-0 meson such as $B^0$ or $K^0$.
	\item $\psi_i$ : a light fermion that allows particle-antiparticle mixing. The decay $\Phi \to \psi_i \psi_i^c$ can mimic the decay of neutral mesons into CP eigenstates. The index $i$ denotes the flavor of $\psi$, and $\psi_i^c$ is the antiparticle of $\psi_i$.
	\item $\chi_i, \xi$ : light fermions used to tag the initial field as $\Phi$ or $\Phi^*$. The decays such as $\Phi \to \chi_i \xi^c$ and $\Phi \to \chi_i^c \xi$ can imitate the semileptonic decays of neutral mesons. The index $i$ denotes the flavor of $\chi$.
\end{enumerate}
The light particles are assumed to be fermions rather than scalars so that (i) the toy model is a renormalizable theory, and (ii) the lowest-order loop corrections to the self-energy are one-loop diagrams. In a non-renormalizable theory, there can be an ambiguity in the choice of the renormalized self-energy, which is never desirable in the discussion at the level of a toy model. Moreover, we assume that the masses of $\psi_i$ and $\xi$ are negligible for simplicity, since introducing non-zero masses to those fermions only complicates the analysis with no practical advantage. Such an assumption is allowed as long as their tree-level masses are small enough. In addition, $\chi_i$ and $\xi$ can be considered to be stable up to the one-loop order in the self-energy, since the decay such as $\chi_i \to (\Phi \xi \to) \chi_j \xi^c \xi$ corresponds to the cutting of a two-loop self-energy diagram of $\chi_i$. Furthermore, we assume that $\chi_i$ and $\xi$ are renormalized in the on-shell scheme, which is required to regard their particles as external states of physical processes.

Note also that $\Phi (x) \coloneqq e^{i H (t - t_0)} \Phi (\mathbf{x}) e^{-i H (t - t_0)}$ is the field in the Heisenberg picture, where $H$ is the Hamiltonian and $\Phi (\mathbf{x}) = \Phi (t_0, \mathbf{x})$ is the field in the Schr\"odinger picture which creates $| \Phi (m_\Phi, \mathbf{p}) \rangle$ from the vacuum $| 0 \rangle$ of the free theory or annihilates $| \Phi^* (m_\Phi, \mathbf{p}) \rangle$. Here, $| \Phi (m_\Phi, \mathbf{p}) \rangle$ and $| \Phi^* (m_\Phi, \mathbf{p}) \rangle$ are the one-particle states of $\Phi$ and $\Phi^* (\coloneqq \Phi^\dag)$, respectively, with mass $m_\Phi$ and three-momentum $\mathbf{p}$. Due to particle-antiparticle mixing, $\Phi (x)$ at $t = x^0 \neq t_0$ does not necessarily create $| \Phi (m_\Phi, \mathbf{p}) \rangle$. For simplicity, let us write $| \Phi \rangle \coloneqq | \Phi (m_\Phi, \mathbf{p}) \rangle$ and $| \Phi^* \rangle \coloneqq | \Phi^* (m_\Phi, \mathbf{p}) \rangle$ in this paper. \\

\begin{figure}[h]
	\centering
	\subfloat[$\Phi \to \chi_i \xi^c$]{
		\includegraphics[width = 28 mm]{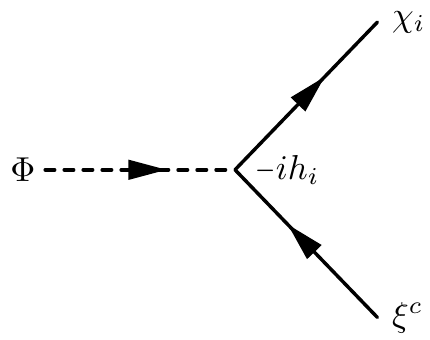}
		\label{fig:VPCXcTree}
	} \qquad
	\subfloat[$\Phi^* \to \chi_i^c \xi$]{
		\includegraphics[width = 28 mm]{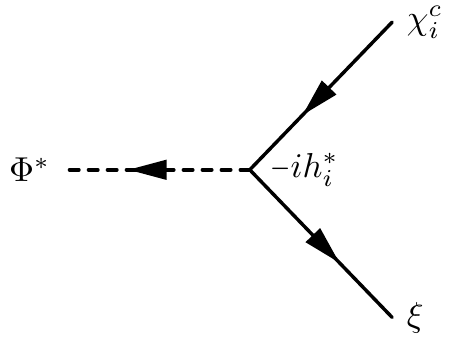}
		\label{fig:VPcCcXTree}
	} \qquad
	\subfloat[$\Phi \to \psi_i \psi_i^c$]{
		\includegraphics[width = 28 mm]{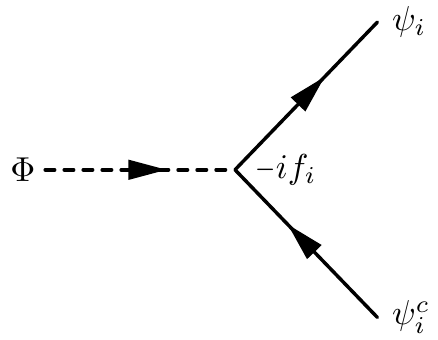}
		\label{fig:VPPsPsTree}
	} \qquad
	\subfloat[$\Phi^* \to \psi_i \psi_i^c$]{
		\includegraphics[width = 28 mm]{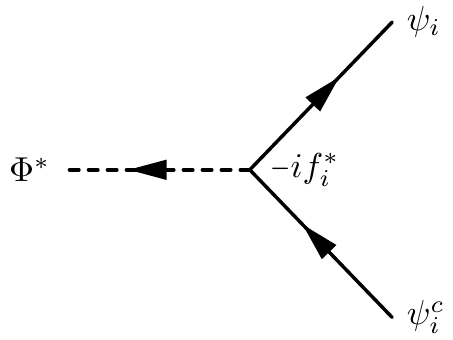}
		\label{fig:VPcPsPsTree}
	}
	\caption{Interactions of the toy model.}
	\label{fig:Vertex}
\end{figure}

In order to carefully derive the renormalized self-energy of $\Phi$, we begin with the bare Lagrangian density of the toy model and renormalize it step-by-step. The Lagrangian density involving $\Phi_0$ is written as
\begin{align}
	\mathcal{L} &= \partial^\mu \Phi_0^\dag \partial_\mu \Phi_0 - m_{\Phi_0}^2 \Phi_0^\dag \Phi_0
		- \sum_i f_{0 i} \overline{\psi_{0 i}} \psi_{0 i} \Phi_0 - \sum_i f_{0 i}^* \overline{\psi_{0 i}} \psi_{0 i} \Phi_0^\dag
		- \sum_i h_{0 i} \overline{\chi_{0 i}} \xi_0 \Phi_0 - \sum_i h_{0 i}^* \overline{\xi_0} \chi_{0 i} \Phi_0^\dag \nonumber \\
		&\qquad - \delta_\Phi' \partial^\mu \Phi_0 \partial_\mu \Phi_0 - \delta_\Phi'^* \partial^\mu \Phi_0^\dag \partial_\mu \Phi_0^\dag - \delta m_\Phi'^2 \Phi_0 \Phi_0 - \delta m_\Phi'^{2 *} \Phi_0^\dag \Phi_0^\dag.
\end{align}
For some given bare and renormalized fields, the field-strength renormalization factors are defined by
\begin{align}
	\Phi_0 \eqqcolon Z_\Phi^\frac{1}{2} \Phi, \qquad
	\psi_0 \eqqcolon Z_\psi^\frac{1}{2} \psi, \qquad
	\chi_0 \eqqcolon Z_\chi^\frac{1}{2} \chi, \qquad
	\xi_0 \eqqcolon Z_\xi^\frac{1}{2} \xi,
\end{align}
where $Z_\psi^\frac{1}{2}$ and $Z_\chi^\frac{1}{2}$ are matrices. We also define the counterterm $\delta_\Phi$ associated with $Z_\Phi$ as
\begin{align}
	Z_\Phi^\frac{1}{2} \eqqcolon 1 + \frac{1}{2} \delta_\Phi.
\end{align}
In addition, for given renormalized mass $m_\Phi$ and coupling $f_i$, the mass renormalization matrix $Z_M$ and vertex counterterms $\delta f_i$, $\delta h_i$ are defined by
\begin{alignat}{2}
	&m_{\Phi_0} \eqqcolon Z_M m_\Phi, \qquad
	&&\delta m_\Phi^2 \coloneqq m_\Phi^2 Z_M^2 |Z_\Phi| - m_\Phi^2, \\
	&(Z_\psi f_0)_i Z_\Phi^\frac{1}{2} \eqqcolon f_i + \delta f_i, \qquad
	&&(Z_\chi^{\frac{1}{2} \dag} h_0)_i Z_\xi^\frac{1}{2} Z_\Phi^\frac{1}{2} \eqqcolon h_i + \delta h_i.
\end{alignat}
The Lagrangian density can now be rewritten in terms of the renormalized fields and couplings as
\begin{align}
	\mathcal{L} &= \partial^\mu \Phi_0^\dag \partial_\mu \Phi_0 - m_{\Phi_0}^2 \Phi_0^\dag \Phi_0
		- \sum_i f_{0 i} \overline{\psi_{0 i}} \psi_{0 i} \Phi_0 - \sum_i f_{0 i}^* \overline{\psi_{0 i}} \psi_{0 i} \Phi_0^\dag
		- \sum_i h_{0 i} \overline{\chi_{0 i}} \xi_0 \Phi_0 - \sum_i h_{0 i}^* \overline{\xi_0} \chi_{0 i} \Phi_0^\dag \nonumber \\
		&\quad + \delta_\Phi' \partial^\mu \Phi_0 \partial_\mu \Phi_0 + \delta_\Phi'^* \partial^\mu \Phi_0^\dag \partial_\mu \Phi_0^\dag + \delta m_\Phi'^2 \Phi_0 \Phi_0 + \delta m_\Phi'^{2 *} \Phi_0^\dag \Phi_0^\dag \nonumber \\[10pt]
	&= |Z_\Phi| \partial^\mu \Phi^\dag \partial_\mu \Phi - m_\Phi^2 Z_M^2 |Z_\Phi| \Phi^\dag \Phi
		- \sum_i (Z_\psi f_0)_i Z_\Phi^\frac{1}{2} \overline{\psi_i} \psi_i \Phi - \sum_i (Z_\psi f_0)_i^* Z_\Phi^{\frac{1}{2} *} \overline{\psi_i} \psi_i \Phi^\dag \nonumber \\
		&\quad - \sum_i (Z_\chi^{\frac{1}{2} \dag} h_0)_i Z_\xi^\frac{1}{2} Z_\Phi^\frac{1}{2} \overline{\chi_i} \xi \Phi - \sum_i (Z_\chi^{\frac{1}{2} \dag} h_0)_i^* Z_\xi^{\frac{1}{2} *} Z_\Phi^{\frac{1}{2} *} \overline{\xi} \chi_i \Phi^\dag \nonumber \\
		&\quad - \delta_\Phi' Z_\Phi \partial^\mu \Phi \partial_\mu \Phi - \delta_\Phi'^* Z_\Phi^* \partial^\mu \Phi^\dag \partial_\mu \Phi^\dag - \delta m_\Phi'^2 Z_\Phi \Phi \Phi - \delta m_\Phi'^{2 *} Z_\Phi^* \Phi^\dag \Phi^\dag \nonumber \\[10pt]
	&= \partial^\mu \Phi^\dag \partial_\mu \Phi - m_\Phi^2 \Phi^\dag \Phi - \sum_i f_i \overline{\psi_i} \psi_i \Phi - \sum_i f_i^* \overline{\psi_i} \psi_i \Phi^\dag - \sum_i h_i \overline{\chi_i} \xi \Phi - \sum_i h_i^* \overline{\xi} \chi_i \Phi^\dag \nonumber \\
		&\quad + \frac{1}{2} (\delta_\Phi^* + \delta_\Phi + \cdots) \partial^\mu \Phi^\dag \partial_\mu \Phi - \delta m_\Phi^2 \Phi^\dag \Phi
			- \sum_i \delta f_i \overline{\psi_i} \psi_i \Phi - \sum_i \delta f_i^* \overline{\psi_i} \psi_i \Phi^\dag - \sum_i \delta h_i \overline{\chi_i} \xi \Phi - \sum_i \delta h_i^* \overline{\xi} \chi_i \Phi^\dag \nonumber \\
		&\quad - \delta_\Phi' (1 + \cdots) \partial^\mu \Phi \partial_\mu \Phi - \delta_\Phi'^* (1 + \cdots) \partial^\mu \Phi^\dag \partial_\mu \Phi^\dag - \delta m_\Phi'^2 (1 + \cdots) \Phi \Phi - \delta m_\Phi'^{2 *} (1 + \cdots) \Phi^\dag \Phi^\dag,
	\label{eq:BareLag}
\end{align}
where the counterterms are explicitly shown only up to the leading terms. The Feynman diagrams of tree-level interactions of this toy model are shown in figure \ref{fig:Vertex}.

\begin{figure}[h]
	\centering
	\subfloat[$(\Sigma_{0 \Phi}^{\psi_i})_{11} (p^2)$]{
		\includegraphics[width = 35.5 mm]{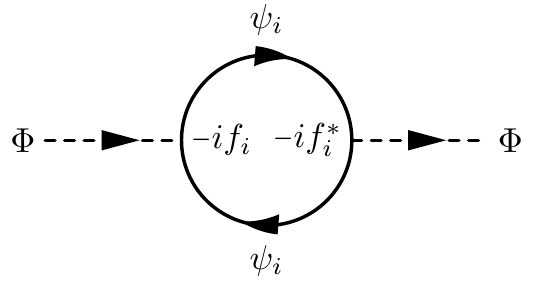}
		\label{fig:SEPPPs}
	}
	\subfloat[$\Sigma_{0 \Phi}^{\chi_i \xi} (p^2)$]{
		\includegraphics[width = 35.5 mm]{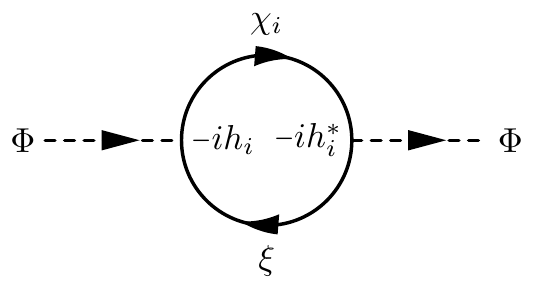}
		\label{fig:SEPPCX}
	}
	\subfloat[$(\Sigma_{0 \Phi}^{\psi_i})_{22} (p^2)$]{
		\includegraphics[width = 35.5 mm]{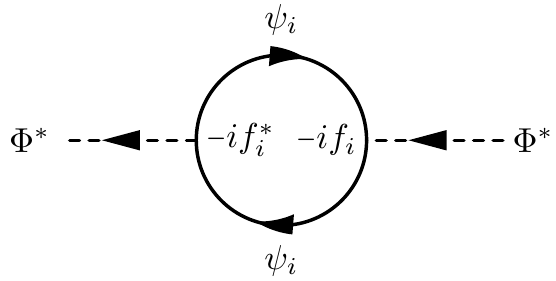}
		\label{fig:SEPcPcPs}
	}
	\subfloat[$\Sigma_{0 \Phi}^{\chi_i \xi} (p^2)$]{
		\includegraphics[width = 35.5 mm]{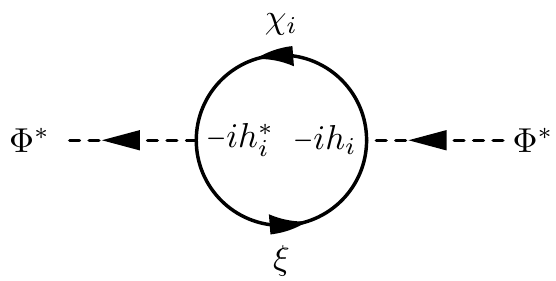}
		\label{fig:SEPcPcCX}
	}
	\caption{One-loop contributions to the diagonal components of $\Sigma_{0 \Phi} (p^2)$.}
	\label{fig:W1LoopDiag}
	\subfloat[$(\Sigma_{0 \Phi}^{\psi_i})_{12} (p^2)$]{
		\includegraphics[width = 35.5 mm]{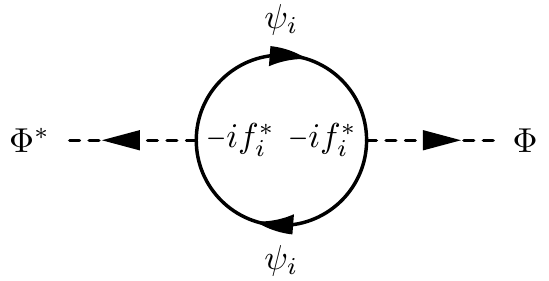}
		\label{fig:SEPcPPs}
	} \qquad
	\subfloat[$(\Sigma_{0 \Phi}^{\psi_i})_{21} (p^2)$]{
		\includegraphics[width = 35.5 mm]{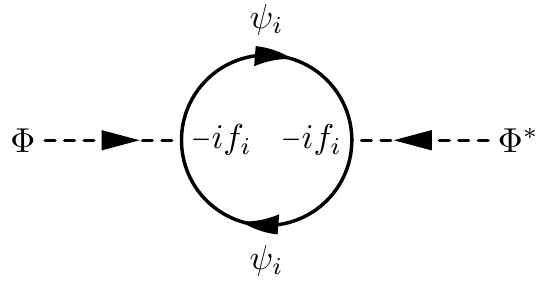}
		\label{fig:SEPPcPs}
	}
	\caption{One-loop contributions to the off-diagonal components of $\Sigma_{0 \Phi} (p^2)$.}
	\label{fig:W1LoopOffDiag}
\end{figure}

\noindent For convenience, let us define
\begin{align}
	\boxed{\Phi_1 \coloneqq \Phi, \qquad
	\Phi_2 \coloneqq \Phi^*.}
\end{align}
The interactions given above generate the non-renormalized self-energy diagrams in figures \ref{fig:W1LoopDiag} and \ref{fig:W1LoopOffDiag}.

\noindent The non-renormalized self-energy matrix of $\Phi$ can be written as
\begin{align}
	(\Sigma_{0 \Phi})_{\beta \alpha} (p^2) = \sum_i \bigg[ (\Sigma_{0 \Phi}^{\psi_i})_{\beta \alpha} (p^2) + \delta_{\beta \alpha} \Sigma_{0 \Phi}^{\chi_i \xi} (p^2) \bigg],
\end{align}
where
\begin{align}
	\Sigma_{0 \Phi}^{\chi_i \xi} (p^2) &= p^2 \Sigma_{0 \Phi}'^{\chi_i \xi} (p^2) + \delta \Sigma_\Phi, \\[5pt]
	\Sigma_{0 \Phi}'^{\chi_i \xi} (p^2) &= \sum_i \frac{|h_i|^2}{16 \pi^2} \bigg[ \frac{2}{\epsilon} + \frac{3}{2} - \log{\frac{m_{\chi_i}^2}{\widetilde{\mu}^2}} - \bigg( 1 - \frac{5 m_{\chi_i}^2}{p^2} + \frac{6 m_{\chi_i}^4}{p^4} - \frac{2 m_{\chi_i}^6}{p^6} \bigg) \log{\frac{m_{\chi_i}^2 - p^2}{m_{\chi_i}^2}} \bigg] + \mathcal{O} (\epsilon), \\
	\delta \Sigma_\Phi &= \minus \sum_i \frac{|h_i|^2}{8 \pi^2} m_{\chi_i}^2 \bigg( \frac{2}{\epsilon} + 1 - \log{\frac{m_{\chi_i}^2}{\widetilde{\mu}^2}} \bigg) + \mathcal{O} (\epsilon),
\end{align}
and
\begin{align}
	\Sigma_{0 \Phi}^{\psi_i} (p^2) = p^2 \Sigma_{0 \Phi}'^{\psi_i} (p^2)
	= \frac{p^2}{16 \pi^2} \left( \begin{array}{cc} |f_i|^2 & f_i^{* 2} \\ f_i^2 & |f_i|^2 \end{array} \right) \bigg( \frac{2}{\epsilon} + \frac{3}{2} - \log{\frac{p^2}{\widetilde{\mu}^2}} + i \pi \bigg) + \mathcal{O} (\epsilon),
\end{align}
where we have used $\log{(\minus 1)} = i \pi$. We have also introduced $\widetilde{\mu}^2 \coloneqq 4 \pi e^{-\gamma_E} \mu^2$ where $\gamma_E$ is the Euler–Mascheroni constant, and $\mu^{2 - d/2} f$ and $\mu^{2 - d/2} h$ are the Yukawa couplings in the $d = 4 - \epsilon$ dimension. The expression of the self-energy as given above is derived in appendix \ref{sec:SECal}. Using the counterterms in equation \ref{eq:BareLag}, we can write the diagonal components of the renormalized self-energy $\Sigma_\Phi (p^2)$ up to the one-loop order as
\begin{align}
	(\Sigma_\Phi)_{11} (p^2) &= p^2 (\Sigma_\Phi')_{11} (p^2) + \delta \Sigma_\Phi - \delta m_\Phi^2
		= p^2 \bigg[ (\Sigma_{0 \Phi}')_{11} (p^2) + \frac{1}{2} (\delta_\Phi^* + \delta_\Phi) \bigg] + \delta \Sigma_\Phi - \delta m_\Phi^2, \\
	(\Sigma_\Phi)_{22} (p^2) &= p^2 (\Sigma_\Phi')_{22} (p^2) + \delta \Sigma_\Phi - \delta m_\Phi^2
		= p^2 \bigg[ (\Sigma_{0 \Phi}')_{22} (p^2) + \frac{1}{2} (\delta_\Phi^* + \delta_\Phi) \bigg] + \delta \Sigma_\Phi - \delta m_\Phi^2 \nonumber \\
		&= (\Sigma_\Phi)_{11} (p^2).
\end{align}
In addition, the off-diagonal components of $\Sigma_\Phi (p^2)$ up to the same precision are written as
\begin{align}
	(\Sigma_\Phi)_{12} (p^2) &= p^2 (\Sigma_\Phi')_{12} (p^2) - \delta m_\Phi'^{2 *}
		= p^2 \bigg[ (\Sigma_{0 \Phi}')_{12} (p^2) + \delta_\Phi'^* \bigg] - \delta m_\Phi'^{2 *}, \\
	(\Sigma_\Phi)_{21} (p^2) &= p^2 (\Sigma_\Phi')_{21} (p^2) - \delta m_\Phi'^2
		= p^2 \bigg[ (\Sigma_{0 \Phi}')_{21} (p^2) + \delta_\Phi' \bigg] - \delta m_\Phi'^2.
\end{align}
We may choose the counterterms
\begin{align}
	\delta_\Phi &= \frac{1}{16 \pi^2} \sum_i \bigg\{ |h_i|^2 \bigg[ \minus \frac{2}{\epsilon} - 2 + \log{\frac{m_{\chi_i}^2}{\widetilde{\mu}^2}} - \frac{m_{\chi_i}^2}{m_\Phi^2}
		+ \bigg( 1 - \frac{m_{\chi_i}^4}{m_\Phi^4} \bigg) \log{\frac{m_\Phi^2 - m_{\chi_i}^2}{m_{\chi_i}^2}} \bigg] \nonumber \\
		&\qquad \qquad \qquad + |f_i|^2 \bigg( \minus \frac{2}{\epsilon} - \frac{3}{2} + \log{\frac{\mu_i^2}{\widetilde{\mu}^2}} \bigg) \bigg\}, \\
	\delta_\Phi' &= \frac{1}{16 \pi^2} \sum_i f_i^2 \bigg( \minus \frac{2}{\epsilon} - \frac{3}{2} + \log{\frac{\mu_i^2}{\widetilde{\mu}^2}} \bigg), \\
	\delta m_\Phi^2 &= \delta \Sigma_\Phi, \qquad
	\delta m_\Phi'^2 = 0
\end{align}
such that
\begin{align}
	\boxed{(\Sigma_\Phi)_{\beta \alpha} (p^2)
	= p^2 (\Sigma_\Phi')_{\beta \alpha} (p^2)
	= p^2 \sum_i \bigg[ (\Sigma_\Phi'^{\psi_i})_{\beta \alpha} (p^2) + \delta_{\beta \alpha} \Sigma_\Phi'^{\chi_i \xi} (p^2) \bigg],}
\end{align}
and
\begin{align}
	\boxed{\Sigma_\Phi'^{\psi_i} (m_\Phi^2) = \frac{1}{16 \pi^2} \left( \begin{array}{cc} |f_i|^2 & f_i^{* 2} \\ f_i^2 & |f_i|^2 \end{array} \right) \bigg[ \minus \log{\bigg( \frac{m_\Phi^2}{\mu_i^2} \bigg)} + i \pi \bigg], \qquad
	\Sigma_\Phi'^{\chi_i \xi} (m_\Phi^2) = i \frac{|h_i|^2}{16 \pi} \bigg( 1 - \frac{m_{\chi_i}^4}{m_\Phi^4} \bigg),}
\end{align}
where another parameter $\mu_i^2$ has been introduced. Note that $\delta_\Phi$ is real-valued. The Lagrangian density can now be rewritten as 
\begin{empheq}[box=\fbox]{align}
	\mathcal{L} &= \partial^\mu \Phi_0^\dag \partial_\mu \Phi_0 - m_{\Phi_0}^2 \Phi_0^\dag \Phi_0 \nonumber \\[5pt]
		&\qquad - \sum_i f_{0 i} \overline{\psi_{0 i}} \psi_{0 i} \Phi_0 - \sum_i f_{0 i}^* \overline{\psi_{0 i}} \psi_{0 i} \Phi_0^\dag
		- \sum_i h_{0 i} \overline{\chi_{0 i}} \xi_0 \Phi_0 - \sum_i h_{0 i}^* \overline{\xi_0} \chi_{0 i} \Phi_0^\dag \nonumber \\
		&\qquad + \delta_\Phi' \partial^\mu \Phi_0 \partial_\mu \Phi_0 + \delta_\Phi'^* \partial^\mu \Phi_0^\dag \partial_\mu \Phi_0^\dag \nonumber \\[10pt]
	&= \partial^\mu \Phi^\dag \partial_\mu \Phi - m_\Phi^2 \Phi^\dag \Phi - \sum_i f_i \overline{\psi_i} \psi_i \Phi - \sum_i f_i^* \overline{\psi_i} \psi_i \Phi^\dag - \sum_i h_i \overline{\chi_i} \xi \Phi - \sum_i h_i^* \overline{\xi} \chi_i \Phi^\dag \nonumber \\
		&\qquad + \delta_\Phi \partial^\mu \Phi^\dag \partial_\mu \Phi - \delta_\Phi' \partial^\mu \Phi \partial_\mu \Phi - \delta_\Phi'^* \partial^\mu \Phi^\dag \partial_\mu \Phi^\dag \nonumber \\[5pt]
		&\qquad - \sum_i \delta f_i \overline{\psi_i} \psi_i \Phi - \sum_i \delta f_i^* \overline{\psi_i} \psi_i \Phi^\dag - \sum_i \delta h_i \overline{\chi_i} \xi \Phi - \sum_i \delta h_i^* \overline{\xi} \chi_i \Phi^\dag + \cdots.
\end{empheq}
For convenience, let us also define the order of perturbative corrections as
\begin{align}
	\boxed{\mathcal{O} (\alpha) \coloneqq \mathcal{O} (\Sigma_\Phi')
	 = \mathcal{O} (f^2 / 16 \pi),}
\end{align}
where $f$ denotes the typical magnitude of $f_i$. Throughout the paper, $|f_i| \gtrsim |h_i|$ will be assumed.

\section{Diagonalization of the propagator matrix}		\label{sec:DiagProp}
\begin{figure}[h]
	\centering
	\subfloat[$(\Delta_\Phi)_{11} (p^2)$]{
		\includegraphics[width = 35 mm]{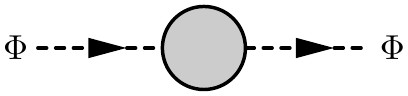}
		\label{fig:Prop11}
	} \
	\subfloat[$(\Delta_\Phi)_{22} (p^2)$]{
		\includegraphics[width = 35 mm]{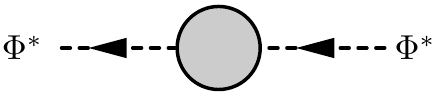}
		\label{fig:Prop22}
	} \
	\subfloat[$(\Delta_\Phi)_{12} (p^2)$]{
		\includegraphics[width = 35 mm]{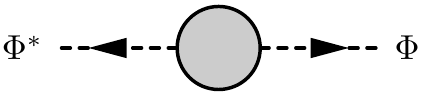}
		\label{fig:Prop12}
	} \
	\subfloat[$(\Delta_\Phi)_{21} (p^2)$]{
		\includegraphics[width = 35 mm]{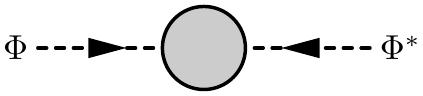}
		\label{fig:Prop21}
	}
	\caption{Components of the dressed propagator matrix. The direction of $p$ is from the left to the right.}
	\label{fig:Prop}
\end{figure}

The dressed propagator is the propagator in which all the quantum corrections are considered, and it is obtained by a geometric summation at a matrix level as follows:
\begin{align}
	i \Delta_\Phi (p^2) &= i (p^2 - M_\Phi^2)^{-1} + i (p^2 - M_\Phi^2)^{-1} [i \Sigma_\Phi (p^2)] i (p^2 - M_\Phi^2)^{-1} \nonumber \\
		&\qquad + i (p^2 - M_\Phi^2)^{-1} [i \Sigma_\Phi (p^2)] i (p^2 - M_\Phi^2)^{-1} [i \Sigma_\Phi (p^2)] i (p^2 - M_\Phi^2)^{-1} + \cdots	\label{eq:PropResum} \\
	&= \sum_{n = 0}^\infty i (p^2 - M_\Phi^2)^{-1} \big\{ [i \Sigma_\Phi (p^2)] i (p^2 - M_\Phi^2)^{-1} \big\}^n \nonumber \\
	&= i \big\{ [1 + \Sigma_\Phi' (p^2)] p^2 - M_\Phi^2 \big\}^{-1},
\end{align}
where $(M_\Phi)_{\beta \alpha} \coloneqq m_{\Phi} \delta_{\beta \alpha}$. The components of the dressed propagator are depicted in figure \ref{fig:Prop}. Note that $(\Delta_\Phi)_{\beta \alpha} (p^2)$ is associated with $\Phi_\alpha \to \Phi_\beta$.
Now we discuss how the propagator matrix can be diagonalized. First we diagonalize the self-energy matrix as follows:
\begin{align}
	\boxed{\widehat{\Sigma}_{\widehat{\Phi}}' (p^2) \coloneqq C (p^2) \Sigma_\Phi' (p^2) C^{-1} (p^2),}
\end{align}
where $C (p^2)$ is a momentum-dependent mixing matrix. Then, we can write
\begin{align}
	P^2 (p^2) \coloneqq m_\Phi^2 C (p^2) [1 + \Sigma_\Phi' (p^2)] ^{-1} C^{-1} (p^2)
	= m_\Phi^2 [1 + \widehat{\Sigma}_{\widehat{\Phi}}' (p^2)]^{-1}.
\end{align}
For simplicity, we introduce a shorthand notation for the component of a diagonal matrix: $(\widehat{\Sigma}')_{\widehat{\alpha}} (p^2) \coloneqq (\widehat{\Sigma}')_{\widehat{\alpha} \widehat{\alpha}} (p^2)$ and $P^2_{\widehat{\alpha}} (p^2) \coloneqq P^2_{\widehat{\alpha} \widehat{\alpha}} (p^2)$, where $\widehat{\alpha}$ denotes the index of physical particles associated with the components of the diagonalized propagator. It is straightforward to show that the matrix $C (p^2)$ which diagonalize the self-energy matrix also diagonalizes the propagator matrix, and the diagonalized propagator $\widehat{\Delta}_{\widehat{\Phi}} (p^2)$ can be written as
\begin{align}
	\boxed{\widehat{\Delta}_{\widehat{\Phi}} (p^2) \coloneqq C (p^2) \Delta_\Phi (p^2) C^{-1} (p^2)
	=  m_\Phi^{-2} P^2 (p^2)[p^2 - P^2 (p^2)]^{-1}.}
\end{align}
\textit{i.e.},
\begin{align}
	(\widehat{\Delta}_{\widehat{\Phi}})_{\widehat{\alpha}} (p^2) = \frac{P_{\widehat{\alpha}}^2 (p^2)}{m_\Phi^2} \frac{1}{p^2 - P_{\widehat{\alpha}}^2 (p^2)}.
	\label{eq:PropMD}
\end{align}
Let us denote the fields of physical particles corresponding to the diagonal components $(\widehat{\Delta}_{\widehat{\Phi}})_{\widehat{\alpha}}$ by $\widehat{\Phi}_{\widehat{\alpha}}$. The cross section of scattering mediated by unstable particles has the Breit-Wigner resonance pattern, and accordingly the complex pole of the propagator is in the form of
\begin{align}
	p_{\widehat{\Phi}_{\widehat{\alpha}}}^2 = m_{\widehat{\Phi}_{\widehat{\alpha}}}^2 - i m_{\widehat{\Phi}_{\widehat{\alpha}}} \Gamma_{\widehat{\Phi}_{\widehat{\alpha}}},
	\label{eq:CPole}
\end{align}
where $m_{\widehat{\Phi}_{\widehat{\alpha}}}$ and $\Gamma_{\widehat{\Phi}_{\widehat{\alpha}}}$ are the pole mass and total decay width of $\widehat{\Phi}_{\widehat{\alpha}}$. The complex mass $p_{\widehat{\Phi}_{\widehat{\alpha}}}$ is a solution of the equation
\begin{align}
	p^2 = P_{\widehat{\alpha}}^2 (p^2)
	= m_\Phi^2 [1 + (\widehat{\Sigma}_{\widehat{\Phi}}')_{\widehat{\alpha}} (p^2)]^{-1}.
\end{align}
Up to $\mathcal{O} (\alpha)$, we can write $P_{\widehat{\alpha}}^2 (p^2) = m_\Phi^2 [1 - (\widehat{\Sigma}_{\widehat{\Phi}}')_{\widehat{\alpha}} (p^2)]$, and thus
\begin{align}
	\text{Re} [(\widehat{\Sigma}_{\widehat{\Phi}}')_{\widehat{\alpha}} (p_{\widehat{\Phi}_{\widehat{\alpha}}}^2)]
		= 1 - \frac{m_{\widehat{\Phi}_{\widehat{\alpha}}}^2}{m_\Phi^2}, \qquad
	\text{Im} [(\widehat{\Sigma}_{\widehat{\Phi}}')_{\widehat{\alpha}} (p_{\widehat{\Phi}_{\widehat{\alpha}}}^2)]
		= \frac{\Gamma_{\widehat{\Phi}_{\widehat{\alpha}}}}{m_\Phi}.
	\label{eq:ImSE}
\end{align}
The residue of the pole is given by
\begin{align}
	R_{\widehat{\Phi}_{\widehat{\alpha}}}
	= |R_{\widehat{\Phi}_{\widehat{\alpha}}}| e^{i \theta_{\widehat{\Phi}_{\widehat{\alpha}}}}
	\coloneqq \lim_{p^2 \to p_{\widehat{\Phi}_{\widehat{\alpha}}}^2} (p^2 - p_{\widehat{\Phi}_{\widehat{\alpha}}}^2) (\widehat{\Delta}_{\widehat{\Phi}})_{\widehat{\alpha}} (p^2)
	= \frac{p_{\widehat{\Phi}_{\widehat{\alpha}}}^2}{m_\Phi^2} \bigg[ 1 - \frac{dP_{\widehat{\alpha}}^2}{dp^2} (p_{\widehat{\Phi}_{\widehat{\alpha}}}^2) \bigg]^{-1},
\end{align}
which is complex-valued and can be written as $R_{\widehat{\Phi}_{\widehat{\alpha}}} = 1 + \mathcal{O} (\alpha)$. Now we can write
\begin{align}
	i (\widehat{\Delta}_{\widehat{\Phi}})_{\widehat{\alpha}} (p^2) = \frac{i R_{\widehat{\Phi}_{\widehat{\alpha}}}}{p^2 - p_{\widehat{\Phi}_{\widehat{\alpha}}}^2} + \cdots.
\end{align}
Defining
\begin{align}
	\boxed{(C_{\widehat{\Phi}^f})_{\beta {\widehat{\alpha}}} \coloneqq |R_{\widehat{\Phi}_{\widehat{\alpha}}}|^\frac{1}{2} (C^{-1})_{\beta {\widehat{\alpha}}} (p_{\widehat{\Phi}_{\widehat{\alpha}}}^2), \qquad
	(C_{\widehat{\Phi}^i})_{\beta {\widehat{\alpha}}} \coloneqq |R_{\widehat{\Phi}_{\widehat{\alpha}}}|^\frac{1}{2} C_{\beta {\widehat{\alpha}}} (p_{\widehat{\Phi}_{\widehat{\alpha}}}^2),}
	\label{eq:MixingMat}
\end{align}
we can also write the components of the non-diagonal propagator as
\begin{empheq}[box=\fbox]{align}
	i (\Delta_\Phi)_{\beta \alpha} (p^2) = \sum_{\widehat{\gamma}} (C_{\widehat{\Phi}^f})_{\beta \widehat{\gamma}} \frac{i e^{i \theta_{\widehat{\Phi}_{\widehat{\gamma}}}}}{p^2 - p_{\widehat{\Phi}_{\widehat{\gamma}}}^2} (C_{\widehat{\Phi}^i})_{\alpha \widehat{\gamma}} + \cdots,
		\label{eq:PropPMDExp}
\end{empheq}
where the ellipsis denotes the higher-order terms of the Laurent expansion. Note that diagonalization discussed here is independent of the expression of the self-energy. Even though the self-energy was renormalized such that $\Sigma_\Phi (p^2) = p^2 \Sigma_\Phi' (p^2)$, there existds no loss of generality since it is always possible with appropriately chosen $M_\Phi$. In other words, the diagonalization of the propagator we have discussed is exact up to the infinite order in perturbation.

\section{Generation of quasiparticles and limitations of quantum mechanics}		\label{sec:Quasi}
Now we discuss the properties of physical particles. The two-point correlation function associated with each component of the non-diagonal dressed propagator can be written as
\begin{align}
	\int \frac{d^4 p}{(2 \pi)^4} e^{-i p \cdot (x - y)} i (\Delta_\Phi)_{\beta \alpha} (p^2)
		= \langle \Omega | \Phi_\beta (x) \Phi_\alpha^\dag (y) | \Omega \rangle \quad (x^0 > y^0),
\end{align}
where $\Phi_1 = \Phi$, $\Phi_2 = \Phi^\dag (= \Phi^*)$. The fields in the correlation function on the right-hand side should have already been time-ordered since each component of the propagator assumes a specific direction of momentum transfer as mentioned before. Using equation \ref{eq:PropPMDExp}, we write
\begin{align}
	\boxed{\int \frac{d^4 p}{(2 \pi)^4} e^{-i p \cdot (x - y)} \frac{i e^{i \theta_{\widehat{\Phi}_{\widehat{\alpha}}}}}{p^2 - p_{\widehat{\Phi}_{\widehat{\alpha}}}^2} + \cdots
	= \langle \Omega | \widehat{\Phi}_{\widehat{\alpha}}^f (x) \widehat{\Phi}_{\widehat{\alpha}}^{i \dag} (y) | \Omega \rangle \quad (x^0 > y^0),}
	\label{eq:CorrD}
\end{align}
where
\begin{align}
	\boxed{\widehat{\Phi}_{\widehat{\alpha}}^i = \sum_\beta (C_{\widehat{\Phi}^i}^{-1})_{\widehat{\alpha} \beta}^* \Phi_\beta, \qquad
	\widehat{\Phi}_{\widehat{\alpha}}^f = \sum_\beta (C_{\widehat{\Phi}^f}^{-1})_{\widehat{\alpha} \beta} \Phi_\beta.}
\end{align}
Note that $\widehat{\Phi}_{\widehat{\alpha}}^f \neq \widehat{\Phi}_{\widehat{\alpha}}^i$ since $C$ in equation \ref{eq:MixingMat} is non-unitary because of the absorptive part of the self-energy. Calculating the Fourier transform in the rest frame and neglecting the subdominant part in equation \ref{eq:CorrD}, we can write
\begin{align}
	e^{-i p_{\widehat{\Phi}_{\widehat{\alpha}}} (x^0 - y^0)}
	\approx e^{-i m_{\widehat{\Phi}_{\widehat{\alpha}}} (x^0 - y^0)} e^{-\frac{1}{2} \Gamma_{\widehat{\Phi}_{\widehat{\alpha}}} (x^0 - y^0)}
	\propto \langle \Omega | \widehat{\Phi}_{\widehat{\alpha}}^f (x) \widehat{\Phi}_{\widehat{\alpha}}^{i \dag} (y) | \Omega \rangle \quad (x^0 > y^0).
	\label{eq:CorrDExp}
\end{align}
Hence, \textit{each of physical particles, which propagates like a free particle in the interacting theory until it decays, emerges as an excitation of $\widehat{\Phi}_{\widehat{\alpha}}^i$ and ends as an excitation of $\widehat{\Phi}_{\widehat{\alpha}}^f~(\widehat{\Phi}_{\widehat{\alpha}}^f \neq \widehat{\Phi}_{\widehat{\alpha}}^i)$}, and it cannot be related to a single renormalized field. Furthermore, $\widehat{\Phi}_{\widehat{\alpha}}^f$ and $\widehat{\Phi}_{\widehat{\alpha}}^i$ can be much different from each other because of the non-perturbative effect in the loop corrections to the fields of unstable particles with small mass differences, which can be understood from the following diagram:
\begin{align*}
	\includegraphics[width = 75 mm]{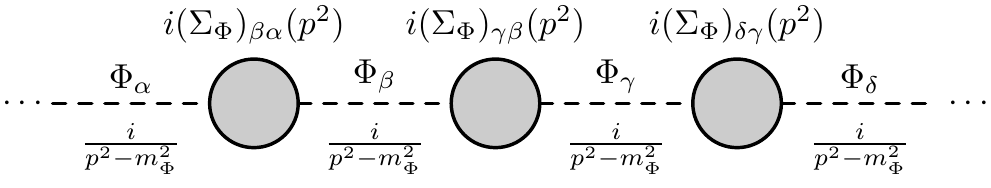}
\end{align*}
When $\widehat{\Phi}_{\widehat{\alpha}}$ is on-shell, we have $p^2 \sim m_\Phi^2$ and the factor $(\Sigma_\Phi)_{\beta \alpha} (p^2) / (p^2 - m_\Phi^2)$ can be hugely enhanced to generate a mixing effect beyond the typical perturbative correction of the theory. The effect of mass differences and the comparison with the case of stable particles are discussed in more details in reference \cite{MixingRen}. Note that, the larger the difference between $\widehat{\Phi}_{\widehat{\alpha}}^f$ and $\widehat{\Phi}_{\widehat{\alpha}}^i$ is, the larger CP asymmetry there exists in the decay of an unstable particle in general. When the CP asymmetry is $\mathcal{O} (1)$, the difference between $C_{\widehat{\Phi}^f}$ and $ C_{\widehat{\Phi}^i}^*$ is very large, \textit{i.e.}, $C_{\widehat{\Phi}^f}^{-1} C_{\widehat{\Phi}^i}^* = 1 + \mathcal{O} (1)$.

It is an emergent phenomenon dynamically generated by interactions, and each of physical particles should be interpreted as a \textit{quasiparticle}. It cannot simply be called a mass eigenstate, not only because it is defined by a specific total decay width as well as mass given by equation \ref{eq:CPole}, but also because the physics of a quasiparticle is much different from that of a mass eigenstate as a one-particle state in the Hilbert space. It should be emphasized that this emergent phenomenon has never been appropriately considered in the literature, where a mass eigenstate has always been regarded as a specific linear combination of a particle and its antiparticle which follows the time evolution of a plane wave with a damping factor, \textit{i.e.}, $e^{-i p_{\widehat{\Phi}_{\widehat{\alpha}}} (t - t_0)} \approx e^{-i m_{\widehat{\Phi}_{\widehat{\alpha}}} (t - t_0)} e^{-(\Gamma_{\widehat{\Phi}_{\widehat{\alpha}}} / 2) (t - t_0)}$ in the rest frame. In other words, the conventional approach of solving the effective Schr\"odinger equation to obtain a time evolution cannot work, since the equation of motion we obtain after diagonalizing the effective Hamiltonian is always $i (d / dt) | \widehat{\Phi}_{\widehat{\alpha}} (t) \rangle = p_{\widehat{\Phi}_{\widehat{\alpha}}} | \widehat{\Phi}_{\widehat{\alpha}} (t) \rangle$ and its solution is nothing but $| \widehat{\Phi}_{\widehat{\alpha}} (t) \rangle = e^{-i p_{\widehat{\Phi}_{\widehat{\alpha}}} (t - t_0)} | \widehat{\Phi}_{\widehat{\alpha}} (t_0) \rangle$, where $| \widehat{\Phi}_{\widehat{\alpha}} (t_0) \rangle$ is the mass eigenstate which is a fixed linear combination of basis states.

\begin{figure}[t]
	\centering
	\includegraphics[width = 40 mm]{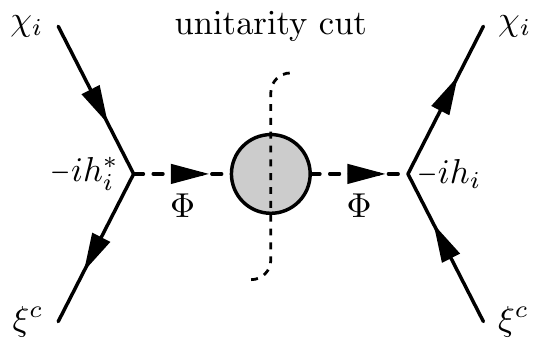}
	\caption{Applying the unitarity cut to the scattering process $\chi_i \xi^c \to \chi_i \xi^c$.}
	\label{fig:DWUnitCut}
\end{figure}

\begin{figure}[t]
	\centering
	\parbox{26 mm}{\includegraphics[width = 26 mm]{FigVPCXcTree}} + 
	\parbox{26 mm}{\includegraphics[width = 26 mm]{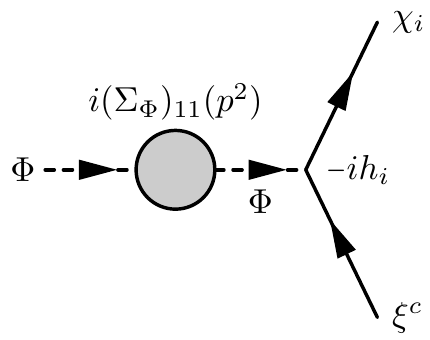}} +
	\parbox{39 mm}{\includegraphics[width = 39 mm]{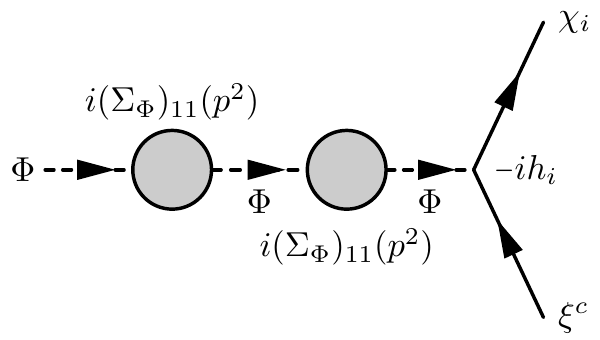}} +
	\parbox{39 mm}{\includegraphics[width = 39 mm]{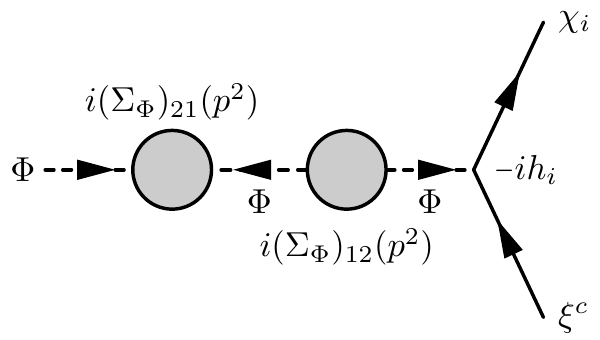}} + $\cdots$
	\caption{Loop corrections to $\Phi \to \chi_i \xi^c$ associated with particle-antiparticle mixing.}
	\label{fig:DWtoCXcAll}
\end{figure}

Another way to see that each of physical particles cannot be associated with a single one-particle state in the Hilbert space is taking the unitarity cut of a scattering process mediated by the unstable particles. Let us consider the diagram in figure \ref{fig:DWUnitCut}.
The unitarity cut generates all possible initial states which produce $| \chi_i \xi^c \rangle$ as the final state. The one-particle state $| \Phi \rangle$, however, does not contribute to the unitarity cut, and only multiparticle states of stable particles such as $| \chi_j \xi^c \rangle$, $| \chi_j^c \xi \rangle$, and $| \psi_j \psi_j^c \rangle$ contribute, as was proved for a single flavor of an unstable particle in reference \cite{UnitUnsP}. To understand that, let us consider the sum of diagrams in figure \ref{fig:DWtoCXcAll}, which are the diagrams we would obtain from the cutting through $\Phi$.
For simplicity, only the loop corrections causing particle-antiparticle mixing are considered here. When the mass difference between physical particles is small, vertex loop corrections are even completely negligible, as we will see later. This sum is proportional to the series
\begin{align}
	1 + \frac{i}{p^2 - m_\Phi^2} i (\Sigma_\Phi)_{11} (p^2) + \sum_\alpha \frac{i}{p^2 - m_\Phi^2} i (\Sigma_\Phi)_{1 \alpha} (p^2) \frac{i}{p^2 - m_\Phi^2} i (\Sigma_\Phi)_{\alpha 1} (p^2) + \cdots.
\end{align}
First note that this approach fails to work if only a finite number of diagrams in figure \ref{fig:DWtoCXcAll} is considered as a perturbative approximation, since the internal propagators of $\Phi$ diverge as $\Phi$ goes on-shell. The standard solution to this problem is summing over all the diagrams in figure \ref{fig:DWtoCXcAll} for off-shell $\Phi$, and taking analytic continuation of the resulting expression to the on-shell region where it is convergent because the physical poles of the dressed propagator is complex-valued. Using equation \ref{eq:PropResum} and $p^2 = m_\Phi^2$ for on-shell $\Phi$, we can rewrite it as
\begin{align}
	1 + \sum_\alpha i (\Delta_\Phi)_{1 \alpha} (p^2) i (\Sigma_\Phi)_{\alpha 1} (p^2)
	&= 1 - \sum_\alpha (\Delta_\Phi)_{1 \alpha} (p^2) \big[ (p^2 - m_\Phi^2) \delta_{\alpha 1} + (\Sigma_\Phi)_{\alpha 1} (p^2) \big] \nonumber \\
	&= 1 - \sum_\alpha (\Delta_\Phi)_{1 \alpha} (p^2) (\Delta_\Phi^{-1})_{\alpha 1} (p^2)
	= 0,
\end{align}
and thus the one-particle state $| \Phi \rangle$ does not contribute to the unitarity cut. In other words, the physical unstable particle cannot simply be related to a single one-particle state in the Hilbert space. Hence, the standpoint on a decay as described in figure \ref{fig:DWtoCXcAll}, \textit{i.e.}, a one-particle state of an unstable particle evolves in time undergoing quantum corrections and then decays into other particles, is not the correct view of a decay, especially when the unstable particles go through particle-antiparticle or flavor mixing. That is to say, it is inappropriate to consider the $S$-matrix element such as $\langle \chi_i \xi^c | \, S \, | \Phi \rangle$ to describe a decay of $\Phi$, although the decay width of an unstable particle is often calculated in such a way, \textit{i.e.}, equation \ref{eq:DWRest}, without rigorous justification. In reference \cite{MixingRen}, the limitation of such an approach with regards to the Lehmann-Symanzik-Zimmermann (LSZ) reduction formula and the difference from the case without mixing are discussed in detail. As a matter of fact, an unstable particle as an asymptotic state at $t = \minus \infty$ should have already decayed before any interaction occurs at $t = t_0$. Only stable particles can be regarded as external states to properly relate the correlation function of a physical process to a non-zero finite $S$-matrix element by the LSZ reduction formula.

We therefore conclude that \textit{quantum mechanics is not the proper non-relativistic limit of the quantum field theory especially in the presence of particle-antiparticle or flavor mixing of unstable particles.} Now we must find a way outside quantum mechanics to deal with this emergent phenomenon.

\section{Derivations of decay widths in the quantum field theory}		\label{sec:DerivDW}
We have seen that the properties of physical particles should be investigated in the quantum field theory, not in quantum mechanics. As quasiparticles dynamically generated by interactions, they cannot be regarded as external states and their properties can be properly studied only from a physical process in which they appear as intermediate states. In this section, we define and calculate some transition rates associated with scattering mediated by unstable particles in various ways, and relate them to their decay widths. We will derive the decay widths, considering what should be on-shell is not $\Phi$ but $\widehat{\Phi}_{\widehat{\alpha}}$, \textit{i.e.}, the fields of physical particles. Furthermore, the ``decay width" such as $\Gamma_{\Phi \to \chi_i \xi^c}$ will also be defined with ``off-shell" $\Phi$ in the sense that $p^2 = m_{\widehat{\Phi}_{\widehat{\alpha}}}^2 \neq m_\Phi^2$.

\subsection{From the self-energy}
The \textit{total decay width} of each of physical particles has been obtained while diagonalizing the dressed propagator, and it is given by
\begin{align}
	\boxed{\Gamma_{\widehat{\Phi}_{\widehat{\alpha}}} = m_\Phi \text{Im} \big[ (\widehat{\Sigma}_{\widehat{\Phi}}')_{\widehat{\alpha}} (m_\Phi^2) \big]
		= m_\Phi \text{Im} \big[ (C \Sigma_\Phi' C^{-1})_{\widehat{\alpha} \widehat{\alpha}} (m_\Phi^2) \big].}
	\label{eq:TotDecWth}
\end{align}
However, it tells nothing about its partial decay width to a specific decay channel. Even though the total decay width is the sum of all the partial decay widths and it is apparent that the contribution of each loop to the self-energy, \textit{e.g.}, $\Sigma_\Phi'^{\chi_i \xi} (m_\Phi^2)$, can be easily identified, it is still not possible to directly find the partial decay width from the self-energy. For example, we might try to replace $\Sigma_\Phi' (m_\Phi^2)$ with $\Sigma_\Phi'^{\chi_i \xi} (m_\Phi^2)$ to find $\Gamma_{\widehat{\Phi}_{\widehat{\alpha}} \to \chi_i \xi^c}$, but it does not work since $(C \Sigma_\Phi'^{\chi_i \xi} C^{-1})_{\widehat{\alpha} \widehat{\alpha}} (m_\Phi^2)$ is no longer diagonal. Only when the self-energy is diagonal, each component can be interpreted as a physical quantity that belongs to each of physical particles. This happens since the mixing matrix $C$ itself is also a complicated function of the self-energy, and thus there does not exist a simple way to identify the contribution of specific final states to $\Gamma_{\widehat{\Phi}_{\widehat{\alpha}}}$. Now we discuss various methods of calculating the \textit{partial decay width} in the quantum field theory.

\subsection{From the on-shell contributions to scattering}

\subsubsection{Overview}
The first method is to read the partial decay width from scattering mediated by physical particles. This method is intuitive and practical, since it is exactly how the unstable particles are observed in an experiment. The strategy is as follows:

\begin{enumerate}
	\item For a specific final multiparticle state of the decay product, consider all possible initial states that can produce it through $s$-channel scattering mediated by unstable particles.
	\item Integrate over all the phase spaces of the initial and final states with delta functions appropriately inserted.
\end{enumerate}

\begin{figure}[t]
	\centering
	\subfloat[$\Phi \to \chi_i \xi^c$]{
		\includegraphics[width = 29.3 mm]{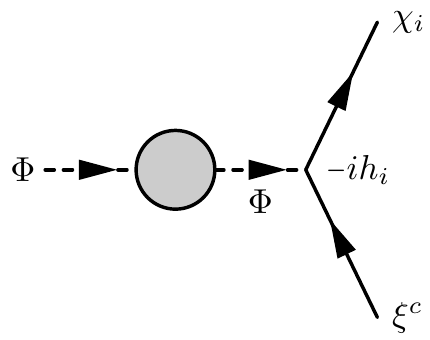}
		\label{fig:DWPCXc}
	} \qquad
	\subfloat[$\Phi^* \to \chi_i \xi^c$]{
		\includegraphics[width = 29.3 mm]{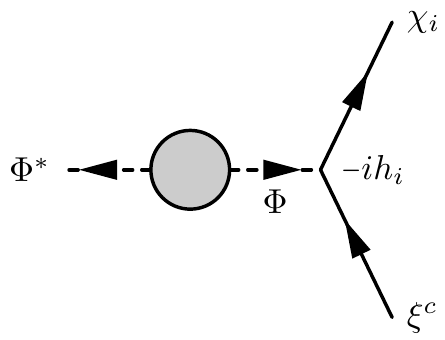}
		\label{fig:DWPcCXc}
	}
	\caption{These diagrams do not contribute to the unitarity cut of $\chi_i \xi^c \to \chi_i \xi^c$.}
	\label{fig:DWtoCXc1}
	\subfloat[$\psi_j \psi_j^c \to \chi_i \xi^c$]{
		\includegraphics[width = 35 mm]{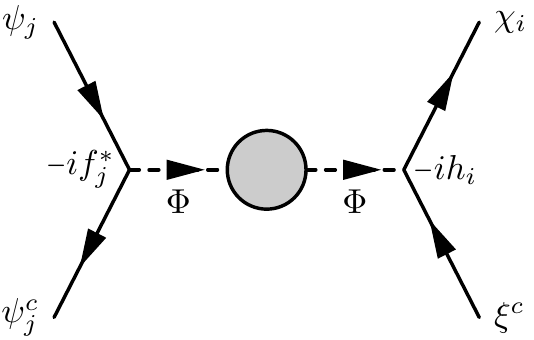}
		\label{fig:DWPsPsPCXc}
	} \
	\subfloat[$\psi_j \psi_j^c \to \chi_i \xi^c$]{
		\includegraphics[width = 35 mm]{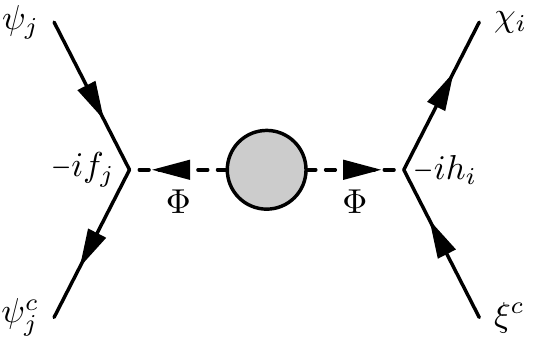}
		\label{fig:DWPsPsPcCXc}
	} \
	\subfloat[$\chi_j \xi^c \to \chi_i \xi^c$]{
		\includegraphics[width = 35 mm]{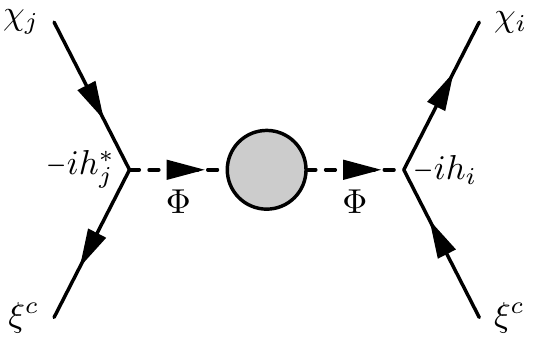}
		\label{fig:DWCXcCXc}
	} \
	\subfloat[$\chi_j^c \xi \to \chi_i \xi^c$]{
		\includegraphics[width = 35 mm]{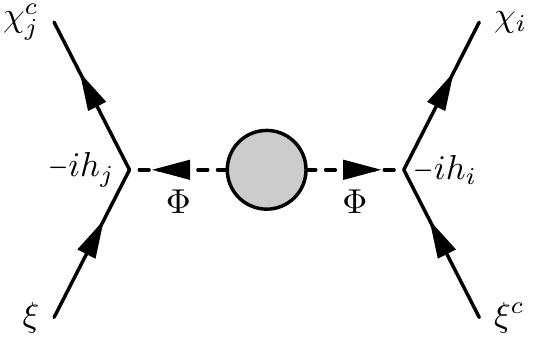}
		\label{fig:DWCcXCXc}
	}
	\caption{Dominant contributions to the decay width associated with final state $| \chi_i \xi^c \rangle$.}
	\label{fig:DWtoCXc2}
\end{figure}

To explicitly calculate the decay width, let us consider the case with decay product $\chi_i \xi^c$. The associated initial states and Feynman diagrams are shown in figures \ref{fig:DWtoCXc1} and \ref{fig:DWtoCXc2}.
The diagrams in figure \ref{fig:DWtoCXc1} tells us how the phase space integration should be performed to obtain the decay width. The correct choice is to apply the phase space integration of the final state given by
\begin{align}
	\int d\Pi_{\chi_i} \int d\Pi_{\xi^c} \, (2 \pi)^3 \delta^3 (\mathbf{p} - \mathbf{p}_{\chi_i} - \mathbf{p}_{\xi^c})
\end{align}
to $|\mathcal{M}|^2$ integrated over the phase space of the initial state
\begin{align}
	\int d\Pi_{\Phi_\alpha} \, (2 \pi)^4 \delta^4 (p_{\Phi_\alpha} - p_{\chi_i} - p_{\xi^c}) \, |\mathcal{M} (\Phi_\alpha \to \chi_i \xi^c)|^2,
	\label{eq:CutP}
\end{align}
where
\begin{align}
	d\Pi_X \coloneqq \prod_{j \in X} \frac{d^3 \textbf{p}_j}{(2 \pi)^3 2 E_j}.
\end{align}
Its straightforward calculation gives
\begin{align}
	\frac{1}{2 E_{\Phi_\alpha}} \int d\Pi_{\chi_i} \int d\Pi_{\xi^c} \, (2 \pi)^4 \delta (E_{\Phi_\alpha} - E_{\chi_i} - E_{\xi^c}) \delta^3 (\mathbf{p} - \mathbf{p}_{\chi_i} - \mathbf{p}_{\xi^c}) \, \big| \mathcal{M} [\Phi_\alpha (\mathbf{p}) \to \chi_i \xi^c] \big|^2.
	\label{eq:DWBoost}
\end{align}
Setting $\mathbf{p} = 0$, we obtain the familiar expression of the partial decay width:
\begin{align}
	\Gamma_{\Phi_\alpha \to \chi_i \xi^c} = \frac{1}{2 m_\Phi} \int d\Pi_{\chi_i} \int d\Pi_{\xi^c} \, (2 \pi)^4 \delta^4 (p_{\chi_i} + p_{\xi^c} - p_{\Phi_\alpha}) \, |\mathcal{M} (\Phi_\alpha \to \chi_i \xi^c)|^2,
	\label{eq:DWRest}
\end{align}
and the expression for $\mathbf{p} \neq 0$ given by equation \ref{eq:DWBoost} is the partial decay width in a boosted frame. Even though we have used the diagrams of figure \ref{fig:DWtoCXc1} to choose the right phase space integration to obtain the decay width, their contributions actually vanish when the loop corrections are fully considered, as mentioned before. The decay width given by equation \ref{eq:DWRest} gives a correct result only in the case without mixing when the loop corrections to the fields of unstable particles are neglected on purpose. Such an approach is never rigorously justifiable, and inconsistent with the LSZ reduction formula, as discussed in reference \cite{MixingRen}.

We therefore claim that the real contributions to the decay width come from the scattering processes in figure \ref{fig:DWtoCXc2}. To be specific, in the case of scattering $\chi_j^c \xi \to \chi_i \xi^c$ in figure \ref{fig:DWCcXCXc}, we apply the integration
\begin{align}
	\int d\Pi_{\chi_i} \int d\Pi_{\xi^c} \, (2 \pi)^3 \delta^3 (\mathbf{p} - \mathbf{p}_{\chi_i} - \mathbf{p}_{\xi^c})
	\label{eq:PhSpInt}
\end{align}
to
\begin{align}
	\frac{1}{4} \int d\Pi_{\chi_j^c} \int d\Pi_\xi \, (2 \pi)^4 \delta^4 (p_{\chi_j^c} + p_\xi - p_{\chi_i} - p_{\xi^c}) \, |\mathcal{M} (\chi_j^c \xi \to \chi_i \xi^c)|^2
	\label{eq:CutCcX}
\end{align}
to calculate its contribution to the associated partial decay width. The factor $1 / 4$ is the spin average over the initial state.

Note that this phase space integration over all the initial states except for the spin average is equivalent to applying the unitarity cut to all possible internal states of the scattering $\chi_i \xi^c \to \chi_i \xi^c$ as shown in figure \ref{fig:DWUnitCut}. The cutting through $\Phi_\alpha$ do not contribute to this unitarity cut, which is why the diagrams in figure \ref{fig:DWtoCXc1} are irrelevant to our purpose, as mentioned above. Only the cutting applied to the intermediate stable multiparticle states need to be considered, and the resulting diagrams are those in figure \ref{fig:DWtoCXc2}. The decay width should therefore be calculated from those diagrams. In this method, the initial and final states of figure \ref{fig:DWUnitCut} are chosen to be the decay product $\chi_i \xi^c$, and the unitarity cut gives all possible initial states of the scattering processes as shown in figure \ref{fig:DWtoCXc2}. This is a convenient choice since the focus is how to calculate the partial decay width to a specific final state. Note the difference from the usual approach to calculate a decay width with the optical theorem, in which the decaying particle is chosen as the initial and final states of the diagram before cutting and the unitarity cut gives all possible decay products.

So far it has been unclear how the diagrams in figure \ref{fig:DWtoCXc2} can be related to the decay widths of unstable particles, as claimed above. In the following sections, we will obtain some  transition rates associated with those diagrams applying the phase space integration discussed above, and show that they can be related to the decay widths of intermediate unstable particles in a specific way.

\subsubsection{Derivation of transition rates}		\label{sec:TransRates}
Now we discuss how to derive the expressions of partial decay widths of unstable particles. In this paper, all the calculations of decay widths will be performed up to the \textit{leading order} in perturbation. Here, the leading order means the followings:
\begin{enumerate}
	\item The \textit{one-loop order} in the \textit{self-energy}.
	\item The \textit{tree level} in the \textit{vertices}.
\end{enumerate}
The reason why these conditions correspond to the leading order will be clarified after we derive the first expression of transition rates, \textit{i.e.}, equation \ref{eq:DWCcXCXc}. Since $R_{\widehat{\Phi}_{\widehat{\alpha}}} = 1$ and $\Sigma_\Phi (p_{\widehat{\Phi}_{\widehat{\alpha}}}^2) = \Sigma_\Phi (m_\Phi^2)$ up to the leading order, we can write
\begin{align}
	\boxed{C_{\widehat{\Phi}^f} = C^{-1} (m_\Phi^2), \qquad
	C_{\widehat{\Phi}^i} = C (m_\Phi^2) \eqqcolon C_\Phi}
\end{align}
and
\begin{align}
	\boxed{\int \frac{d^4 p}{(2 \pi)^4} e^{-i p \cdot (x - y)} \frac{i}{p^2 - p_{\widehat{\Phi}_{\widehat{\alpha}}}^2} + \cdots
	= \langle \Omega | \widehat{\Phi}_{\widehat{\alpha}}^f (x) \widehat{\Phi}_{\widehat{\alpha}}^{i \dag} (y) | \Omega \rangle \quad (x^0 > y^0)}
\end{align} \\

In order to calculate the unitarity cut, it is convenient to use the $\mathcal{T}$ matrix, which is the non-trivial part of the $S$ matrix:
\begin{align}
	S = 1 + i \mathcal{T}.
\end{align}
For transition $i \to f$ $(i \neq f)$, the $\mathcal{T}$-matrix element is written as
\begin{align}
	\langle f | \mathcal{T} | i \rangle = (2 \pi)^4 \delta^4 (p_i - p_f) \mathcal{M} (i \to f).
\end{align}
Denoting the one- and multi-particle states in the Fock space by $| X^a \rangle$ where $a$ indicates the internal degrees of freedom of $X$, we can write the completeness relation as
\begin{align}
	1 = | 0 \rangle \langle 0 | + \sum_{a, X} \int d\Pi_X | X^a \rangle \langle X^a |,
	\label{eq:ComRel}
\end{align}
where $| 0 \rangle$ is a vacuum state. The unitarity of $S$, \textit{i.e.}, $1 = S S^\dag$, implies
\begin{align}
	i (\mathcal{T}^\dag - \mathcal{T}) = \mathcal{T} \mathcal{T}^\dag.
	\label{eq:UnitCon}
\end{align}

In this section, we consider only the right-hand side of equation \ref{eq:UnitCon}. Cutting through the intermediate states of $\chi_i \xi^c \to \chi_i \xi^c$ is equivalent to inserting the completeness relation given by equation \ref{eq:ComRel} as follows:
\begin{align}
	\langle &\chi_{i, r} \xi^c_s | \mathcal{T} \mathcal{T}^\dag | \chi_{i, r} \xi^c_s \rangle
	= \sum_{X, a} \int d\Pi_X \, \langle \chi_{i, r} \xi^c_s | \mathcal{T} | X^a \rangle \langle X^a | \mathcal{T}^\dag | \chi_{i, r} \xi^c_s \rangle \nonumber \\
	&= (2 \pi)^4 \delta^4 (0)
		\bigg\{ \sum_{j, k, l} \int d\Pi_{\psi_j} \int d\Pi_{\psi_j^c} \, (2 \pi)^4 \delta^4 (p_{\psi_j^c} + p_{\psi_j} - p_{\chi_i} - p_{\xi^c}) \, \big| \mathcal{M} (\psi_{j, k} \psi_{j, l}^c \to \chi_{i, r} \xi^c_s) \big|^2 \nonumber \\
		&\qquad + \sum_{j, k, l} \int d\Pi_{\chi_j}' \int d\Pi_{\xi^c}' \, (2 \pi)^4 \delta^4 (p_{\chi_j}' + p_{\xi^c}' - p_{\chi_i} - p_{\xi^c}) \, \big| \mathcal{M} [\chi_{j, k} (p_{\chi_j}') \xi^c_l (p_{\xi^c}') \to \chi_{i, r} \xi^c_s] \big|^2 \nonumber \\
		&\qquad + \sum_{j, k, l} \int d\Pi_{\chi_j^c} \int d\Pi_\xi \, (2 \pi)^4 \delta^4 (p_{\chi_j^c} + p_\xi - p_{\chi_i} - p_{\xi^c}) \, \big| \mathcal{M} (\chi^c_{j, k} \xi_l \to \chi_{i, r} \xi^c_s) \big|^2 \bigg\} \nonumber \\
		&\qquad + \cdots,
	\label{eq:UnitRH}
\end{align}
where $r, s, k, l$ are the spin indices of light fermions. The factor $(2 \pi)^4 \delta^4 (0)$ is generated due to the fact that we are considering total time $T$ and volume $V$, \textit{i.e.}, $(2 \pi)^4 \delta^4 (0) = TV$, and it is canceled out in observable quantities \cite{Peskin&Schroeder}, which we do not discuss in detail. The ellipsis denotes subdominant contributions from the multiparticles states other than those written above. Those multiparticle states come from cutting the self-energy diagrams of unstable particles with more than one loop, and thus they can be neglected up to the one-loop order. As mentioned above, we integrate over the phase space of $\chi_{i, r} \xi^c_s$ as follows:
\begin{align}
	\boxed{\int d\Pi_{\chi_i} \int d\Pi_{\xi^ c} \, (2 \pi)^3 \delta^3 (\mathbf{p} - \mathbf{p}_{\chi_i} - \mathbf{p}_{\xi^c}) \sum_{r, s} \langle \chi_{i, r} \xi^c_s | \mathcal{T} \mathcal{T}^\dag | \chi_{i, r} \xi^c_s \rangle.}
	\label{eq:DWUnitRH}
\end{align}

First we consider the third term in equation \ref{eq:UnitRH} with initial state $| \chi_j^c \xi \rangle$. Rearranging the phase space integration and delta functions, we can write
\begin{align}
	\frac{1}{4} \sum_j &\int d\Pi_{\chi_i} \int d\Pi_{\xi^c} \int d\Pi_{\chi_j^c} \int d\Pi_\xi \, (2 \pi)^3 \delta^3 (\mathbf{p} - \mathbf{p}_{\chi_i} - \mathbf{p}_{\xi^c}) (2 \pi)^4 \delta^4 (p_{\chi_j^c} + p_\xi - p_{\chi_i} - p_{\xi^c}) \nonumber \\
		&\qquad \qquad \sum_{r, s, k, l} \big| \mathcal{M} (\chi^c_{j, k} \xi_l \to \chi_{i, r} \xi^c_s) \big|^2 \nonumber \\
	&= \boxed{\frac{1}{4} \sum_j \int d\Pi_{\chi_j^c} \int d\Pi_\xi \, (2 \pi)^3 \delta^3 (\mathbf{p} - \mathbf{p}_{\chi_j^c} - \mathbf{p}_\xi) \, \sigma' (\chi_j^c \xi \to \chi_i \xi^c),}
	\label{eq:DWUnitRHDiv}
\end{align}
where we have defined
\begin{align}
	\sigma' (\chi_j^c \xi \to \chi_i \xi^c) &\coloneqq \int d\Pi_{\chi_i} \int d\Pi_{\xi^c} \, (2 \pi)^4 \delta^3 (\mathbf{p} - \mathbf{p}_{\chi_i} - \mathbf{p}_{\xi^c}) \delta (E_{\chi_j^c} + E_\xi - E_{\chi_i} - E_{\xi^c}) \nonumber \\
			&\qquad \qquad \qquad \sum_{r, s, k, l} \big| \mathcal{M} (\chi^c_{j, k} \xi_l \to \chi_{i, r} \xi^c_s) \big|^2.
\end{align}
This is a dimensionless quantity related to the scattering cross section $\sigma$ as follows:
\begin{align}
	\sigma' (\chi_j^c \xi \to \chi_i \xi^c) = 2 E_{\chi_j^c} \, 2 E_\xi \big| \mathbf{v}_{\chi_j^c} - \mathbf{v}_\xi \big| \, 4 \, \sigma (\chi_j^c \xi \to \chi_i \xi^c),
\end{align}
where the factor 4 is relevant to the spin average of initial state $\chi_j^c \xi$ in $\sigma$. The expression given by equation \ref{eq:DWUnitRHDiv} is in fact divergent in the limit $E = E_{\chi_j^c} + E_\xi \to \infty$ in the center-of-momentum (CM) frame, and such a limit is a highly off-shell contribution which should be irrelevant to the decay widths of intermediate particles. We have to consider only the on-shell contributions to equation \ref{eq:DWUnitRHDiv} to relate them to decay widths, and will discuss how to do it soon.

The scattering amplitude for $\chi^c_{j, k} \xi_l \to \chi_{i, r} \xi^c_s$ is written as
\begin{align}
	i \mathcal{M} (\chi^c_{j, k} \xi_l \to \chi_{i, r} \xi^c_s)
	&= \overline{u_{\chi_i}^r} (\mathbf{p}_{\chi_i}) v_\xi^s (\mathbf{p}_{\xi^c}) (\minus i h_i) [i \Delta_{12} (p^2)] (\minus i h_j) \overline{v_{\chi_j}^k} (\mathbf{p}_{\chi_j}') v_\xi^l (\mathbf{p}_{\xi^c}') \nonumber \\
	&= \minus i \overline{u_{\chi_i}^r} (\mathbf{p}_{\chi_i}) v_\xi^s (\mathbf{p}_{\xi^c}) \overline{v_{\chi_j}^k} (\mathbf{p}_{\chi_j}') v_\xi^l (\mathbf{p}_{\xi^c}')
		\sum_{\widehat{\alpha}} \frac{Q^{\chi_j^c \xi}_{\widehat{\Phi}_{\widehat{\alpha}} \to \chi_i \xi^c}}{p^2 - p_{\widehat{\Phi}_{\widehat{\alpha}}}^2} + \cdots,
\end{align}
where we have defined a coefficient
\begin{align}
	Q^{\chi_j^c \xi}_{\widehat{\Phi}_{\widehat{\alpha}} \to \chi_i \xi^c}
	\coloneqq h_i h_j (C_\Phi^{-1})_{1 \widehat{\alpha}} (C_\Phi)_{2 \widehat{\alpha}}.
\end{align}
A straightforward calculation in the CM frame gives
\begin{align}
	\sigma'_\text{CM} (\chi_j^c \xi \to \chi_i \xi^c)
	= \frac{1}{2 \pi} \sum_{\widehat{\alpha}, \widehat{\beta}} Q^{\chi_j \xi^c *}_{\widehat{\Phi}_{\widehat{\beta}} \to \chi_i^c \xi} Q^{\chi_j \xi^c}_{\widehat{\Phi}_{\widehat{\alpha}} \to \chi_i^c \xi}
		\frac{(E^2 - m_{\chi_j}^2) (E^2 + m_{\chi_i}^2) (E^2 - m_{\chi_i}^2)}{E^2 (E^2 - p_{\widehat{\Phi}_{\widehat{\beta}}}^{* 2}) (E^2 - p_{\widehat{\Phi}_{\widehat{\alpha}}}^2)},
\end{align}
where $E$ denotes the total energy. Defining
\begin{align}
	m_{\chi_{ij}} \coloneqq \left\{ \begin{array}{ll} m_{\chi_i}, & \ m_{\chi_i} \geq m_{\chi_j}, \\
		m_{\chi_j}, & \ m_{\chi_i} < m_{\chi_j}, \end{array} \right.
\end{align}
we integrate the energy-dependent part over the phase space of the initial state to obtain
\begin{align}
	\int &d\Pi_{\chi_j^c} \int d\Pi_\xi \, (2 \pi)^3 \delta^3 (\mathbf{p}_{\chi_j^c} + \mathbf{p}_\xi)
		\frac{(E^2 - m_{\chi_j}^2) (E^2 + m_{\chi_i}^2) (E^2 - m_{\chi_i}^2)}{E^2 (E^2 - p_{\widehat{\Phi}_{\widehat{\beta}}}^{* 2}) (E^2 - p_{\widehat{\Phi}_{\widehat{\alpha}}}^2)} \nonumber \\
	&= \frac{1}{16 \pi^2} \int_{m_{\chi_{ij}}}^\infty dE \frac{(E^2 + m_{\chi_j}^2) (E^2 - m_{\chi_j}^2) (E^2 + m_{\chi_i}^2) (E^2 - m_{\chi_i}^2)}{E^4 (E^2 - p_{\widehat{\Phi}_{\widehat{\beta}}}^{* 2}) (E^2 - p_{\widehat{\Phi}_{\widehat{\alpha}}}^2)}
		\label{eq:DWScatCXcIntE} \\
	&= \minus \frac{1}{16 \pi^2 (p_{\widehat{\Phi}_{\widehat{\beta}}}^{* 2} - p_{\widehat{\Phi}_{\widehat{\alpha}}}^2)}
		\int_{m_{\chi_{ij}}}^\infty dE \frac{1}{E^4} (E^2 + m_{\chi_j}^2) (E^2 - m_{\chi_j}^2) (E^2 + m_{\chi_i}^2) (E^2 - m_{\chi_i}^2) \nonumber \\
		&\qquad \qquad \qquad \qquad \qquad \quad \bigg[ \frac{E^2 - m_{\widehat{\Phi}_{\widehat{\alpha}}}^2 - i m_{\widehat{\Phi}_{\widehat{\alpha}}} \Gamma_{\widehat{\Phi}_{\widehat{\alpha}}}}{(E^2 - m_{\widehat{\Phi}_{\widehat{\alpha}}}^2)^2 + (m_{\widehat{\Phi}_{\widehat{\alpha}}} \Gamma_{\widehat{\Phi}_{\widehat{\alpha}}})^2} - \frac{E^2 - m_{\widehat{\Phi}_{\widehat{\beta}}}^2 + i m_{\widehat{\Phi}_{\widehat{\beta}}} \Gamma_{\widehat{\Phi}_{\widehat{\beta}}}}{(E^2 - m_{\widehat{\Phi}_{\widehat{\beta}}}^2)^2 + (m_{\widehat{\Phi}_{\widehat{\beta}}} \Gamma_{\widehat{\Phi}_{\widehat{\beta}}})^2} \bigg].
		\label{eq:DWScatCXcOS}
\end{align}
Note that the contributions of $E^2 - m_{\widehat{\Phi}_{\widehat{\gamma}}}^2$ in the numerators diverge after the integration over $E$, which makes $\sigma'$ diverges as mentioned above. Those terms are, however, off-shell contributions which vanish for physical particles, \textit{i.e.}, for $E^2 = m_{\widehat{\Phi}_{\widehat{\gamma}}}^2$. We neglect those off-shell contributions, and call this procedure an \textit{on-shell prescription}. Applying the narrow-width approximation to the remaining part:
\begin{align}
	\lim_{\frac{\Gamma_{\widehat{\Phi}_{\widehat{\gamma}}}}{m_{\widehat{\Phi}_{\widehat{\gamma}}}} \to 0} \frac{1}{(E^2 - m_{\widehat{\Phi}_{\widehat{\gamma}}}^2)^2 + (m_{\widehat{\Phi}_{\widehat{\gamma}}} \Gamma_{\widehat{\Phi}_{\widehat{\gamma}}})^2} = \frac{\pi}{m_{\widehat{\Phi}_{\widehat{\gamma}}} \Gamma_{\widehat{\Phi}_{\widehat{\gamma}}}} \delta(E^2 - m_{\widehat{\Phi}_{\widehat{\gamma}}}^2),
	\label{eq:NWA}
\end{align}
we obtain
\begin{align}
	\int &d\Pi_{\chi_j^c} \int d\Pi_\xi \, (2 \pi)^3 \delta^3 (\mathbf{p}_{\chi_j^c} + \mathbf{p}_\xi)
		\frac{(E^2 - m_{\chi_j}^2) (E^2 + m_{\chi_i}^2) (E^2 - m_{\chi_i}^2)}{E^2 (E^2 - p_{\widehat{\Phi}_{\widehat{\beta}}}^{* 2}) (E^2 - p_{\widehat{\Phi}_{\widehat{\alpha}}}^2)} \Bigg|_\text{OS} \\
	&= \frac{i}{32 \pi (p_{\widehat{\Phi}_{\widehat{\beta}}}^{* 2} - p_{\widehat{\Phi}_{\widehat{\alpha}}}^2)}
		\Bigg[ m_{\widehat{\Phi}_{\widehat{\alpha}}}^3 \bigg( 1 - \frac{m_{\chi_i}^4}{m_{\widehat{\Phi}_{\widehat{\alpha}}}^4} \bigg) \bigg( 1 - \frac{m_{\chi_j}^4}{m_{\widehat{\Phi}_{\widehat{\alpha}}}^4} \bigg)
			+ m_{\widehat{\Phi}_{\widehat{\beta}}}^3 \bigg( 1 - \frac{m_{\chi_i}^4}{m_{\widehat{\Phi}_{\widehat{\beta}}}^4} \bigg) \bigg( 1 - \frac{m_{\chi_j}^4}{m_{\widehat{\Phi}_{\widehat{\beta}}}^4} \bigg) \Bigg] \\
	&= \frac{i}{32 \pi (p_{\widehat{\Phi}_{\widehat{\beta}}}^* - p_{\widehat{\Phi}_{\widehat{\alpha}}})}
		m_\Phi^2 \bigg( 1 - \frac{m_{\chi_i}^4}{m_\Phi^4} \bigg) \bigg( 1 - \frac{m_{\chi_j}^4}{m_\Phi^4} \bigg),
\end{align}
where the subscript $\text{OS}$ implies the on-shell prescription has been imposed, and the last equality is correct up to the leading order because $m_{\widehat{\Phi}_{\widehat{\alpha}}} \sim m_\Phi [1 + \mathcal{O} (\alpha)]$. Note also that the narrow-width approximation works well up to working precision even for large couplings such as $f \sim \mathcal{O} (10^{-1})$, and the derivation can therefore be considered to be generally valid since $f \lesssim \mathcal{O} (10^{-5})$ for neutral mesons, as we will see in specific examples in section \ref{sec:Example}.

We have thus obtained an expression of a transition rate associated with scattering $\chi_j^c \xi \to \chi_i \xi^c$:
\begin{empheq}[box=\fbox]{align}
	\sum_{j, \widehat{\alpha}} \Gamma^\text{scat}_{\chi_j^c \xi \to \widehat{\Phi}_{\widehat{\alpha}} \to \chi_i \xi^c}
	&\coloneqq \frac{1}{4} \sum_j \int d\Pi_{\chi_j^c} \int d\Pi_\xi \, (2 \pi)^3 \delta^3 (\mathbf{p}_{\chi_j^c} + \mathbf{p}_\xi) \,
		\sigma'_\text{CM} (\chi_j^c \xi \to \chi_i \xi^c) \bigg|_\text{OS} \\
	&= \frac{i}{2^8 \pi^2} \sum_{j, \widehat{\alpha}, \widehat{\beta}}
		\frac{Q^{\chi_j^c \xi *}_{\widehat{\Phi}_{\widehat{\beta}} \to \chi_i \xi^c} Q^{\chi_j^c \xi}_{\widehat{\Phi}_{\widehat{\alpha}} \to \chi_i \xi^c}}{p_{\widehat{\Phi}_{\widehat{\beta}}}^* - p_{\widehat{\Phi}_{\widehat{\alpha}}}} m_\Phi^2 \bigg( 1 - \frac{m_{\chi_i}^4}{m_\Phi^4} \bigg) \bigg( 1 - \frac{m_{\chi_j}^4}{m_\Phi^4} \bigg).
	\label{eq:DWCcXCXc}
\end{empheq}
This expression is correct up to the leading order in perturbation, the meaning of which needs some clarification. Let us first consider the denominator $p_{\widehat{\Phi}_{\widehat{\alpha}}}^* - p_{\widehat{\Phi}_{\widehat{\alpha}}}$. Since $p_{\widehat{\Phi}_{\widehat{\alpha}}}^* - p_{\widehat{\Phi}_{\widehat{\alpha}}} = i \Gamma_{\widehat{\Phi}_{\widehat{\alpha}}}$ up to the leading order and $\Gamma_{\widehat{\Phi}_{\widehat{\alpha}}}$ is calculated from the self-energy, the leading contributions to $p_{\widehat{\Phi}_{\widehat{\beta}}}^* - p_{\widehat{\Phi}_{\widehat{\alpha}}}$ come from the one-loop diagrams of the self-energy. In addition, the numerator containing $Q^{\chi_j^c \xi}_{\widehat{\Phi}_{\widehat{\alpha}} \to \chi_i \xi^c}$ has already been calculated up to the leading order, \textit{i.e.}, up to the tree-level vertices. In other words, the leading-order contributions to $\sum_{j, \widehat{\alpha}} \Gamma^\text{scat}_{\chi_j^c \xi \to \widehat{\Phi}_{\widehat{\alpha}} \to \chi_i \xi^c}$ consist of the one-loop diagrams of the self-energy and the tree-level diagrams of the vertices. If we want to improve the precision of calculation to the next-leading order, the one-loop corrections to the vertices as well as the two-loop corrections to the self-energy must be taken into account for consistency in perturbative calculation.

In equation \ref{eq:DWCcXCXc}, the contribution of $\widehat{\beta} \neq \widehat{\alpha}$ comes from the interference between two propagators of physical particles, which can clearly be seen in the step from equation \ref{eq:DWScatCXcIntE} to equation \ref{eq:DWScatCXcOS}. The factor $1 / (p_{\widehat{\Phi}_{\widehat{\beta}}}^{* 2} - p_{\widehat{\Phi}_{\widehat{\alpha}}}^2)$ in equation \ref{eq:DWScatCXcOS} is the off-shell contribution of a propagator. In other words, the interference is caused by the on-shell and off-shell propagators of physical particles in the energy-momentum domain. In addition, the narrow-width approximation here is simply a calculation method, and does not imply that the contribution of every propagator is treated as a delta function, in which case no interference between physical particles can occur. Note that, before one propagator is replaced by the delta function, the off-shell contribution of the other propagator became a multiplicative coefficient of the form $1 / (p_{\widehat{\Phi}_{\widehat{\beta}}}^{* 2} - p_{\widehat{\Phi}_{\widehat{\alpha}}}^2)$. \\

We can proceed similar calculations for all the other initial and final states, and define 
\begin{align}
	\boxed{\sum_{\widehat{\alpha}} \Gamma^\text{scat}_{\widehat{\Phi}_{\widehat{\alpha}} \to Y}
		\coloneqq \sum_{X, \widehat{\alpha}} \Gamma^\text{scat}_{X \to \widehat{\Phi}_{\widehat{\alpha}} \to Y},}
	\label{eq:DWScat}
\end{align}
\textit{e.g.},
\begin{align}
	\sum_{\widehat{\alpha}} \Gamma^\text{scat}_{\widehat{\Phi}_{\widehat{\alpha}} \to \chi_i \xi^c}
		= \sum_{j, \widehat{\alpha}} \Gamma^\text{scat}_{\chi_j \xi^c \to \widehat{\Phi}_{\widehat{\alpha}} \to \chi_i \xi^c}
			+ \sum_{j, \widehat{\alpha}} \Gamma^\text{scat}_{\chi_j^c \xi \to \widehat{\Phi}_{\widehat{\alpha}} \to \chi_i \xi^c}
			+ \sum_{j, \widehat{\alpha}} \Gamma^\text{scat}_{\psi_j \psi_j^c \to \widehat{\Phi}_{\widehat{\alpha}} \to \chi_i \xi^c}.
\end{align}
The expression of every transition rate will be presented in section \ref{sec:DWSum}. It is tempting to move a step forward and remove $\sum_{\widehat{\alpha}}$ to obtain an expression associated with each of physical particles. However, it is logically flawed, and not even allowed because $\Gamma^\text{scat}_{X \to \widehat{\Phi}_{\widehat{\alpha}} \to Y}$ and $\Gamma^\text{scat}_{\widehat{\Phi}_{\widehat{\alpha}} \to Y}$ are complex-valued in general. This means that $\Gamma^\text{scat}_{\widehat{\Phi}_{\widehat{\alpha}} \to Y}$ itself cannot be regarded as a physical transition rate such as a partial decay width because it must be real-valued to allow such an interpretation.

\subsubsection{Summary of $\Gamma^\text{scat}_{X \to \widehat{\Phi}_{\widehat{\alpha}} \to Y}$}		\label{sec:DWSum}
Here, $\Gamma^\text{scat}_{X \to \widehat{\Phi}_{\widehat{\alpha}} \to Y}$ for all possible $X$, $Y$ are provided. Every expression is obtained up to the leading order in perturbation. \\

\paragraph{$\boxed{\chi_j \xi^c \to \chi_i \xi^c}$}
When the initial state is $| \chi_j \xi^c \rangle$, we have
\begin{align}
	\sigma'_\text{CM} [\chi_j (p_{\chi_j}') \xi^c (p_{\xi^c}') \to \chi_i \xi^c]
	= \frac{1}{2 \pi} \sum_{\widehat{\alpha}, \widehat{\beta}} Q^{\chi_j \xi^c *}_{\widehat{\Phi}_{\widehat{\beta}} \to \chi_i \xi^c} Q^{\chi_j \xi^c}_{\widehat{\Phi}_{\widehat{\alpha}} \to \chi_i \xi^c}
		\frac{(E^2 - m_{\chi_j}^2) (E^2 + m_{\chi_i}^2) (E^2 - m_{\chi_i}^2)}{E^2 (E^2 - p_{\widehat{\Phi}_{\widehat{\beta}}}^{* 2}) (E^2 - p_{\widehat{\Phi}_{\widehat{\alpha}}}^2)},
\end{align}
where
\begin{align}
	Q^{\chi_j \xi^c}_{\widehat{\Phi}_{\widehat{\alpha}} \to \chi_i \xi^c} \coloneqq h_i h_j^* (C_\Phi^{-1})_{1 {\widehat{\alpha}}} (C_\Phi)_{{\widehat{\alpha}} 1}.
\end{align}
Hence,
\begin{align}
	\sum_{j, \widehat{\alpha}} \Gamma^\text{scat}_{\chi_j \xi^c \to \widehat{\Phi}_{\widehat{\alpha}} \to \chi_i \xi^c}
	&\coloneqq \frac{1}{4} \sum_j \int d\Pi_{\chi_j}' \int d\Pi_{\xi^c}' \, (2 \pi)^3 \delta^3 (\mathbf{p}_{\chi_j}' + \mathbf{p}_{\xi^c}') \,
		\sigma'_\text{CM} [\chi_j (p_{\chi_j}') \xi^c (p_{\xi^c}') \to \chi_i \xi^c] \bigg|_\text{OS} \\ 
	&= \frac{i}{2^8 \pi^2} \sum_{j, \widehat{\alpha}, \widehat{\beta}}
		\frac{Q^{\chi_j \xi^c *}_{\widehat{\Phi}_{\widehat{\beta}} \to \chi_i \xi^c} Q^{\chi_j \xi^c}_{\widehat{\Phi}_{\widehat{\alpha}} \to \chi_i \xi^c}}{p_{\widehat{\Phi}_{\widehat{\beta}}}^* - p_{\widehat{\Phi}_{\widehat{\alpha}}}} m_\Phi^2 \bigg( 1 - \frac{m_{\chi_i}^4}{m_\Phi^4} \bigg) \bigg( 1 - \frac{m_{\chi_j}^4}{m_\Phi^4} \bigg).
	\label{eq:DWScatCXc}
\end{align} \\

\paragraph{$\boxed{\chi_j^c \xi \to \chi_i \xi^c}$}
For the initial state $| \chi_j^c \xi \rangle$, we obtain
\begin{align}
	\sigma'_\text{CM} (\chi_j^c \xi \to \chi_i \xi^c)
	= \frac{1}{2 \pi} \sum_{\widehat{\alpha}, \widehat{\beta}} Q^{\chi_j \xi^c *}_{\widehat{\Phi}_{\widehat{\beta}} \to \chi_i^c \xi} Q^{\chi_j \xi^c}_{\widehat{\Phi}_{\widehat{\alpha}} \to \chi_i^c \xi}
		\frac{(E^2 - m_{\chi_j}^2) (E^2 + m_{\chi_i}^2) (E^2 - m_{\chi_i}^2)}{E^2 (E^2 - p_{\widehat{\Phi}_{\widehat{\beta}}}^{* 2}) (E^2 - p_{\widehat{\Phi}_{\widehat{\alpha}}}^2)}.
\end{align}
where
\begin{align}
	Q^{\chi_j^c \xi}_{\widehat{\Phi}_{\widehat{\alpha}} \to \chi_i \xi^c} \coloneqq h_i h_j (C_\Phi^{-1})_{1 {\widehat{\alpha}}} (C_\Phi)_{{\widehat{\alpha}} 2}.
\end{align}
Hence,
\begin{align}
	\sum_{j, \widehat{\alpha}} \Gamma^\text{scat}_{\chi_j^c \xi \to \widehat{\Phi}_{\widehat{\alpha}} \to \chi_i \xi^c}
	&\coloneqq \frac{1}{4} \sum_j \int d\Pi_{\chi_j^c} \int d\Pi_\xi \, (2 \pi)^3 \delta^3 (\mathbf{p}_{\chi_j^c} + \mathbf{p}_\xi) \
		\sigma'_\text{CM} (\chi_j^c \xi \to \chi_i \xi^c) \bigg|_\text{OS} \\
	&= \frac{i}{2^8 \pi^2} \sum_{j, \widehat{\alpha}, \widehat{\beta}}
		\frac{Q^{\chi_j^c \xi *}_{\widehat{\Phi}_{\widehat{\beta}} \to \chi_i \xi^c} Q^{\chi_j^c \xi}_{\widehat{\Phi}_{\widehat{\alpha}} \to \chi_i \xi^c}}{p_{\widehat{\Phi}_{\widehat{\beta}}}^* - p_{\widehat{\Phi}_{\widehat{\alpha}}}} m_\Phi^2 \bigg( 1 - \frac{m_{\chi_i}^4}{m_\Phi^4} \bigg) \bigg( 1 - \frac{m_{\chi_j}^4}{m_\Phi^4} \bigg).
\end{align} \\

\paragraph{$\boxed{\psi_j \psi_j^c \to \chi_i \xi^c}$}
When the initial state is $| \psi_j \psi_j^c \rangle$, we find
\begin{align}
	\sigma'_\text{CM} (\psi_j \psi_j^c \to \chi_i \xi^c)
	&= \frac{1}{2 \pi} \sum_{\widehat{\alpha}, \widehat{\beta}} Q^{\psi_j \psi_j^c *}_{\widehat{\Phi}_{\widehat{\beta}} \to \chi_i \xi^c} Q^{\psi_j \psi_j^c}_{\widehat{\Phi}_{\widehat{\alpha}} \to \chi_i \xi^c}
		\frac{(E^2 + m_{\chi_i}^2) (E^2 - m_{\chi_i}^2)}{(E^2 - p_{\widehat{\Phi}_{\widehat{\beta}}}^{* 2}) (E^2 - p_{\widehat{\Phi}_{\widehat{\alpha}}}^2)},
\end{align}
where
\begin{align}
	Q^{\psi_j \psi_j^c}_{\widehat{\Phi}_{\widehat{\alpha}} \to \chi_i \xi^c}
		\coloneqq h_i f_j^* (C_\Phi^{-1})_{1 {\widehat{\alpha}}} (C_\Phi)_{{\widehat{\alpha}} 1}
			+ h_i f_j (C_\Phi^{-1})_{1 {\widehat{\alpha}}} (C_\Phi)_{{\widehat{\alpha}} 2}.
\end{align}
Hence,
\begin{align}
	\sum_{j, \widehat{\alpha}} \Gamma^\text{scat}_{\psi_j \psi_j^c \to \widehat{\Phi}_{\widehat{\alpha}} \to \chi_i \xi^c}
	&\coloneqq \frac{1}{4} \sum_j \int d\Pi_{\psi_j} \int d\Pi_{\psi_j^c} \, (2 \pi)^3 \delta^3 (\mathbf{p}_{\psi_j} + \mathbf{p}_{\psi_j^c}) \,
		\sigma'_\text{CM} (\psi_j \psi_j^c \to \chi_i \xi^c) \bigg|_\text{OS} \\
	&= \frac{i}{2^8 \pi^2} \sum_{j, \widehat{\alpha}, \widehat{\beta}}
		\frac{Q^{\psi_j \psi_j^c *}_{\widehat{\Phi}_{\widehat{\beta}} \to \chi_i \xi^c} Q^{\psi_j \psi_j^c}_{\widehat{\Phi}_{\widehat{\alpha}} \to \chi_i \xi^c}}{p_{\widehat{\Phi}_{\widehat{\beta}}}^* - p_{\widehat{\Phi}_{\widehat{\alpha}}}} m_\Phi^2 \bigg( 1 - \frac{m_{\chi_i}^4}{m_\Phi^4} \bigg).
\end{align} \\

\paragraph{$\boxed{\chi_j \xi^c \to \chi_i^c \xi}$}
In a similar way, we can also obtain
\begin{align}
	\sum_{j, \widehat{\alpha}} \Gamma^\text{scat}_{\chi_j \xi^c \to \widehat{\Phi}_{\widehat{\alpha}} \to \chi_i^c \xi}
	&\coloneqq \frac{1}{4} \sum_j \int d\Pi_{\chi_j} \int d\Pi_{\xi^c} \, (2 \pi)^3 \delta^3 (\mathbf{p}_{\chi_j} + \mathbf{p}_{\xi^c}) \,
			\sigma'_\text{CM} (\chi_j \xi^c \to \chi_i^c \xi) \bigg|_\text{OS} \\
	&= \frac{i}{2^8 \pi^2} \sum_{j, \widehat{\alpha}, \widehat{\beta}}
		\frac{Q^{\chi_j \xi^c *}_{\widehat{\Phi}_{\widehat{\beta}} \to \chi_i^c \xi} Q^{\chi_j \xi^c}_{\widehat{\Phi}_{\widehat{\alpha}} \to \chi_i^c \xi}}{p_{\widehat{\Phi}_{\widehat{\beta}}}^* - p_{\widehat{\Phi}_{\widehat{\alpha}}}} m_\Phi^2 \bigg( 1 - \frac{m_{\chi_i}^4}{m_\Phi^4} \bigg) \bigg( 1 - \frac{m_{\chi_j}^4}{m_\Phi^4} \bigg),
\end{align}
where
\begin{align}
	Q^{\chi_j \xi^c}_{\widehat{\Phi}_{\widehat{\alpha}} \to \chi_i^c \xi} \coloneqq h_i^* h_j^* (C_\Phi^{-1})_{2 {\widehat{\alpha}}} (C_\Phi)_{{\widehat{\alpha}} 1}.
\end{align} \\

\paragraph{$\boxed{\chi_j^c \xi \to \chi_i^c \xi}$}
For the initial state $| \chi_j^c \xi \rangle$, we have
\begin{align}
	\sum_{j, \widehat{\alpha}} \Gamma^\text{scat}_{\chi_j^c \xi \to \widehat{\Phi}_{\widehat{\alpha}} \to \chi_i^c \xi}
	&\coloneqq \frac{1}{4} \sum_j \int d\Pi_{\chi_j^c}' \int d\Pi_\xi' \, (2 \pi)^3 \delta^3 (\mathbf{p}_{\chi_j^c}' + \mathbf{p}_\xi') \,
			\sigma'_\text{CM} [\chi_j^c (p_{\chi_j^c}') \xi (p_\xi') \to \chi_i^c \xi] \bigg|_\text{OS} \\ 
	&= \frac{i}{2^8 \pi^2} \sum_{j, \widehat{\alpha}, \widehat{\beta}}
		\frac{Q^{\chi_j^c \xi *}_{\widehat{\Phi}_{\widehat{\beta}} \to \chi_i^c \xi} Q^{\chi_j^c \xi}_{\widehat{\Phi}_{\widehat{\alpha}} \to \chi_i^c \xi}}{p_{\widehat{\Phi}_{\widehat{\beta}}}^* - p_{\widehat{\Phi}_{\widehat{\alpha}}}} m_\Phi^2 \bigg( 1 - \frac{m_{\chi_i}^4}{m_\Phi^4} \bigg) \bigg( 1 - \frac{m_{\chi_j}^4}{m_\Phi^4} \bigg),
\end{align}
where
\begin{align}
	Q^{\chi_j^c \xi}_{\widehat{\Phi}_{\widehat{\alpha}} \to \chi_i^c \xi} \coloneqq h_i^* h_j (C_\Phi^{-1})_{2 {\widehat{\alpha}}} (C_\Phi)_{{\widehat{\alpha}} 2}.
\end{align} \\

\paragraph{$\boxed{\psi_j \psi_j^c \to \chi_i^c \xi}$}
For the initial state $| \psi_j \psi_j^c \rangle$, we obtain
\begin{align}
	\sum_{j, \widehat{\alpha}} \Gamma^\text{scat}_{\psi_j \psi_j^c \to \widehat{\Phi}_{\widehat{\alpha}} \to \chi_i^c \xi}
	&\coloneqq \frac{1}{4} \sum_j \int d\Pi_{\psi_j} \int d\Pi_{\psi_j^c} \, (2 \pi)^3 \delta^3 (\mathbf{p}_{\psi_j} + \mathbf{p}_{\psi_j^c}) \,
		\sigma'_\text{CM} (\psi_j \psi_j^c \to \chi_i^c \xi) \bigg|_\text{OS} \\
	&= \frac{i}{2^8 \pi^2} \sum_{j, \widehat{\alpha}, \widehat{\beta}}
		\frac{Q^{\psi_j \psi_j^c *}_{\widehat{\Phi}_{\widehat{\beta}} \to \chi_i^c \xi} Q^{\psi_j \psi_j^c}_{\widehat{\Phi}_{\widehat{\alpha}} \to \chi_i^c \xi}}{p_{\widehat{\Phi}_{\widehat{\beta}}}^* - p_{\widehat{\Phi}_{\widehat{\alpha}}}} m_\Phi^2 \bigg( 1 - \frac{m_{\chi_i}^4}{m_\Phi^4} \bigg),
\end{align}
where
\begin{align}
	Q^{\psi_j \psi_j^c}_{\widehat{\Phi}_{\widehat{\alpha}} \to \chi_i^c \xi}
		\coloneqq h_i^* f_j^* (C_\Phi^{-1})_{2 {\widehat{\alpha}}} (C_\Phi)_{{\widehat{\alpha}} 1}
			+ h_i^* f_j (C_\Phi^{-1})_{2 {\widehat{\alpha}}} (C_\Phi)_{{\widehat{\alpha}} 2}.
\end{align} \\

\paragraph{$\boxed{\chi_j \xi^c \to \psi_i \psi_i^c}$}
Moreover, we can also find
\begin{align}
	\sum_{j, \widehat{\alpha}} \Gamma^\text{scat}_{\chi_j \xi^c \to \widehat{\Phi}_{\widehat{\alpha}} \to \psi_i \psi_i^c}
	&\coloneqq \frac{1}{4} \sum_j \int d\Pi_{\chi_j} \int d\Pi_{\xi^c} \, (2 \pi)^3 \delta^3 (\mathbf{p}_{\chi_j} + \mathbf{p}_{\xi^c}) \,
			\sigma'_\text{CM} (\chi_j \xi^c \to \psi_i \psi_i^c) \bigg|_\text{OS} \\
	&= \frac{i}{2^8 \pi^2} \sum_{j, \widehat{\alpha}, \widehat{\beta}}
		\frac{Q^{\chi_j \xi^c *}_{\widehat{\Phi}_{\widehat{\beta}} \to \psi_i \psi_i^c} Q^{\chi_j \xi^c}_{\widehat{\Phi}_{\widehat{\alpha}} \to \psi_i \psi_i^c}}{p_{\widehat{\Phi}_{\widehat{\beta}}}^* - p_{\widehat{\Phi}_{\widehat{\alpha}}}} m_\Phi^2 \bigg( 1 - \frac{m_{\chi_j}^4}{m_\Phi^4} \bigg),
	\label{eq:DWScatCXcPP}
\end{align}
where
\begin{align}
	Q^{\chi_j \xi^c}_{\widehat{\Phi}_{\widehat{\alpha}} \to \psi_i \psi_i^c} \coloneqq f_i h_j^* (C_\Phi^{-1})_{1 {\widehat{\alpha}}} (C_\Phi)_{{\widehat{\alpha}} 1} + f_i^* h_j^* (C_\Phi^{-1})_{2 {\widehat{\alpha}}} (C_\Phi)_{{\widehat{\alpha}} 1}.
	\label{eq:DWQCXcPP}
\end{align} \\

\paragraph{$\boxed{\chi_j^c \xi \to \psi_i \psi_i^c}$}
For the initial states $| \chi_j^c \xi \rangle$, we obtain
\begin{align}
	\sum_{j, \widehat{\alpha}} \Gamma^\text{scat}_{\chi_j^c \xi \to \widehat{\Phi}_{\widehat{\alpha}} \to \psi_i \psi_i^c}
	&\coloneqq \frac{1}{4} \sum_j \int d\Pi_{\chi_j^c} \int d\Pi_\xi \, (2 \pi)^3 \delta^3 (\mathbf{p}_{\chi_j^c} + \mathbf{p}_\xi) \,
			\sigma'_\text{CM} (\chi_j^c \xi \to \psi_i \psi_i^c) \bigg|_\text{OS} \\ 
	&= \frac{i}{2^8 \pi^2} \sum_{j, \widehat{\alpha}, \widehat{\beta}}
		\frac{Q^{\chi_j^c \xi *}_{\widehat{\Phi}_{\widehat{\beta}} \to \chi_i^c \xi} Q^{\chi_j^c \xi}_{\widehat{\Phi}_{\widehat{\alpha}} \to \chi_i^c \xi}}{p_{\widehat{\Phi}_{\widehat{\beta}}}^* - p_{\widehat{\Phi}_{\widehat{\alpha}}}} m_\Phi^2 \bigg( 1 - \frac{m_{\chi_j}^4}{m_\Phi^4} \bigg),
\end{align}
where
\begin{align}
	Q^{\chi_j^c \xi}_{\widehat{\Phi}_{\widehat{\alpha}} \to \psi_i \psi_i^c} \coloneqq f_i h_j (C_\Phi^{-1})_{1 {\widehat{\alpha}}} (C_\Phi)_{{\widehat{\alpha}} 2} + f_i^* h_j (C_\Phi^{-1})_{2 {\widehat{\alpha}}} (C_\Phi)_{{\widehat{\alpha}} 2}.
\end{align} \\

\paragraph{$\boxed{\psi_j \psi_j^c \to \psi_i \psi_i^c}$}
When the initial state is $| \psi_j \psi_j^c \rangle$, we have
\begin{align}
	\sum_{j, \widehat{\alpha}} \Gamma^\text{scat}_{\psi_j \psi_j^c \to \widehat{\Phi}_{\widehat{\alpha}} \to \psi_i \psi_i^c}
	&\coloneqq \frac{1}{4} \sum_j \int d\Pi_{\psi_j} \int d\Pi_{\psi_j^c} \, (2 \pi)^3 \delta^3 (\mathbf{p}_{\psi_j} + \mathbf{p}_{\psi_j^c}) \,
		\sigma'_\text{CM} (\psi_j \psi_j^c \to \psi_i \psi_i^c) \bigg|_\text{OS} \\
	&= \frac{i}{2^8 \pi^2} \sum_{j, \widehat{\alpha}, \widehat{\beta}}
		\frac{Q^{\psi_j \psi_j^c *}_{\widehat{\Phi}_{\widehat{\beta}} \to \psi_i \psi_i^c} Q^{\psi_j \psi_j^c}_{\widehat{\Phi}_{\widehat{\alpha}} \to \psi_i \psi_i^c}}{p_{\widehat{\Phi}_{\widehat{\beta}}}^* - p_{\widehat{\Phi}_{\widehat{\alpha}}}} m_\Phi^2,
\end{align}
where
\begin{align}
	Q^{\psi_j \psi_j^c}_{\widehat{\Phi}_{\widehat{\alpha}} \to \psi_i \psi_i^c}
		\coloneqq f_i f_j^* (C_\Phi^{-1})_{1 {\widehat{\alpha}}} (C_\Phi)_{{\widehat{\alpha}} 1}
			+ f_i f_j (C_\Phi^{-1})_{1 {\widehat{\alpha}}} (C_\Phi)_{{\widehat{\alpha}} 2}
			+ f_i^* f_j^* (C_\Phi^{-1})_{2 {\widehat{\alpha}}} (C_\Phi)_{{\widehat{\alpha}} 1}
			+ f_i^* f_j (C_\Phi^{-1})_{2 {\widehat{\alpha}}} (C_\Phi)_{{\widehat{\alpha}} 2}.
\end{align}

\subsection{From the time-dependent on-shell contributions to scattering}		\label{sec:DWtdep}
In practical applications, time-dependent expressions of differential decay widths are frequently used. To express the transition rates we have obtained as an integral over time, we can apply the Fourier transform to equation \ref{eq:DWScatCXcIntE} which is an integral over the total energy. To discard the off-shell divergent contribution, we rewrite
\begin{align}
	\frac{(E^2 + m_{\chi_i}^2) (E^2 - m_{\chi_i}^2)}{E^2 (E^2 - p_{\widehat{\Phi}_{\widehat{\alpha}}}^2)} \to \frac{(m_{\widehat{\Phi}_{\widehat{\alpha}}}^2 + m_{\chi_i}^2) (m_{\widehat{\Phi}_{\widehat{\alpha}}}^2 - m_{\chi_i}^2)}{m_{\widehat{\Phi}_{\widehat{\alpha}}}^2 (E^2 - p_{\widehat{\Phi}_{\widehat{\alpha}}}^2)},
\end{align}
which is the \textit{on-shell prescription} in this case. The Fourier transform of its energy-dependent part is written as
\begin{align}
	F_{\widehat{\alpha}} (t - t_0) \coloneqq \int_{-\infty}^\infty \frac{dE}{2 \pi} \, e^{-i E (t - t_0)} \frac{1}{E^2 - p_{\widehat{\Phi}_{\widehat{\alpha}}}^2}
	= \minus \frac{i}{2 p_{\widehat{\Phi}_{\widehat{\alpha}}}} e^{-i p_{\widehat{\Phi}_{\widehat{\alpha}}} |t - t_0|}.
	\label{eq:PWFT}
\end{align}
Since $E = m_{\chi_{ij}}$ is an off-shell contribution which is well outside the resonance peak unless the relevant couplings are large, we may neglect its effect and rewrite
\begin{align}
	&\int_{m_{\chi_{ij}}}^\infty dE \frac{1}{(E^2 - p_{\widehat{\Phi}_{\widehat{\beta}}}^{* 2}) (E^2 - p_{\widehat{\Phi}_{\widehat{\alpha}}}^2)}
	= \int_0^\infty dE \frac{1}{(E^2 - p_{\widehat{\Phi}_{\widehat{\beta}}}^{* 2}) (E^2 - p_{\widehat{\Phi}_{\widehat{\alpha}}}^2)}
	= \frac{1}{2} \int_{-\infty}^\infty dE \frac{1}{(E^2 - p_{\widehat{\Phi}_{\widehat{\beta}}}^{* 2}) (E^2 - p_{\widehat{\Phi}_{\widehat{\alpha}}}^2)} \nonumber \\
	&= \frac{1}{2} \int_{-\infty}^\infty dE \int_{-\infty}^\infty dt' \, e^{-i E (t' - t_0)} F_{\widehat{\beta}} (t' - t_0) \int_{-\infty}^\infty dt \, e^{i E (t - t_0)} F_{\widehat{\alpha}} (t - t_0)
	= \frac{\pi}{4 p_{\widehat{\Phi}_{\widehat{\beta}}}^* p_{\widehat{\Phi}_{\widehat{\alpha}}}} \int_{-\infty}^\infty dt \, e^{i (p_{\widehat{\Phi}_{\widehat{\beta}}}^* - p_{\widehat{\Phi}_{\widehat{\alpha}}}) |t - t_0|} \nonumber \\
	&= \frac{\pi}{2 p_{\widehat{\Phi}_{\widehat{\beta}}}^* p_{\widehat{\Phi}_{\widehat{\alpha}}}} \int_{t_0}^\infty dt \, e^{i (p_{\widehat{\Phi}_{\widehat{\beta}}}^* - p_{\widehat{\Phi}_{\widehat{\alpha}}}) (t - t_0)}
	= \frac{i \pi}{2 p_{\widehat{\Phi}_{\widehat{\beta}}}^* p_{\widehat{\Phi}_{\widehat{\alpha}}} (p_{\widehat{\Phi}_{\widehat{\beta}}}^* - p_{\widehat{\Phi}_{\widehat{\alpha}}})}.
	\label{eq:IntTime}
\end{align}
We can therefore write the transition rate in the form of an integration over time:
\begin{empheq}[box=\fbox]{align}
	\sum_{j, \widehat{\alpha}} \Gamma^t_{\chi_j^c \xi \to \widehat{\Phi}_{\widehat{\alpha}} \to \chi_i \xi^c}
	&\coloneqq \frac{1}{2^8 \pi^2} \sum_{j, \widehat{\alpha}, \widehat{\beta}}
			Q^{\chi_j^c \xi *}_{\widehat{\Phi}_{\widehat{\beta}} \to \chi_i \xi^c} Q^{\chi_j^c \xi}_{\widehat{\Phi}_{\widehat{\alpha}} \to \chi_i \xi^c}
			m_\Phi^2 \bigg( 1 - \frac{m_{\chi_i}^4}{m_\Phi^4} \bigg) \bigg( 1 - \frac{m_{\chi_j}^4}{m_\Phi^4} \bigg) \nonumber \\
			&\qquad \qquad \int_{t_0}^\infty dt \, e^{i (p_{\widehat{\Phi}_{\widehat{\beta}}}^* - p_{\widehat{\Phi}_{\widehat{\alpha}}}) (t - t_0)}
		\label{eq:DWtCXcQFT} \\ 
	&= \frac{i}{2^8 \pi^2} \sum_{j, \widehat{\alpha}, \widehat{\beta}}
		\frac{Q^{\chi_j^c \xi *}_{\widehat{\Phi}_{\widehat{\beta}} \to \chi_i \xi^c} Q^{\chi_j^c \xi}_{\widehat{\Phi}_{\widehat{\alpha}} \to \chi_i \xi^c}}{p_{\widehat{\Phi}_{\widehat{\beta}}}^* - p_{\widehat{\Phi}_{\widehat{\alpha}}}} m_\Phi^2 \bigg( 1 - \frac{m_{\chi_i}^4}{m_\Phi^4} \bigg) \bigg( 1 - \frac{m_{\chi_j}^4}{m_\Phi^4} \bigg),
		\label{eq:DWtCXc}
\end{empheq}
where
\begin{align}
	Q^{\chi_j^c \xi}_{\widehat{\Phi}_{\widehat{\alpha}} \to \chi_i \xi^c} \coloneqq h_i h_j (C_\Phi^{-1})_{1 {\widehat{\alpha}}} (C_\Phi)_{{\widehat{\alpha}} 2}
\end{align}
as defined before. Equation \ref{eq:DWtCXc} is identical to equation \ref{eq:DWScatCXc}, which is an almost trivial result since they are simply related to each other by the Fourier transform as long as the on-shell prescriptions are mutually consistent. The expression given by equation \ref{eq:DWtCXcQFT} is useful when we need to track the time evolution of the decay, as is often done in experiments or in theoretical analyses with the time-dependent solution of the effective Schr\"odinger equation.

Note that the contribution of $\widehat{\beta} \neq \widehat{\alpha}$ in equation \ref{eq:DWtCXc} generates oscillation in time, and it is caused by the interference between the time evolution factors of two physical particles, $e^{-i p_{\widehat{\Phi}_{\widehat{\beta}}} (t - t_0)}$ and $e^{-i p_{\widehat{\Phi}_{\widehat{\alpha}}} (t - t_0)}$. The spacetime representation of a scattering amplitude is obtained by the Fourier transform of its energy-momentum representation which involves integration over the whole energy and momentum, and thus the interference can occur between two amplitudes in the time domain which have dominant contributions at different energies. \\

If the decay widths of two physical particles are well-separated as in the neutral kaon system, we can always choose an intermediate time $t = t_1 > t_0$ after which apparently only the longer-living particle survives. Integrating equation \ref{eq:DWtCXcQFT} over $t_1 < t < \infty$, it is possible to obtain an expression associated only with the longer-living one. We will investigate a kaon-like system in such a way in section \ref{sec:Kaon}. Now we determine such an instant $t = t_1$, assuming $\Gamma_{\widehat{\Phi}_{\widehat{\alpha}}} \ll \Gamma_{\widehat{\Phi}_{\widehat{\beta}}}$. The time integral for $t > t_1$ is written as
\begin{align}
	\int_{t_1}^\infty dt \, e^{i (p_{\widehat{\Phi}_{\widehat{\beta}}}^* - p_{\widehat{\Phi}_{\widehat{\alpha}}}) (t - t_0)}
	&= \frac{i}{p_{\widehat{\Phi}_{\widehat{\beta}}}^* - p_{\widehat{\Phi}_{\widehat{\alpha}}}} e^{i (p_{\widehat{\Phi}_{\widehat{\beta}}}^* - p_{\widehat{\Phi}_{\widehat{\alpha}}}) (t_1 - t_0)} \nonumber \\
	&= \frac{i}{p_{\widehat{\Phi}_{\widehat{\beta}}}^* - p_{\widehat{\Phi}_{\widehat{\alpha}}}} e^{i (m_{\widehat{\Phi}_{\widehat{\beta}}} - m_{\widehat{\Phi}_{\widehat{\alpha}}}) (t_1 - t_0)} e^{-\frac{1}{2} (\Gamma_{\widehat{\Phi}_{\widehat{\beta}}} + \Gamma_{\widehat{\Phi}_{\widehat{\alpha}}}) (t_1 - t_0)}
	\label{eq:Tint}
\end{align}
up to the leading order. In order to be able to neglect the contribution of $\widehat{\Phi}_{\widehat{\beta}}$ after $t = t_1 > t_0$ up to working precision, we have to choose $t_1$ such that
\begin{align}
	e^{-\Gamma_{\widehat{\Phi}_{\widehat{\beta}}} (t_1 - t_0)} \lesssim \mathcal{O} (\alpha)
	\ll e^{-\Gamma_{\widehat{\Phi}_{\widehat{\alpha}}} (t_1 - t_0)},
\end{align}
\textit{i.e.},
\begin{align}
	\minus \frac{\hbar \log{[\mathcal{O} (\alpha)]}}{\Gamma_{\widehat{\Phi}_{\widehat{\beta}}}} \lesssim t_1 - t_0
	\ll \minus \frac{\hbar \log{[\mathcal{O} (\alpha)]}}{\Gamma_{\widehat{\Phi}_{\widehat{\alpha}}}},
\end{align}
where $\hbar$ is written for clarity. We can always find $t_1$ which satisfies this condition as long as $\Gamma_{\widehat{\Phi}_{\widehat{\alpha}}}$ and $\Gamma_{\widehat{\Phi}_{\widehat{\beta}}}$ are well-separated. For convenience, we explicitly define $t_1$ as
\begin{align}
	t_1 - t_0 \coloneqq \minus \frac{\hbar}{\Gamma_{\widehat{\Phi}_{\widehat{\beta}}}} \log{\bigg[ \frac{1}{2 \pi} \sum_i (|f_i|^2 + |h_i|^2) \bigg]},
	\label{eq:T1}
\end{align}
and write the transition rate for $t > t_1$ as
\begin{empheq}[box=\fbox]{align}
	\sum_{j, \widehat{\alpha}} \Gamma^{t > t_1}_{\chi_j^c \xi \to \widehat{\Phi}_{\widehat{\alpha}} \to \chi_i \xi^c}
	&\coloneqq \frac{1}{2^8 \pi^2} \sum_{j, \widehat{\alpha}, \widehat{\beta}}
			Q^{\chi_j^c \xi *}_{\widehat{\Phi}_{\widehat{\beta}} \to \chi_i \xi^c} Q^{\chi_j^c \xi}_{\widehat{\Phi}_{\widehat{\alpha}} \to \chi_i \xi^c}
			m_\Phi^2 \bigg( 1 - \frac{m_{\chi_i}^4}{m_\Phi^4} \bigg) \bigg( 1 - \frac{m_{\chi_j}^4}{m_\Phi^4} \bigg) \nonumber \\
			&\qquad \qquad \int_{t_1}^\infty dt \, e^{i (p_{\widehat{\Phi}_{\widehat{\beta}}}^* - p_{\widehat{\Phi}_{\widehat{\alpha}}}) (t - t_0)},
\end{empheq}
which is useful in the $K^0$-$\overline{K^0}$ system since $\Gamma_{K^0_S} \gg \Gamma_{K^0_L}$. Note that, when $\Gamma_{\widehat{\Phi}_{\widehat{\alpha}}} \ll \Gamma_{\widehat{\Phi}_{\widehat{\beta}}}$, the interference between $\widehat{\Phi}_{\widehat{\beta}}$ and $\widehat{\Phi}_{\widehat{\alpha}}$ decays as fast as the contribution only of $\widehat{\Phi}_{\widehat{\alpha}}$ so that we must lose a part of $\widehat{\Phi}_{\widehat{\alpha}}$ after $t = t_1$. In other words, we can never observe each of physical particles, \textit{i.e.}, each quasiparticle, as it is.

\subsection{From the optical theorem}		\label{sec:DWOpt}
Now we discuss a method to calculate some alternative transition rates from the left-hand side of equation \ref{eq:UnitCon}. It is explicitly written as
\begin{align}
	\boxed{\int d\Pi_{\chi_i} \int d\Pi_{\xi^c} \, (2 \pi)^3 \delta^3 (\mathbf{p} - \mathbf{p}_{\chi_i} - \mathbf{p}_{\xi^c}) \sum_{r, s} \langle \chi_{i, r} \xi^c_s | i (\mathcal{T}^\dag - \mathcal{T}) | \chi_{i, r} \xi^c_s \rangle,}
	\label{eq:DWUnitLH}
\end{align}
which corresponds to applying integration \ref{eq:PhSpInt} over the phase space of the final state to the diagram in figure \ref{fig:DWUnitCut} without the unitarity cut and taking its imaginary part. Since
\begin{align}
	\langle \chi_{i, r} \xi^c_s | i (\mathcal{T}^\dag - \mathcal{T}) | \chi_{i, r} \xi^c_s \rangle = (2 \pi)^4 \delta^4 (0) \, 2 \, \text{Im} [\mathcal{M} (\chi_{i, r} \xi^c_s \to \chi_{i, r} \xi^c_s)],
\end{align}
the scattering amplitude is written as
\begin{align}
	i \mathcal{M} (\chi_{i, r} \xi^c_s \to \chi_{i, r} \xi^c_s)
	&= \overline{u_{\chi_i}^r} (\mathbf{p}_{\chi_i}) v_\xi^s (\mathbf{p}_\xi)
		(\minus i h_i) [i \Delta_{11} (p^2)] (\minus i h_i^*)
		\overline{v_\xi^s} (\mathbf{p}_\xi) u_{\chi_i}^r (\mathbf{p}_{\chi_i}) \nonumber \\
	&= \minus i \overline{u_{\chi_i}^r} (\mathbf{p}_{\chi_i}) v_\xi^s (\mathbf{p}_\xi) \overline{v_\xi^s} (\mathbf{p}_\xi) u_{\chi_i}^r (\mathbf{p}_{\chi_i}) \sum_{\widehat{\alpha}} \frac{Q^{\widehat{\Phi}_{\widehat{\alpha}}}_{\chi_i \xi^c}}{p^2 - p_{\widehat{\Phi}_{\widehat{\alpha}}}^2} + \cdots,
\end{align}
where we have defined a coefficient
\begin{align}
	Q^{\widehat{\Phi}_{\widehat{\alpha}}}_{\chi_i \xi^c} \coloneqq |h_i|^2 (C_\Phi^{-1})_{1 {\widehat{\alpha}}} (C_\Phi)_{{\widehat{\alpha}} 1}.
\end{align}
Using $\sum_{r, s} \overline{u_{\chi_i}^r} v_\xi^s \overline{v_\xi^s} u_{\chi_i}^r = 4 (p_\xi \cdot p_{\chi_i})$, we can write in the CM frame
\begin{align}
	\sum_{r, s} \text{Im} [\mathcal{M} (\chi_{i, r} \xi^c_s \to \chi_{i, r} \xi^c_s)] &= \minus 4 (p_\xi \cdot p_{\chi_i}) \sum_{\widehat{\alpha}} \text{Im} \bigg[ \frac{Q^{\widehat{\Phi}_{\widehat{\alpha}}}_{\chi_i \xi^c}}{p^2 - p_{\widehat{\Phi}_{\widehat{\alpha}}}^2} \bigg] \nonumber \\
	&= 2 (E^2 - m_{\chi_i}^2) \sum_{\widehat{\alpha}} \frac{\minus \text{Im} \big[ Q^{\widehat{\Phi}_{\widehat{\alpha}}}_{\chi_i \xi^c} \big] (E^2 - m_{\widehat{\Phi}_{\widehat{\alpha}}}^2) + \text{Re} \big[ Q^{\widehat{\Phi}_{\widehat{\alpha}}}_{\chi_i \xi^c} \big] m_{\widehat{\Phi}_{\widehat{\alpha}}} \Gamma_{\widehat{\Phi}_{\widehat{\alpha}}}}{(E^2 - m_{\widehat{\Phi}_{\widehat{\alpha}}}^2)^2 + (m_{\widehat{\Phi}_{\widehat{\alpha}}} \Gamma_{\widehat{\Phi}_{\widehat{\alpha}}})^2}.
	\label{eq:ImM}
\end{align}
As in the previous derivation, we need an \textit{on-shell prescription} to remove the divergent off-shell contribution. In the right-hand side of equation \ref{eq:ImM}, the term $E^2 - m_{\widehat{\Phi}_{\widehat{\alpha}}}^2$ in the numerator is such an off-shell part, and let us therefore neglect it. Applying the narrow-width approximation of equation \ref{eq:NWA} to the remaining part, we obtain
\begin{empheq}[box=\fbox]{align}
	\sum_{\widehat{\alpha}} \Gamma^\text{opt}_{\widehat{\Phi}_{\widehat{\alpha}} \to \chi_i \xi^c}
	&\coloneqq \int d\Pi_{\chi_i} \int d\Pi_{\xi^c} \, (2 \pi)^3 \delta^3 (\mathbf{p}_{\chi_i} + \mathbf{p}_{\xi^c}) \sum_{r, s} \text{Im} [\mathcal{M}_\text{OS} (\chi_{i, r} \xi^c_s \to \chi_{i, r} \xi^c_s)], \\
	&= \frac{m_\Phi}{16 \pi} \bigg( 1 - \frac{m_{\chi_i}^4}{m_\Phi^4} \bigg) \sum_{\widehat{\alpha}} \text{Re} \big[ Q^{\widehat{\Phi}_{\widehat{\alpha}}}_{\chi_i \xi^c} \big],
	\label{eq:DWOptCXc}
\end{empheq}
where the second equality is correct up to the leading order in perturbation. Similar calculations can be done for the other final states as well. For $\widehat{\Phi}_{\widehat{\alpha}} \to \chi_i^c \xi$, we obtain
\begin{align}
	\sum_{\widehat{\alpha}} \Gamma^\text{opt}_{\widehat{\Phi}_{\widehat{\alpha}} \to \chi_i^c \xi}
	&\coloneqq \int d\Pi_{\chi_i^c} \int d\Pi_\xi \, (2 \pi)^3 \delta^3 (\mathbf{p}_{\chi_i^c} + \mathbf{p}_\xi) \sum_{r, s} \text{Im} [\mathcal{M}_\text{OS} (\chi^c_{i, r} \xi_s \to \chi^c_{i, r} \xi_s)] \\
	&= \frac{m_\Phi}{16 \pi} \bigg( 1 - \frac{m_{\chi_i}^4}{m_\Phi^4} \bigg) \sum_{\widehat{\alpha}} \text{Re} \big[ Q^{\widehat{\Phi}_{\widehat{\alpha}}}_{\chi_i^c \xi} \big],
\end{align}
where
\begin{align}
	Q^{\widehat{\Phi}_{\widehat{\alpha}}}_{\chi_i^c \xi} \coloneqq |h_i|^2 (C_\Phi^{-1})_{2 {\widehat{\alpha}}} (C_\Phi)_{{\widehat{\alpha}} 2}.
\end{align}
In addition, for $\widehat{\Phi}_{\widehat{\alpha}} \to \psi_i \psi_i^c$, we have
\begin{align}
	\sum_{\widehat{\alpha}} \Gamma^\text{opt}_{\widehat{\Phi}_{\widehat{\alpha}} \to \psi_i \psi_i^c}
	&\coloneqq \int d\Pi_\psi \int d\Pi_{\psi^c} \ (2 \pi)^3 \delta^3 (\mathbf{p}_\psi + \mathbf{p}_{\psi^c}) \sum_{r, s} \text{Im} [\mathcal{M}_\text{OS} (\psi_{i, r} \psi_{i, s}^c \to \psi_{i, r} \psi_{i, s}^c)] \\
	&= \frac{m_\Phi}{16 \pi} \sum_{\widehat{\alpha}} \text{Re} \big[ Q^{\widehat{\Phi}_{\widehat{\alpha}}}_{\psi_i \psi_i^c} \big],
\end{align}
where
\begin{align}
	Q^{\widehat{\Phi}_{\widehat{\alpha}}}_{\psi_i \psi_i^c}
	&\coloneqq |f_i|^2 (C_\Phi^{-1})_{1 {\widehat{\alpha}}} (C_\Phi)_{{\widehat{\alpha}} 1} + |f_i|^2 (C_\Phi^{-1})_{2 {\widehat{\alpha}}} (C_\Phi)_{{\widehat{\alpha}} 2}.
\end{align}
If we take off $\sum_{\widehat{\alpha}}$ from $\sum_{\widehat{\alpha}} \Gamma^\text{opt}_{\widehat{\Phi}_{\widehat{\alpha}} \to Y}$, the resulting expression is a real number. However, it does not necessarily mean that $\Gamma^\text{opt}_{\widehat{\Phi}_{\widehat{\alpha}} \to Y}$ can be interpreted as a quantity particularly related to $\widehat{\Phi}_{\widehat{\alpha}}$, because what we have obtained through rigorous mathematical steps is the sum and it tells nothing about each contribution. The quantity $\sum_{\widehat{\alpha}} \Gamma^\text{opt}_{\widehat{\Phi}_{\widehat{\alpha}} \to Y}$ should be regarded as the final result free from any logical loophole.

\subsection{Comparison of transition rates}		\label{sec:DWComp}
We have obtained two apparently different candidates to which the partial decay width of $\widehat{\Phi}_{\widehat{\alpha}}$ is going to be related, \textit{i.e.}, $\sum_{\widehat{\alpha}} \Gamma^\text{scat}_{\widehat{\Phi}_{\widehat{\alpha}} \to Y}$ and $\sum_{\widehat{\alpha}} \Gamma^\text{opt}_{\widehat{\Phi}_{\widehat{\alpha}} \to Y}$. The expression $\sum_{\widehat{\alpha}} \Gamma^t_{\widehat{\Phi}_{\widehat{\alpha}} \to Y}$ is identical to $\sum_{\widehat{\alpha}} \Gamma^\text{scat}_{\widehat{\Phi}_{\widehat{\alpha}} \to Y}$, which is almost trivial since they are related to each other by the Fourier transform as long as the on-shell prescriptions are mutually consistent. Even though $\sum_{\widehat{\alpha}} \Gamma^\text{scat}_{\widehat{\Phi}_{\widehat{\alpha}} \to Y}$ and $\sum_{\widehat{\alpha}} \Gamma^\text{opt}_{\widehat{\Phi}_{\widehat{\alpha}} \to Y}$ are calculated from the right- and left-hand sides of unitarity condition \ref{eq:UnitCon}, respectively, it is not guaranteed that they are identical since the on-shell prescriptions are differently imposed in their derivations. Note that the unitarity condition assures the equality between two expressions when off-shell as well as on-shell contributions to both of them are taken into account. In particular, there is a clear difference in $\sum_{\widehat{\alpha}} \Gamma^\text{scat}_{\widehat{\Phi}_{\widehat{\alpha}} \to Y}$ and $\sum_{\widehat{\alpha}} \Gamma^\text{opt}_{\widehat{\Phi}_{\widehat{\alpha}} \to Y}$: the interference between physical particles exists in the former while no intereference in the latter.

\begin{figure}[h]
	\centering
	\includegraphics[width = 70 mm]{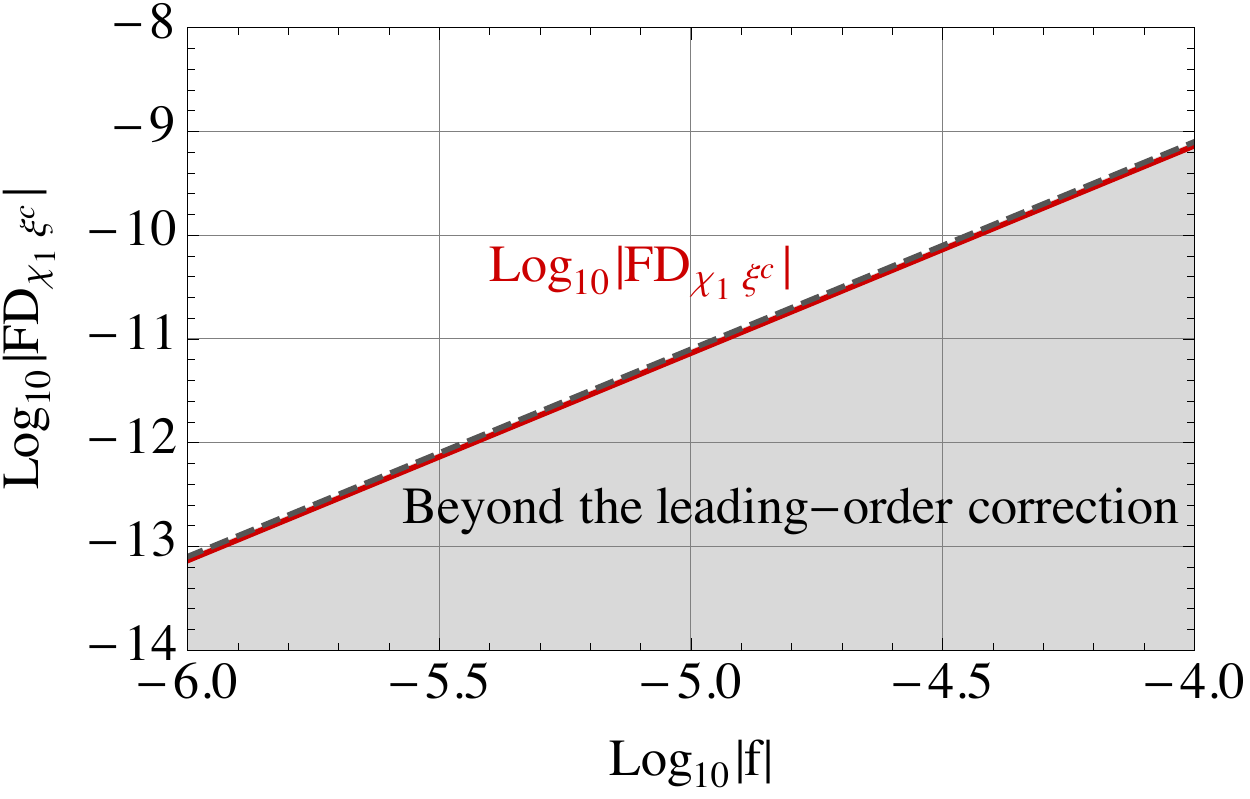}
	\caption{Fractional difference of $\sum_{\widehat{\alpha}} \Gamma^\text{scat}_{\widehat{\Phi}_{\widehat{\alpha}} \to \chi_1 \xi^c}$ and $\sum_{\widehat{\alpha}} \Gamma^\text{opt}_{\widehat{\Phi}_{\widehat{\alpha}} \to \chi_1 \xi^c}$. The difference is approximately of the next-leading order in perturbation, which implies $\sum_{\widehat{\alpha}} \Gamma^\text{scat}_{\widehat{\Phi}_{\widehat{\alpha}} \to \chi_1 \xi^c}$ and $\sum_{\widehat{\alpha}} \Gamma^\text{opt}_{\widehat{\Phi}_{\widehat{\alpha}} \to \chi_1 \xi^c}$ are identical up to the leading order.}
	\label{fig:DWFracDiff}
\end{figure}

In spite of the apparent difference, we claim that they are actually identical:
\begin{align}
	\boxed{\sum_{\widehat{\alpha}} \Gamma_{\widehat{\Phi}_{\widehat{\alpha}} \to Y}
	\coloneqq \sum_{\widehat{\alpha}} \Gamma^\text{scat}_{\widehat{\Phi}_{\widehat{\alpha}} \to Y}
	= \sum_{\widehat{\alpha}} \Gamma^\text{opt}_{\widehat{\Phi}_{\widehat{\alpha}} \to Y}.}
	\label{eq:DWScatOpt}
\end{align}
Moreover, we also claim that they are consistent with the total decay widths obtained from the self-energy in the following sense:
\begin{align}
	\boxed{\sum_{\widehat{\alpha}} \Gamma_{\widehat{\Phi}_{\widehat{\alpha}}}
	= \sum_{\widehat{\alpha}, Y} \Gamma_{\widehat{\Phi}_{\widehat{\alpha}} \to Y}.}
	\label{eq:DWTotScatOpt}
\end{align}
These claims can be verified by explicit numerical evaluations. In order to verify equation \ref{eq:DWScatOpt}, let us consider their fractional difference defined by
\begin{align}
	\text{FD}_Y \coloneqq \left| \frac{\sum_{\widehat{\alpha}} \Gamma^\text{opt}_{\widehat{\Phi}_{\widehat{\alpha}} \to Y} - \sum_{\widehat{\alpha}} \Gamma^\text{scat}_{\widehat{\Phi}_{\widehat{\alpha}} \to Y}}{\sum_{\widehat{\alpha}} \Gamma^\text{scat}_{\widehat{\Phi}_{\widehat{\alpha}} \to Y}} \right|.
\end{align}
If $\text{FD}_{\chi_1 \xi^c} \lesssim \mathcal{O} (\alpha)$, the difference of $\sum_{\widehat{\alpha}} \Gamma^\text{scat}_{\widehat{\Phi}_{\widehat{\alpha}} \to Y}$ and $\sum_{\widehat{\alpha}} \Gamma^\text{opt}_{\widehat{\Phi}_{\widehat{\alpha}} \to Y}$ is beyond the leading order in perturbation. Let us consider the case of $Y = \chi_1 \xi^c$, which will be discussed in section \ref{sec:Example} as a numerical example of imitating the semileptonic decay of $B^0$. We choose the scale of parameter $f$ to be in the range $-6 \leq \log_{10} {|f|} \leq -4$ and set the values of $f_i$ as $f_1 = f$ and $\log_{10} {|f_2|} = \log_{10} {|f_1|} + 0.0036$. All the other parameters are chosen as given in table \ref{tab:BInp}. The result is shown in figure \ref{fig:DWFracDiff}. The fractional difference $\text{FD}_{\chi_1 \xi^c}$ is close to the dotted line of $f^2 / 4 \pi \sim \mathcal{O} (\alpha)$ in the plot, and we therefore conclude that $\sum_{\widehat{\alpha}} \Gamma^\text{scat}_{\widehat{\Phi}_{\widehat{\alpha}} \to \chi_1 \xi^c}$ and $\sum_{\widehat{\alpha}} \Gamma^\text{opt}_{\widehat{\Phi}_{\widehat{\alpha}} \to \chi_1 \xi^c}$ are identical up to the leading order. 
The verification of equation \ref{eq:DWTotScatOpt} is similar. Since the difference between $\sum_{\widehat{\alpha}} \Gamma_{\widehat{\Phi}_{\widehat{\alpha}}}$ and $\sum_{\widehat{\alpha}, Y} \Gamma_{\widehat{\Phi}_{\widehat{\alpha}} \to Y}$ was simply zero in all the cases numerically checked, we do not present any figures about their comparison. Note that equations \ref{eq:DWScatOpt} and \ref{eq:DWTotScatOpt} have been checked and confirmed for many other numerical examples of various decay products.

\subsection{Partial decay widths}
Now we discuss how the partial decay widths of $\Phi$ and $\Phi^*$ can be defined from the transition rates we have derived so far. Assume that the initial field is tagged by $\chi_j^c \xi$ or $\chi_j \xi^c$ using all possible $j$. Then, the partial decay widths can be defined as follows:
\begin{align}
	\boxed{\Gamma_{\Phi \to Y} \coloneqq \sum_{j, \widehat{\alpha}} \Gamma^\text{scat}_{\chi_j \xi^c \to \widehat{\Phi}_{\widehat{\alpha}} \to Y}, \qquad
	\Gamma_{\Phi^* \to Y} \coloneqq \sum_{j, \widehat{\alpha}} \Gamma^\text{scat}_{\chi_j^c \xi \to \widehat{\Phi}_{\widehat{\alpha}} \to Y}.}
	\label{eq:PDW}
\end{align}
In the absence of particle-antiparticle mixing, \textit{i.e.}, in the limit of $f_i = 0$, these expressions reproduce the well-known formulas up to the leading order:
\begin{align}
	\Gamma_{\Phi \to \chi_i \xi^c} &= \Gamma_{\Phi^* \to \chi_i^c \xi} = \frac{|h_i|^2}{16 \pi} m_\Phi \bigg( 1 - \frac{m_{\chi_i}^4}{m_\Phi^4} \bigg), \\
	\Gamma_{\Phi \to \chi_i^c \xi} &= \Gamma_{\Phi^* \to \chi_i \xi^c} = \Gamma_{\Phi \to \psi_i \psi_i^c} = \Gamma_{\Phi^* \to \psi_i \psi_i^c} = 0,
\end{align}
since, in that case, we have $\widehat{\Phi}_1 = \Phi$, $\widehat{\Phi}_2 = \Phi^*$, $C_\Phi = 1$, $\Gamma_{\widehat{\Phi}_1} = \Gamma_{\Phi \to \chi_i \xi^c}$, and $\Gamma_{\widehat{\Phi}_2} = \Gamma_{\Phi^* \to \chi_i^c \xi}$. Hence, the definitions of the partial decay widths given by equation \ref{eq:PDW} are reasonable.

Note that, in equation \ref{eq:PDW}, the initial states such as $| \psi_j \psi_j^c \rangle$ that couple to both $\Phi$ and $\Phi^*$ are excluded. Any scattering process whose initial states are $| \psi_j \psi_j^c \rangle$ must have contributions of both $\psi_j \psi_j^c \to \Phi$ and $\psi_j \psi_j^c \to \Phi^*$, and they interfere with each other. Hence, defining the partial decay widths of $\Phi$ considering $\psi_j \psi_j^c \to \Phi$ while neglecting $\psi_j \psi_j^c \to \Phi^*$ does not make sense, since such a quantity is never directly related to any physical observables. In a real experiment for observing $\Phi \to \chi_i \xi^c$ or $\Phi^* \to \chi_i \xi^c$, the initial field should be tagged. Moreover, it is non-trivial that $\Gamma_{\Phi \to Y}$ and $\Gamma_{\Phi^* \to Y}$ need tagging as defined above, because the total decay widths require that all possible initial multiparticle states be taken into account, as shown while deriving equations \ref{eq:DWScat}, \ref{eq:DWScatOpt}, and \ref{eq:DWTotScatOpt}. Note also that, when $\widehat{\Phi}_{\widehat{\alpha}}$ is on-shell, the field $\Phi$ must be ``off-shell" in the sense that $p^2 \neq m_\Phi^2$. As mentioned before, the one-particle states $| \Phi \rangle$ and $| \Phi^* \rangle$ associated with $\Phi (\mathbf{x})$ are irrelevant to the decay of unstable physical particles, and the ``decay widths" of $| \Phi \rangle$ and $| \Phi^* \rangle$ must be defined from physical particles. This is the correct approach, since $\Phi \to \chi_i \xi^c$ is intended experimentally as well as theoretically to be associated with a decay of something that propagates as a plane wave (or its variation due to a weak external field as in a collider) and what propagates as such is nothing but the physical particle.

When we actually calculate a ratio of two partial decay widths up to the leading order, it does not really matter whether all possible $j$ is considered or not, and we can in fact use only one of them, \textit{e.g.}, $\chi_1^c \xi$ or $\chi_1 \xi^c$ for tagging. This is because the factor relevant to the initial state is common in all the expressions, and they can therefore be factorized to be canceled in a ratio of those decay widths. To be specific, we can define a common factor incorporating the overall multiplicative factor as well as the couplings and phase space factors of the initial states:
\begin{align}
	\mathcal{N} \coloneqq \frac{m_\Phi^2}{2^8 \pi^2} \sum_j |h_j|^2 \bigg( 1 - \frac{m_{\chi_j}^4}{m_\Phi^4} \bigg),
\end{align}
to write the partial decay widths as
\begin{align}
	\Gamma_{\Phi \to \chi_i \xi^c} &= i \mathcal{N} |h_i|^2 \bigg( 1 - \frac{m_{\chi_i}^4}{m_\Phi^4} \bigg)
		\sum_{\widehat{\alpha}, \widehat{\beta}} \frac{(C_\Phi^{-1})_{1 {\widehat{\beta}}}^* (C_\Phi)_{{\widehat{\beta}} 1}^* (C_\Phi^{-1})_{1 {\widehat{\alpha}}} (C_\Phi)_{{\widehat{\alpha}} 1}}{p_{\widehat{\Phi}_{\widehat{\beta}}}^* - p_{\widehat{\Phi}_{\widehat{\alpha}}}},
		\label{eq:PDWPCXc} \\
	\Gamma_{\Phi^* \to \chi_i^c \xi} &= i \mathcal{N} |h_i|^2 \bigg( 1 - \frac{m_{\chi_i}^4}{m_\Phi^4} \bigg)
		\sum_{\widehat{\alpha}, \widehat{\beta}} \frac{(C_\Phi^{-1})_{2 {\widehat{\beta}}}^* (C_\Phi)_{{\widehat{\beta}} 2}^* (C_\Phi^{-1})_{2 {\widehat{\alpha}}} (C_\Phi)_{{\widehat{\alpha}} 2}}{p_{\widehat{\Phi}_{\widehat{\beta}}}^* - p_{\widehat{\Phi}_{\widehat{\alpha}}}},
		\label{eq:PDWPCcX} \\
	\Gamma_{\Phi \to \chi_i^c \xi} &= i \mathcal{N} |h_i|^2 \bigg( 1 - \frac{m_{\chi_i}^4}{m_\Phi^4} \bigg)
		\sum_{\widehat{\alpha}, \widehat{\beta}} \frac{(C_\Phi^{-1})_{2 {\widehat{\beta}}}^* (C_\Phi)_{{\widehat{\beta}} 1}^* (C_\Phi^{-1})_{2 {\widehat{\alpha}}} (C_\Phi)_{{\widehat{\alpha}} 1}}{p_{\widehat{\Phi}_{\widehat{\beta}}}^* - p_{\widehat{\Phi}_{\widehat{\alpha}}}},
		\label{eq:PDWPcCcX} \\
	\Gamma_{\Phi^* \to \chi_i \xi^c} &= i \mathcal{N} |h_i|^2 \bigg( 1 - \frac{m_{\chi_i}^4}{m_\Phi^4} \bigg)
		\sum_{\widehat{\alpha}, \widehat{\beta}} \frac{(C_\Phi^{-1})_{1 {\widehat{\beta}}}^* (C_\Phi)_{{\widehat{\beta}} 2}^* (C_\Phi^{-1})_{1 {\widehat{\alpha}}} (C_\Phi)_{{\widehat{\alpha}} 2}}{p_{\widehat{\Phi}_{\widehat{\beta}}}^* - p_{\widehat{\Phi}_{\widehat{\alpha}}}},
		\label{eq:PDWPcCXc} \\[10pt]
	\Gamma_{\Phi \to \psi_i \psi_i^c} &= i \mathcal{N}
		\sum_{\widehat{\alpha}, \widehat{\beta}} \frac{1}{p_{\widehat{\Phi}_{\widehat{\beta}}}^* - p_{\widehat{\Phi}_{\widehat{\beta}}}} \big[ f_i (C_\Phi^{-1})_{1 {\widehat{\beta}}} (C_\Phi)_{{\widehat{\beta}} 1} + f_i^* (C_\Phi^{-1})_{2 {\widehat{\beta}}} (C_\Phi)_{{\widehat{\beta}} 1} \big]^* \nonumber \\
			&\qquad \qquad \qquad \qquad \qquad \big[ f_i (C_\Phi^{-1})_{1 {\widehat{\alpha}}} (C_\Phi)_{{\widehat{\alpha}} 1} + f_i^* (C_\Phi^{-1})_{2 {\widehat{\alpha}}} (C_\Phi)_{{\widehat{\alpha}} 1} \big],
		\label{eq:PDWPPsPs} \\[10pt]
	\Gamma_{\Phi^* \to \psi_i \psi_i^c} &= i \mathcal{N}
		\sum_{\widehat{\alpha}, \widehat{\beta}} \frac{1}{p_{\widehat{\Phi}_{\widehat{\beta}}}^* - p_{\widehat{\Phi}_{\widehat{\beta}}}} \big[ f_i (C_\Phi^{-1})_{1 {\widehat{\beta}}} (C_\Phi)_{{\widehat{\beta}} 2} + f_i^* (C_\Phi^{-1})_{2 {\widehat{\beta}}} (C_\Phi)_{{\widehat{\beta}} 2} \big]^* \nonumber \\
			&\qquad \qquad \qquad \qquad \qquad \big[ f_i (C_\Phi^{-1})_{1 {\widehat{\alpha}}} (C_\Phi)_{{\widehat{\alpha}} 2} + f_i^* (C_\Phi^{-1})_{2 {\widehat{\alpha}}} (C_\Phi)_{{\widehat{\alpha}} 2} \big].
		\label{eq:PDWPcPsPs}
\end{align}
Note that this factorization is correct only up to the leading order. We can also define the transition probability, for example, as follows:
\begin{align}
	\mathcal{P}_{\Phi^* \to \Phi} &\coloneqq \mathcal{N}' \sum_i \Gamma_{\Phi^* \to \chi_i \xi^c}
		\propto i \sum_{\widehat{\alpha}, \widehat{\beta}} \frac{(C_\Phi^{-1})_{1 {\widehat{\beta}}}^* (C_\Phi)_{{\widehat{\beta}} 2}^* (C_\Phi^{-1})_{1 {\widehat{\alpha}}} (C_\Phi)_{{\widehat{\alpha}} 2}}{p_{\widehat{\Phi}_{\widehat{\beta}}}^* - p_{\widehat{\Phi}_{\widehat{\alpha}}}},
	\label{eq:TransPropPcP}
\end{align}
where $\mathcal{N}'$ is a normalization constant.\footnote{This is an expression consistent with equation 187 of reference \cite{OscNeutMesonQFT} since $C_\Phi = V$.}

\subsection{CP asymmetry}		\label{sec:CPAsym}
The CP asymmetry in the decay of a meson is a quantity calculated from partial decay widths. For example, the CP asymmetry in the semileptonic decay $B^0 \to D^- \ell^+ \nu_\ell$ is defined as
\begin{align}
	\mathcal{A}_\text{CP}^\text{SL} \coloneqq \frac{\Gamma_{\overline{B^0} \to D^+ \ell^- \overline{\nu_\ell}} - \Gamma_{B^0 \to D^- \ell^+ \nu_\ell}}{\Gamma_{\overline{B^0} \to D^+ \ell^- \overline{\nu_\ell}} + \Gamma_{B^0 \to D^- \ell^+ \nu_\ell}}.
\end{align}
In an experiment designed to observe this quantity, the initial field can be set to $B^0$ or $\overline{B^0}$ by an interaction that only couples to either of them, \textit{i.e.}, the initial field can be tagged as we want.

Using the definitions of partial decay widths given by equation \ref{eq:PDW}, we can define the CP asymmetry in the decays of $\Phi \to Y$ and $\Phi^* \to Y^c$ as follows:
\begin{empheq}[box=\fbox]{align}
	\mathcal{A}_\text{CP}
	&\coloneqq \frac{\Gamma_{\Phi^* \to Y^c} - \Gamma_{\Phi \to Y}}{\Gamma_{\Phi^* \to Y^c} + \Gamma_{\Phi \to Y}} \\
	&= \frac{\sum_{j, \widehat{\alpha}} \Gamma^\text{scat}_{\chi_j^c \xi \to \widehat{\Phi}_{\widehat{\alpha}} \to Y^c} - \sum_{j, \widehat{\alpha}} \Gamma^\text{scat}_{\chi_j \xi^c \to \widehat{\Phi}_{\widehat{\alpha}} \to Y}}
		{\sum_{j, \widehat{\alpha}} \Gamma^\text{scat}_{\chi_j^c \xi \to \widehat{\Phi}_{\widehat{\alpha}} \to Y^c} + \sum_{j, \widehat{\alpha}} \Gamma^\text{scat}_{\chi_j \xi^c \to \widehat{\Phi}_{\widehat{\alpha}} \to Y}}.
	\label{eq:CPAsym}
\end{empheq}
The CP asymmetry up to the leading order in the flavor-specific decay such as the semileptonic decay, for example, is written as
\begin{align}
	\mathcal{A}_\text{CP}
	&\coloneqq \frac{\Gamma_{\Phi^* \to \chi_i \xi^c} - \Gamma_{\Phi \to \chi_i^c \xi}}{\Gamma_{\Phi^* \to \chi_i \xi^c} + \Gamma_{\Phi \to \chi_i^c \xi}} \nonumber \\
	&= \frac{\sum_{\widehat{\alpha}, \widehat{\beta}} \frac{1}{p_{\widehat{\Phi}_{\widehat{\beta}}}^* - p_{\widehat{\Phi}_{\widehat{\alpha}}}}
			\big[ (C_\Phi^{-1})_{1 {\widehat{\beta}}}^* (C_\Phi)_{{\widehat{\beta}} 2}^* (C_\Phi^{-1})_{1 {\widehat{\alpha}}} (C_\Phi)_{{\widehat{\alpha}} 2}
				- (C_\Phi^{-1})_{2 {\widehat{\beta}}}^* (C_\Phi)_{{\widehat{\beta}} 1}^* (C_\Phi^{-1})_{2 {\widehat{\alpha}}} (C_\Phi)_{{\widehat{\alpha}} 1} \big]}
		{\sum_{\widehat{\alpha}, \widehat{\beta}} \frac{1}{p_{\widehat{\Phi}_{\widehat{\beta}}}^* - p_{\widehat{\Phi}_{\widehat{\alpha}}}}
			\big[ (C_\Phi^{-1})_{1 {\widehat{\beta}}}^* (C_\Phi)_{{\widehat{\beta}} 2}^* (C_\Phi^{-1})_{1 {\widehat{\alpha}}} (C_\Phi)_{{\widehat{\alpha}} 2}
				+ (C_\Phi^{-1})_{2 {\widehat{\beta}}}^* (C_\Phi)_{{\widehat{\beta}} 1}^* (C_\Phi^{-1})_{2 {\widehat{\alpha}}} (C_\Phi)_{{\widehat{\alpha}} 1} \big]}.
	\label{eq:CPAsymFS}
\end{align}

Now we briefly discuss the CP asymmetry in the direct decay. In the toy model presented in this paper, there does not exist CP violation in the direct decay, \textit{i.e.}, $\Gamma_{\Phi \to \chi_i \xi^c} = \Gamma_{\Phi^* \to \chi_i^c \xi}$. We can, of course, extend the toy model to allow CP violation in the direct decay by introducing an additional field. It is evident that the new field should not be a complex scalar since it does would induce a vertex loop correction as is the case only with $\Phi$ and $\Phi^*$. It also had better not be a real scalar field, since it would mix with $\Phi$ and $\Phi^*$ to complicate the analysis. One of the simplest way is introducing a spin-one field such as a gauge boson $A^\mu$ which couples to $\chi_j^c \xi$ by $\overline{\chi_j} \gamma_\mu \xi A^\mu$. However, CP violation in the direct decay is beyond the scope of this paper, since it requires loop corrections to the vertices, \textit{i.e.}, the next-leading-order effects in perturbation. To keep consistency in perturbation in a neutral meson system, calculating decay widths up to the one-loop correction to the vertices requires calculating two-loop contributions to the self-energy, as mentioned before. We will come back to this issue in section \ref{sec:BeyLO}, where the derivation of decay widths beyond the leading order will be briefly discussed.

\section{Comparison with the analysis in quantum mechanics}		\label{sec:QM}
In this section, we calculate $\Gamma_{\Phi^* \to \chi_i \xi^c}$ using the conventional method based on quantum mechanics, and compare the result with equation \ref{eq:PDWPcCXc}. It turns out that there exists a discrepancy which sometimes results in a big difference in the values of decay widths. \\

Following the standard approach presented in references \cite{BPhysTeva, MesonMixing&CKM}, we introduce time-dependent states $| \Phi (t) \rangle$ and $| \Phi^* (t) \rangle$, which satisfy the effective Schr\"odinger equation
\begin{align}
	i \frac{d}{dt} \left( \begin{array}{c} | \Phi (t) \rangle \\ | \Phi^* (t) \rangle \end{array} \right)
	= H_\text{eff} \left( \begin{array}{c} | \Phi (t) \rangle \\ | \Phi^* (t) \rangle \end{array} \right).
	\label{eq:SchEq}
\end{align}
Here, the effective Hamiltonian $H_\text{eff}$ is written in terms of the self-energy up to the one-loop order as
\begin{align}
	H_\text{eff} \coloneqq m_\Phi \bigg[ 1 - \frac{1}{2} \Sigma_\Phi' (m_\Phi^2) \bigg],
\end{align}
and it is diagonalized as
\begin{align}
	C_\Phi H_\text{eff} C_\Phi^{-1}
	= m_\Phi \bigg[ 1 - \frac{1}{2} (C \Sigma_\Phi' C^{-1}) (m_\Phi^2) \bigg]
	= m_\Phi \bigg[ 1 - \frac{1}{2} \widehat{\Sigma}_{\widehat{\Phi}}' (m_\Phi^2) \bigg]
	= \left( \begin{array}{cc} p_{\widehat{\Phi}_1} & 0 \\ 0 & p_{\widehat{\Phi}_2} \end{array} \right),
\end{align}
where $p_{\widehat{\Phi}_{\widehat{\alpha}}} = m_{\widehat{\Phi}_{\widehat{\alpha}}} - i \Gamma_{\widehat{\Phi}_{\widehat{\alpha}}} / 2$. Hence, defining the time-dependent mass eigenstates by
\begin{align}
	\left( \begin{array}{c} | \widehat{\Phi}_1 (t) \rangle \\ | \widehat{\Phi}_2 (t) \rangle \end{array} \right)
		\coloneqq C_\Phi \left( \begin{array}{c} | \Phi (t) \rangle \\ | \Phi^* (t) \rangle \end{array} \right)
\end{align}
and solving the decoupled Schr\"odinger equation for each mass eigenstate, we obtain
\begin{align}
	| \widehat{\Phi}_{\widehat{\alpha}} (t) \rangle = e^{-i p_{\widehat{\Phi}_{\widehat{\alpha}}} (t - t_0)} \ | \widehat{\Phi}_{\widehat{\alpha}} \rangle.
	\label{eq:TEvol}
\end{align}
Here, $| \widehat{\Phi}_{\widehat{\alpha}} \rangle = | \widehat{\Phi}_{\widehat{\alpha}} (t_0) \rangle$ is the mass eigenstate that can be expressed as a fixed linear combination of $| \Phi \rangle$ and $| \Phi^* \rangle$, and it evolves in time as a plane wave with a damping factor. Setting the initial states as
\begin{align}
	| \Phi (t_0) \rangle = | \Phi \rangle, \qquad
	| \Phi^* (t_0) \rangle = | \Phi^* \rangle,
\end{align}
\textit{i.e.},
\begin{align}
	| \widehat{\Phi}_1 \rangle = (C_\Phi)_{11} | \Phi \rangle + (C_\Phi)_{12} | \Phi^* \rangle, \qquad
	| \widehat{\Phi}_2 \rangle = (C_\Phi)_{21} | \Phi \rangle + (C_\Phi)_{22} | \Phi^* \rangle,
	\label{eq:MassEigen}
\end{align}
we can write
\begin{align}
	\left( \begin{array}{c} | \Phi (t) \rangle \\ | \Phi^* (t) \rangle \end{array} \right)
	&= C_\Phi^{-1} \left( \begin{array}{cc} e^{-i p_{\widehat{\Phi}_1} (t - t_0)} & 0 \\
			0 & e^{-i p_{\widehat{\Phi}_2} (t - t_0)} \end{array} \right) C_\Phi
		\left( \begin{array}{c} | \Phi \rangle \\ | \Phi^* \rangle \end{array} \right) \\
	&= \sum_{\widehat{\alpha}} e^{-i p_{\widehat{\Phi}_{\widehat{\alpha}}} (t - t_0)}
		\left( \begin{array}{cc} (C_\Phi^{-1})_{1 \widehat{\alpha}} (C_\Phi)_{\widehat{\alpha} 1} &
			(C_\Phi^{-1})_{1 \widehat{\alpha}} (C_\Phi)_{\widehat{\alpha} 2} \\
			(C_\Phi^{-1})_{2 \widehat{\alpha}} (C_\Phi)_{\widehat{\alpha} 1} &
			(C_\Phi^{-1})_{2 \widehat{\alpha}} (C_\Phi)_{\widehat{\alpha} 2} \end{array} \right)
		\left( \begin{array}{c} | \Phi \rangle \\ | \Phi^* \rangle \end{array} \right),
\end{align}
and thus
\begin{align}
	| \Phi (t) \rangle &= \sum_{\widehat{\alpha}} e^{-i p_{\widehat{\Phi}_{\widehat{\alpha}}} (t - t_0)} \big[ (C_\Phi^{-1})_{1 \widehat{\alpha}} (C_\Phi)_{\widehat{\alpha} 1} | \Phi \rangle 
		+ (C_\Phi^{-1})_{1 \widehat{\alpha}} (C_\Phi)_{\widehat{\alpha} 2} | \Phi^* \rangle \big], \label{eq:Pt} \\
	| \Phi^* (t) \rangle &= \sum_{\widehat{\alpha}} e^{-i p_{\widehat{\Phi}_{\widehat{\alpha}}} (t - t_0)} \big[ (C_\Phi^{-1})_{2 \widehat{\alpha}} (C_\Phi)_{\widehat{\alpha} 1} | \Phi \rangle
		+ (C_\Phi^{-1})_{2 \widehat{\alpha}} (C_\Phi)_{\widehat{\alpha} 2} | \Phi^* \rangle \big]. \label{eq:Pct}
\end{align}
In other words, $| \Phi (t) \rangle$ and $| \Phi^* (t) \rangle$ are the states at time $t$ which evolved from $| \Phi \rangle$ and $| \Phi^* \rangle$ at $t = t_0$, respectively. Since the potential in the Schr\"odinger picture is given by
\begin{align}
	V = \int d^3 \mathbf{x} \, \bigg[ \sum_i f_i \overline{\psi_i} (\mathbf{x}) \psi_i (\mathbf{x}) \Phi (\mathbf{x}) + \sum_i h_i \overline{\chi_i} (\mathbf{x}) \xi (\mathbf{x}) \Phi (\mathbf{x}) + \text{H.c.} \bigg],
\end{align}
we can write
\begin{alignat}{2}
	&\minus i \langle \chi_i \xi^c | V | \Phi \rangle \propto \minus i h_i \, \delta^3 (\mathbf{p} - \mathbf{p}_{\chi_i} - \mathbf{p}_{\xi^c}), \qquad
	&&\langle \chi_i^c \xi | V | \Phi \rangle = 0, \\
	&\minus i \langle \chi_i^c \xi | V | \Phi^* \rangle \propto \minus i h_i^* \, \delta^3 (\mathbf{p} - \mathbf{p}_{\chi_i^c} - \mathbf{p}_\xi), \qquad
	&&\langle \chi_i \xi^c | V | \Phi^* \rangle = 0,
\end{alignat}
where $\mathbf{p}$ is the total three-momentum, \textit{i.e.}, the momentum of $\Phi$ and $\Phi^*$. Hence,
\begin{align}
	\minus i \langle \chi_i \xi^c | V | \Phi (t) \rangle &\propto \minus i \sum_{\widehat{\alpha}} e^{-i p_{\widehat{\Phi}_{\widehat{\alpha}}} (t - t_0)} Q^{\Phi (t)}_{\widehat{\Phi}_{\widehat{\alpha}} \to \chi_i \xi^c} \, \delta^3 (\mathbf{p} - \mathbf{p}_{\chi_i} - \mathbf{p}_{\xi^c}), \label{eq:PtCXc} \\
	\minus i \langle \chi_i^c \xi | V | \Phi (t) \rangle &\propto \minus i \sum_{\widehat{\alpha}} e^{-i p_{\widehat{\Phi}_{\widehat{\alpha}}} (t - t_0)} Q^{\Phi (t)}_{\widehat{\Phi}_{\widehat{\alpha}} \to \chi_i^c \xi} \, \delta^3 (\mathbf{p} - \mathbf{p}_{\chi_i^c} - \mathbf{p}_\xi), \label{eq:PtCcX} \\
	\minus i \langle \chi_i \xi^c | V | \Phi^* (t) \rangle &\propto \minus i \sum_{\widehat{\alpha}} e^{-i p_{\widehat{\Phi}_{\widehat{\alpha}}} (t - t_0)} Q^{\Phi^* (t)}_{\widehat{\Phi}_{\widehat{\alpha}} \to \chi_i \xi^c} \, \delta^3 (\mathbf{p} - \mathbf{p}_{\chi_i} - \mathbf{p}_{\xi^c}), \label{eq:PctCXc} \\
	\minus i \langle \chi_i^c \xi | V | \Phi^* (t) \rangle &\propto \minus i \sum_{\widehat{\alpha}} e^{-i p_{\widehat{\Phi}_{\widehat{\alpha}}} (t - t_0)} Q^{\Phi^* (t)}_{\widehat{\Phi}_{\widehat{\alpha}} \to \chi_i^c \xi} \, \delta^3 (\mathbf{p} - \mathbf{p}_{\chi_i^c} - \mathbf{p}_\xi), \label{eq:PctCcX}
\end{align}
where
\begin{alignat}{2}
	&Q^{\Phi (t)}_{\widehat{\Phi}_{\widehat{\alpha}} \to \chi_i \xi^c} \coloneqq h_i (C_\Phi^{-1})_{1 {\widehat{\alpha}}} (C_\Phi)_{{\widehat{\alpha}} 1}, \qquad
	&&Q^{\Phi (t)}_{\widehat{\Phi}_{\widehat{\alpha}} \to \chi_i^c \xi} \coloneqq h_i^* (C_\Phi^{-1})_{1 {\widehat{\alpha}}} (C_\Phi)_{{\widehat{\alpha}} 2}, \\
	&Q^{\Phi^* (t)}_{\widehat{\Phi}_{\widehat{\alpha}} \to \chi_i \xi^c} \coloneqq h_i (C_\Phi^{-1})_{2 {\widehat{\alpha}}} (C_\Phi)_{{\widehat{\alpha}} 1}, \qquad
	&&Q^{\Phi^* (t)}_{\widehat{\Phi}_{\widehat{\alpha}} \to \chi_i^c \xi} \coloneqq h_i^* (C_\Phi^{-1})_{2 {\widehat{\alpha}}} (C_\Phi)_{{\widehat{\alpha}} 2}.
\end{alignat}

As a specific example, let us consider $\Phi^* \to \chi_i \xi^c$ whose initial state $| \Phi^* \rangle$ is tagged by $| \chi_j^c \xi \rangle$ at $t = t_0$. To calculate the decay widths, we have to integrate over the phase space of $| \chi_i \xi^c \rangle$, through which the delta function can also be appropriately dealt with. However, it requires the spin sum over spinors, which is beyond the current quantum mechanical approach. In fact, we only need the coefficient of the delta function in equation \ref{eq:PctCXc}, since we are interested in a ratio of decay widths in which the phase space factor is canceled out. Hence, introducing a common factor $\mathcal{N}_{\Phi^* \to \chi_i \xi^c} $, we can write the decay width as 
\begin{align}
	\Gamma^\text{QM}_{\Phi^* \to \chi_i \xi^c} &= \mathcal{N}_{\Phi^* \to \chi_i \xi^c}  \int_{t_0}^\infty dt \, \bigg| \sum_{\widehat{\alpha}} e^{-i p_{\widehat{\Phi}_{\widehat{\alpha}}} (t - t_0)} Q^{\Phi^* (t)}_{\widehat{\Phi}_{\widehat{\alpha}} \to \chi_i \xi^c} \bigg|^2 \nonumber \\
	&= \mathcal{N}_{\Phi^* \to \chi_i \xi^c} \sum_{\widehat{\alpha}, \widehat{\beta}} \big[ Q^{\Phi^* (t)}_{\widehat{\Phi}_{\widehat{\beta}} \to \chi_i \xi^c} \big]^* Q^{\Phi^* (t)}_{\widehat{\Phi}_{\widehat{\alpha}} \to \chi_i \xi^c}
		\int_{t_0}^\infty dt \, e^{i (p_{\widehat{\Phi}_{\widehat{\beta}}}^* - p_{\widehat{\Phi}_{\widehat{\alpha}}}) (t - t_0)} \nonumber \\
	&= i \mathcal{N}_{\Phi^* \to \chi_i \xi^c} \sum_{\widehat{\alpha}, \widehat{\beta}} \frac{\big[ Q^{\Phi^* (t)}_{\widehat{\Phi}_{\widehat{\beta}} \to \chi_i \xi^c} \big]^* Q^{\Phi^* (t)}_{\widehat{\Phi}_{\widehat{\alpha}} \to \chi_i \xi^c}}{p_{\widehat{\Phi}_{\widehat{\beta}}}^* - p_{\widehat{\Phi}_{\widehat{\alpha}}}}.
	\label{eq:DWtCXcQM}
\end{align}
Now we compare this expression with equation \ref{eq:DWtCXcQFT} obtained in the quantum field theory. First we can identify
\begin{align}
	\mathcal{N}_{\Phi^* \to \chi_i \xi^c} \coloneqq \frac{1}{2^8 \pi^2} \sum_j |h_j|^2 m_\Phi^2 \bigg( 1 - \frac{m_{\chi_i}^4}{m_\Phi^4} \bigg) \bigg( 1 - \frac{m_{\chi_j}^4}{m_\Phi^4} \bigg)
\end{align}
to incorporate all the information on the interaction $\chi_j^c \xi \to \Phi^*$ as well as the phase space factor of $| \chi_i \xi^c \rangle$. The transition probability can also be written as\footnote{This expression is consistent with equation 12 of reference \cite{OscNeutMesonQFT}. In that paper, $U$ was defined by $| \Phi_\alpha \rangle \eqqcolon \sum_{\widehat{\beta}} U_{\alpha \widehat{\beta}} | \widehat{\Phi}_{\widehat{\beta}} \rangle$, and thus equation \ref{eq:MassEigen} shows that $C_\Phi = U^{-1}$. Hence, equation 12 and 187 in reference \cite{OscNeutMesonQFT} are in fact mutually inconsistent since $C_\Phi = V$ in the latter. Note that reference \cite{OscNeutMesonQFT} drew the opposite conclusion that they are consistent claiming $V = U^\mathsf{T}$ without justification.}
\begin{align}
	\mathcal{P}^\text{QM}_{\Phi^* \to \Phi} &\coloneqq \mathcal{N}' \sum_i \Gamma^\text{QM}_{\Phi^* \to \chi_i \xi^c}
		\propto i \sum_{\widehat{\alpha}, \widehat{\beta}} \frac{(C_\Phi^{-1})_{2 {\widehat{\beta}}}^* (C_\Phi)_{{\widehat{\beta}} 1}^* (C_\Phi^{-1})_{2 {\widehat{\alpha}}} (C_\Phi)_{{\widehat{\alpha}} 1}}{p_{\widehat{\Phi}_{\widehat{\beta}}}^* - p_{\widehat{\Phi}_{\widehat{\alpha}}}}.
	\label{eq:TransPropPcPQM}
\end{align}
Note that there exists a difference between equations \ref{eq:DWtCXcQFT} and \ref{eq:DWtCXcQM} or between equations \ref{eq:TransPropPcP} and \ref{eq:TransPropPcPQM}, which originates from the difference in $Q^{\Phi^* (t)}_{\widehat{\Phi}_{\widehat{\alpha}} \to \chi_i \xi^c}$ and $Q^{\chi_j^c \xi}_{\widehat{\Phi}_{\widehat{\alpha}} \to \chi_i \xi^c}$: $\beta$ and $\gamma$ in $(C_\Phi^{-1})_{\beta {\widehat{\alpha}}} (C_\Phi)_{{\widehat{\alpha}} \gamma}$ are in the opposite order. \\

In fact, any approach to find a solution solving an equation of motion cannot succeed, since such a method is supposed to obtain a time evolution as given by equation \ref{eq:TEvol} for each of physical particles. That solution is correct only when there exists a mass eigenstate, which is a specific linear combination of $| \Phi \rangle$ and $| \Phi^* \rangle$, such that it evolves in time as a plane wave while keeping its identity. In the derivation based on the quantum field theory, however, we have seen that there does not exist a one-particle state in the Hilbert space corresponding to each of physical unstable particles.

The discrepancy in the ordering of indices can sometimes cause big differences in the values of the decay widths and their ratios. As an example, let us consider $\Phi \to \psi_i \psi_i^c$, all the contributions to which can be grouped into $\Phi \to \cdots \to \Phi \to \psi_i \psi_i^c$ and $\Phi \to \cdots \to \Phi^* \to \psi_i \psi_i^c$. In the derivation using the quantum field theory where we have obtained equations \ref{eq:DWScatCXcPP} and \ref{eq:DWQCXcPP}, their coefficients are $(C_\Phi^{-1})_{1 {\widehat{\alpha}}} (C_\Phi)_{{\widehat{\alpha}} 1}$ and $(C_\Phi^{-1})_{2 {\widehat{\alpha}}} (C_\Phi)_{{\widehat{\alpha}} 1}$, respectively. On the other hand, in the derivations based on quantum mechanics, the coefficients are $(C_\Phi^{-1})_{1 {\widehat{\alpha}}} (C_\Phi)_{{\widehat{\alpha}} 1}$ and $(C_\Phi^{-1})_{1 {\widehat{\alpha}}} (C_\Phi)_{{\widehat{\alpha}} 2}$, respectively. Since there exists an interference between those two contributions, the different ordering of the indices can sometimes give a significantly different result. We will see such an example in section \ref{sec:Example}.

\section{Examples}		\label{sec:Example}
In this section, we discuss two numerical examples in which the toy model imitates real meson systems. In the first example, we will see that the model can indeed closely mimic the $B^0$-$\overline{B^0}$ system. It is a simple test of the formalism developed in this paper. The second example is the $K^0$-$\overline{K^0}$ system, through which the nature of physical particles will be more clearly understood. In both examples, we will indeed see that the contribution of each of physical particles to a transition rate is sometimes complex-valued.

\subsection{$B^0$-$\overline{B^0}$}
Here, we choose the model parameters such that it imitates the $B^0$-$\overline{B^0}$ system. The input parameters of the toy model are presented in table \ref{tab:BInp}, and the corresponding output parameters are given in table \ref{tab:BOutp}. The mixing matrix $C_\Phi$ is found to be
\begin{align}
	C_\Phi = \left( \begin{array}{cc} 0.519268 + i \, 0.480066 & \minus 0.707034 \\ 0.519268 + i \, 0.480066 & 0.707034 \end{array} \right)
	\eqqcolon \left( \begin{array}{cc} p & q \\ p & \minus q \end{array} \right),
\end{align}
which is a non-unitary matrix:
\begin{align}
	C_\Phi^\dag C_\Phi = \left( \begin{array}{cc} 2 |p|^2 & 0 \\ 0 & 2 |q|^2 \end{array} \right)
	= \left( \begin{array}{cc} 1.00021 & 0 \\ 0 & 0.999794 \end{array} \right).
\end{align}
The components of $C_\Phi$ are chosen to satisfy $|p|^2 + |q|^2 = 1$. Note that the CP asymmetry defined by equation \ref{eq:CPAsym} indeed gives a value that is consistent with the observed values of the CP asymmetry in the semileptonic decay $B^0 \to D^- \mu^+ \nu_\mu X$ or $B^0 \to D^{* -} \mu^+ \nu_\mu X$, where $X$ denotes any additional decay products. The predicted values in the Standard Model (SM) are also provided for comparison. The deviation from unitarity of $C_\Phi$ is of the order of the CP asymmetry, $\mathcal{O} (10^{-4})$, which is much larger than the typical perturbative correction, $\mathcal{O} (\alpha) \sim \mathcal{O} (10^{-12})$.

\begin{figure}[h]
	\centering
	\subfloat[$R_{\psi_i \psi_i^c}$]{
		\includegraphics[width = 70 mm]{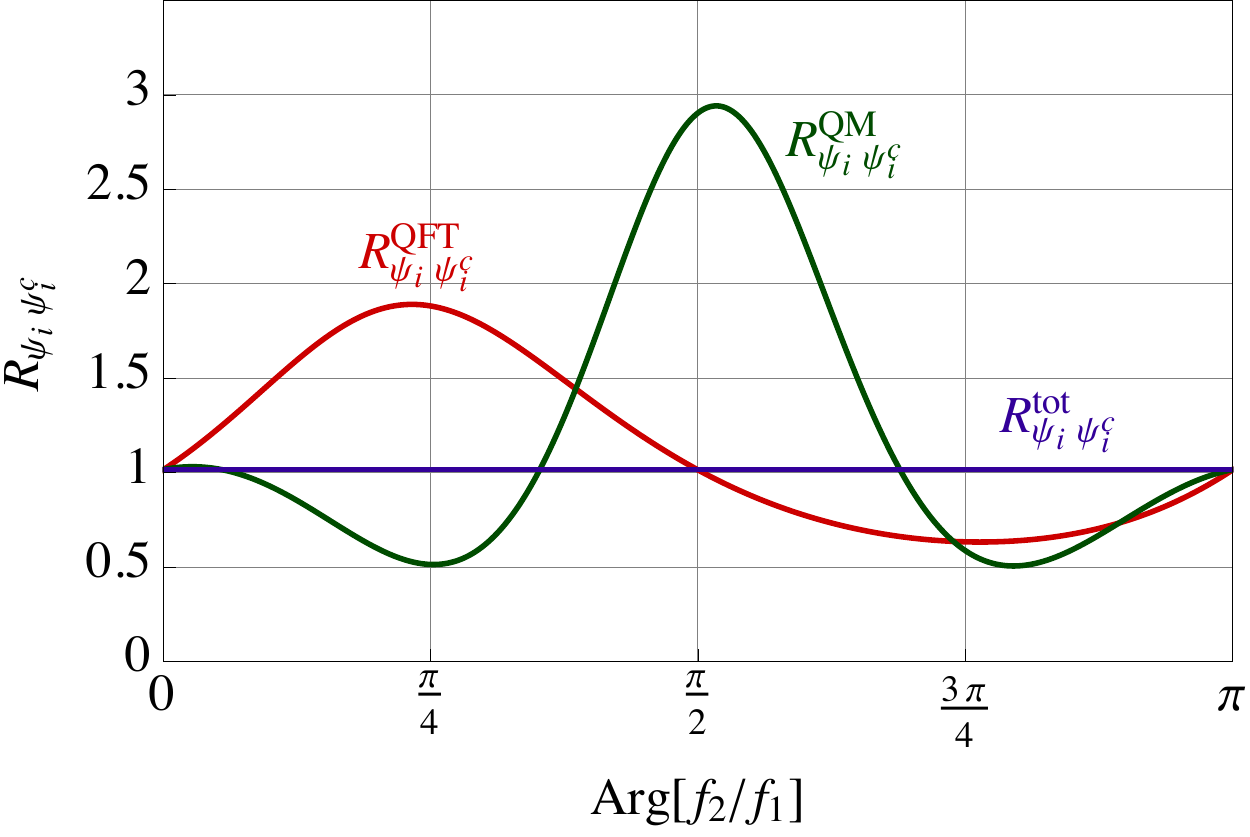}
		\label{fig:DWRatio}
	} \qquad
	\subfloat[CP asymmetry in $\Phi^* \to \chi_1 \xi^c$ and $\Phi \to \chi_1^c \xi$]{
		\includegraphics[width = 70 mm]{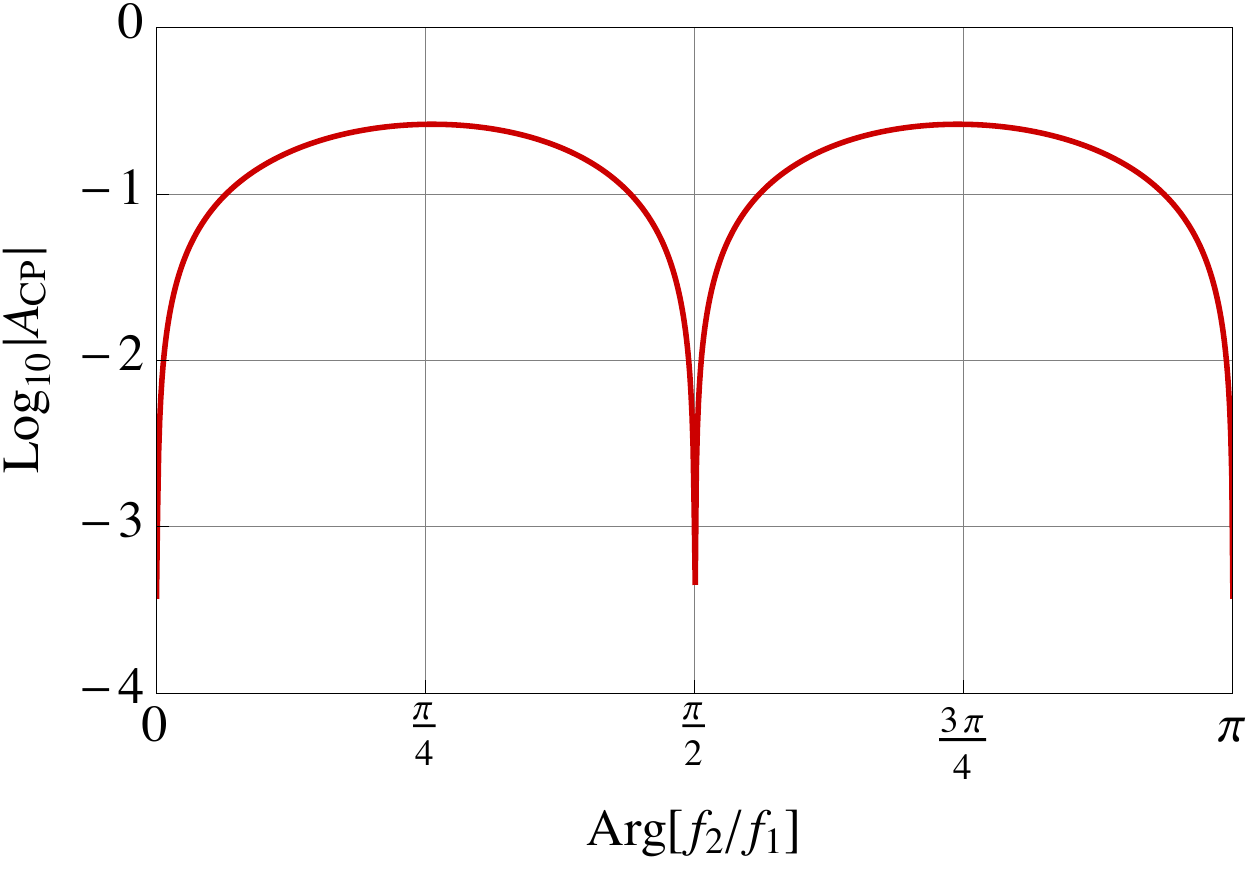}
		\label{fig:CPAsym}
	}
	\caption{Results obtained by varying the relative phase of $f_1$ and $f_2$.}
	\label{fig:DWB}
\end{figure}

The transition rates calculated from the input parameters are presented in table \ref{tab:BDW}. The values of $\Gamma^\text{scat}_{\widehat{\Phi}_{\widehat{\alpha}} \to Y}$ are sometimes complex, and thus $\Sigma_{\widehat{\alpha}}$ cannot be arbitrarily removed from $\Sigma_{\widehat{\alpha}} \Gamma^\text{scat}_{\widehat{\Phi}_{\widehat{\alpha}} \to Y}$ to define a partial decay width of $\widehat{\Phi}_{\widehat{\alpha}}$. Also note $\Gamma^\text{scat}_{\widehat{\Phi}_{\widehat{\alpha}} \to \psi_1 \psi_1^c} \neq \Gamma^\text{opt}_{\widehat{\Phi}_{\widehat{\alpha}} \to \psi_1 \psi_1^c}$. Nevertheless, we still have the identities: $\sum_{\widehat{\alpha}} \Gamma^\text{scat}_{\widehat{\Phi}_{\widehat{\alpha}} \to Y} = \sum_{\widehat{\alpha}} \Gamma^\text{opt}_{\widehat{\Phi}_{\widehat{\alpha}} \to Y}$ and $\sum_{\widehat{\alpha}} \Gamma_{\widehat{\Phi}_{\widehat{\alpha}}} = \sum_{\widehat{\alpha}, Y} \Gamma^\text{scat}_{\widehat{\Phi}_{\widehat{\alpha}} \to Y} = \sum_{\widehat{\alpha}, Y} \Gamma^\text{opt}_{\widehat{\Phi}_{\widehat{\alpha}} \to Y}$. We can also see that the ratio of decay widths obtained by the method of this paper is sometimes consistent with the value from the quantum mechanical approach, \textit{e.g.}, $\Gamma_{\Phi \to \chi_2 \xi^c} / \Gamma_{\Phi \to \chi_1 \xi^c}$, while it is sometimes different, \textit{e.g.}, $\Gamma_{\Phi \to \psi_2 \psi_2^c} / \Gamma_{\Phi \to \psi_1 \psi_1^c}$.
\begin{table}[h]
	\center
	\begin{tabular}{|c|c|}
		\hline
		Input parameter for $B^0$-$\overline{B^0}$ & Value \\ \Xhline{2\arrayrulewidth}
		$m_\Phi$ & 5279.63 MeV \cite{PDG} \\ \hline
		$m_{\chi_1}$ & 2900 MeV \\ \hline
		$m_{\chi_2}$ & 4500 MeV \\ \hline
		$f_1$ & $10^{-6} \, e^{i 0.119 \pi}$ \\ \hline
		$f_2$ & $10^{-5.9964} \, e^{-i 0.38124 \pi}$ \\ \hline
		$h_i$ & $10^{-5.632}$ \\ \hline
		$\mu_1$ & $m_\Phi$ \\ \hline
		$\mu_2$ & $10^{4.92} \ m_\Phi$ \\ \hline
	\end{tabular}
	\caption{Model parameters for $\Phi \to \chi_i \xi^c$ to mimic the semileptonic decay $B^0 \to D^- \mu^+ \nu_\mu X$ or $B^0 \to D^{* -} \mu^+ \nu_\mu X$. Note that $\mu_i$ is an argument of a logarithmic function, and $\mu_2$ therefore corresponds to a contribution of around $\mathcal{O} (10 \, \alpha)$ to the self-energy. The masses $m_{\chi_i}$ are arbitrarily chosen.}
	\label{tab:BInp}
\end{table}
\begin{table}[h]
	\footnotesize
	\center
	\begin{tabular}{|c|c|c|c|}
		\hline
		Output parameter for $B^0$-$\overline{B^0}$ & Value & SM prediction & Observed value \\ \Xhline{2\arrayrulewidth}
		$p_{\widehat{\Phi}_L} = p_{\widehat{\Phi}_1}$ & $5279.63 - i \, 5.01761 \cdot 10^{-10}$ MeV & $\cdot$ & $\cdot$ \\ \hline
		$p_{\widehat{\Phi}_H} = p_{\widehat{\Phi}_2}$ & $5279.63 - i \, 5.00005 \cdot 10^{-10}$ MeV & $\cdot$ & $\cdot$ \\ \hline
		$\Gamma_{\widehat{\Phi}_L} = \Gamma_{\widehat{\Phi}_1}$ & $1.00352 \cdot 10^{-9}$ MeV & $\cdot$ & $\cdot$ \\ \hline
		$\Gamma_{\widehat{\Phi}_H} = \Gamma_{\widehat{\Phi}_2}$ & $1.00001 \cdot 10^{-9}$ MeV & $\cdot$ & $\cdot$ \\ \hline
		$\sum_{\widehat{\alpha}} \Gamma_{\widehat{\Phi}_{\widehat{\alpha}}}$ & $2.00353 \cdot 10^{-9}$ MeV & $\cdot$ & $(2.000 \pm 0.006) \cdot 10^{-9}$ MeV \cite{PDG} \\ \hline
		$\Delta m_{\widehat{\Phi}} = m_{\widehat{\Phi}_H} - m_{\widehat{\Phi}_L}$ & $7.70184 \cdot 10^{-10}$ MeV & $\cdot$ & $(7.70 \pm 0.04) \cdot 10^{-10}$ MeV \cite{PDG} \\ \hline
		$\Delta \Gamma_{\widehat{\Phi}} = \Gamma_{\widehat{\Phi}_L} - \Gamma_{\widehat{\Phi}_H}$ & $3.51193 \cdot 10^{-12}$ MeV & $\cdot$ & $\minus (0.2 \pm 1.0) \cdot 10^{-11}$ MeV \cite{PDG} \\ \hline
		$\text{Re} [\epsilon] / (1 + |\epsilon|^2)$ & $\minus 1.02804 \cdot 10^{-4}$ & $\cdot$ & $\minus (0.5 \pm 0.4) \cdot 10^{-3}$ \cite{PDG} \\ \hline
		$\text{arg} [\minus M_{12} / \Gamma_{12}]$ & $\minus 5.15415^\circ$ & $\minus {5.2^\circ}_{-2.1^\circ}^{+1.5^\circ}$ \cite{BsMix} & $\cdot$ \\ \hline
		$\mathcal{A}_\text{CP}$ & $\minus 4.11299 \cdot 10^{-4}$ & $\minus 4.8_{-1.2}^{+1.0} \cdot 10^{-4}$ \cite{BsMix} & $\minus (0.02 \pm 0.19 \pm 0.30) \cdot 10^{-2}$ \cite{SemiCPAsymB0} \\ \hline
	\end{tabular}
	\caption{Output parameters corresponding to the input parameters given in table \ref{tab:BInp}. The observed values available in the references are in fact $(1 / 2) \sum_{\widehat{\alpha}} \Gamma_{\widehat{\Phi}_{\widehat{\alpha}}} \approx \Gamma_{B^0} = (1.520 \pm 0.004)$ ps$^{-1}$, $\Delta m_{\widehat{\Phi}} / \Gamma_{\widehat{\Phi}_L} \approx \Delta m_{B^0} / \Gamma_{B^0} = 0.770 \pm 0.004$, and $\Delta \Gamma_{\widehat{\Phi}} / \Gamma_{\widehat{\Phi}_L} \approx \Delta \Gamma_{B^0} / \Gamma_{B^0} = -(0.2 \pm 1.0) \cdot 10^{-2}$. In addition, $\text{Re} [\epsilon] / (1 + |\epsilon|^2) = \text{Re} [\epsilon_{B^0}] / (1 + |\epsilon_{B^0}|^2) \approx (|q / p|^2 - 1) / 4$ has been used.}
	\label{tab:BOutp}
	\small
	\center
	\begin{tabular}{|c|c|}
		\hline
		Output parameter for $B^0$-$\overline{B^0}$ & Value \\ \Xhline{2\arrayrulewidth}
		$\sum_{j, \widehat{\alpha}} \Gamma^\text{scat}_{\chi_j \xi^c \to \widehat{\Phi}_{\widehat{\alpha}} \to \chi_1 \xi^c}$ & $3.33787 \cdot 10^{-10}$ MeV \\ \hline
		$\sum_{j, \widehat{\alpha}} \Gamma^\text{scat}_{\chi_j^c \xi \to \widehat{\Phi}_{\widehat{\alpha}} \to \chi_1 \xi^c}$ & $7.61146 \cdot 10^{-11}$ MeV \\ \hline
		$\sum_{j, \widehat{\alpha}} \Gamma^\text{scat}_{\psi_j \psi_j^c \to \widehat{\Phi}_{\widehat{\alpha}} \to \chi_1 \xi^c}$ & $1.09955 \cdot 10^{-10}$ MeV \\ \hline
		$\sum_{\widehat{\alpha}} \Gamma^\text{scat}_{\widehat{\Phi}_{\widehat{\alpha}} \to \chi_1 \xi^c} = \sum_{\widehat{\alpha}} \Gamma^\text{opt}_{\widehat{\Phi}_{\widehat{\alpha}} \to \chi_1 \xi^c}$ & $5.19857 \cdot 10^{-10}$ MeV \\ \hline
		$\sum_{\widehat{\alpha}} \Gamma^\text{scat}_{\widehat{\Phi}_{\widehat{\alpha}} \to \chi_2 \xi^c} = \sum_{\widehat{\alpha}} \Gamma^\text{opt}_{\widehat{\Phi}_{\widehat{\alpha}} \to \chi_2 \xi^c}$ & $2.70083 \cdot 10^{-10}$ MeV \\ \hline
		$\sum_{\widehat{\alpha}} \Gamma^\text{scat}_{\widehat{\Phi}_{\widehat{\alpha}} \to \chi_1^c \xi} = \sum_{\widehat{\alpha}} \Gamma^\text{opt}_{\widehat{\Phi}_{\widehat{\alpha}} \to \chi_1^c \xi}$ & $5.19857 \cdot 10^{-9}$ MeV \\ \hline
		$\sum_{\widehat{\alpha}} \Gamma^\text{scat}_{\widehat{\Phi}_{\widehat{\alpha}} \to \chi_2^c \xi} = \sum_{\widehat{\alpha}} \Gamma^\text{opt}_{\widehat{\Phi}_{\widehat{\alpha}} \to \chi_2^c \xi}$ & $2.70083 \cdot 10^{-10}$ MeV \\ \hline
		$\sum_{\widehat{\alpha}} \Gamma^\text{scat}_{\widehat{\Phi}_{\widehat{\alpha}} \to \psi_1 \psi_1^c} = \sum_{\widehat{\alpha}} \Gamma^\text{opt}_{\widehat{\Phi}_{\widehat{\alpha}} \to \psi_1 \psi_1^c}$ & $2.10070 \cdot 10^{-10}$ MeV \\ \hline
		$\sum_{\widehat{\alpha}} \Gamma^\text{scat}_{\widehat{\Phi}_{\widehat{\alpha}} \to \psi_2 \psi_2^c} = \sum_{\widehat{\alpha}} \Gamma^\text{opt}_{\widehat{\Phi}_{\widehat{\alpha}} \to \psi_2 \psi_2^c}$ & $2.13581 \cdot 10^{-10}$ MeV \\ \hline
		$\sum_{\widehat{\alpha}} \Gamma_{\widehat{\Phi}_{\widehat{\alpha}}} = \sum_{\widehat{\alpha}, Y} \Gamma^\text{scat}_{\widehat{\Phi}_{\widehat{\alpha}} \to Y} = \sum_{\widehat{\alpha}, Y} \Gamma^\text{opt}_{\widehat{\Phi}_{\widehat{\alpha}} \to Y}$ & $2.00353 \cdot 10^{-9}$ MeV \\ \Xhline{2\arrayrulewidth}
		$\Gamma^\text{scat}_{\widehat{\Phi}_{\widehat{\alpha}} \to \chi_1 \xi^c} = \Gamma^\text{scat}_{\widehat{\Phi}_{\widehat{\alpha}} \to \chi_1^c \xi}$ & $2.59928 \cdot 10^{-10}$ MeV \\ \hline
		$\Gamma^\text{scat}_{\widehat{\Phi}_1 \to \psi_1 \psi_1^c}$ & $1.17276 \cdot 10^{-16} + i \, 3.25827 \cdot 10^{-17}$ MeV \\ \hline
		$\Gamma^\text{scat}_{\widehat{\Phi}_2 \to \psi_1 \psi_1^c}$ & $2.10070 \cdot 10^{-10} - i \, 3.25827 \cdot 10^{-17}$ MeV \\ \hline
		$\Gamma^\text{opt}_{\widehat{\Phi}_{\widehat{\alpha}} \to \chi_1 \xi^c} = \Gamma^\text{opt}_{\widehat{\Phi}_{\widehat{\alpha}} \to \chi_1^c \xi}$ & $2.59928 \cdot 10^{-10}$ MeV \\ \hline
		$\Gamma^\text{opt}_{\widehat{\Phi}_1 \to \psi_1 \psi_1^c}$ & $1.05035 \cdot 10^{-10}$ MeV \\ \hline
		$\Gamma^\text{opt}_{\widehat{\Phi}_2 \to \psi_1 \psi_1^c}$ & $1.05035 \cdot 10^{-10}$ MeV \\ \hline
		$\sum_{\widehat{\alpha}} \Gamma^\text{scat}_{\widehat{\Phi}_{\widehat{\alpha}} \to \psi_1 \psi_1^c} = \sum_{\widehat{\alpha}} \Gamma^\text{opt}_{\widehat{\Phi}_{\widehat{\alpha}} \to \psi_1 \psi_1^c}$ & $2.10070 \cdot 10^{-10}$ MeV \\ \Xhline{2\arrayrulewidth}
		$\frac{\Gamma_{\Phi \to \chi_2 \xi^c}}{\Gamma_{\Phi \to \chi_1 \xi^c}} = \frac{\sum_{j, \widehat{\alpha}} \Gamma^\text{scat}_{\chi_j \xi^c \to \widehat{\Phi}_{\widehat{\alpha}} \to \chi_2 \xi^c}}{\sum_{j, \widehat{\alpha}} \Gamma^\text{scat}_{\chi_j \xi^c \to \widehat{\Phi}_{\widehat{\alpha}} \to \chi_1 \xi^c}} = \frac{\Gamma^\text{QM}_{\Phi \to \chi_2 \xi^c}}{\Gamma^\text{QM}_{\Phi \to \chi_1 \xi^c}}$ & $0.519533$ \\ \hline
		$\frac{\Gamma_{\Phi \to \psi_2 \psi_2^c}}{\Gamma_{\Phi \to \psi_1 \psi_1^c}} = \frac{\sum_{j, \widehat{\alpha}} \Gamma^\text{scat}_{\chi_j \xi^c \to \widehat{\Phi}_{\widehat{\alpha}} \to \psi_2 \psi_2^c}}{\sum_{j, \widehat{\alpha}} \Gamma^\text{scat}_{\chi_j \xi^c \to \widehat{\Phi}_{\widehat{\alpha}} \to \psi_1 \psi_1^c}}$ & $1.01390$ \\ \hline
		$\frac{\Gamma^\text{QM}_{\Phi \to \psi_2 \psi_2^c}}{\Gamma^\text{QM}_{\Phi \to \psi_1 \psi_1^c}}$ & $2.90553$ \\ \hline
	\end{tabular}
	\caption{Transition rates corresponding to the input parameters given in table \ref{tab:BInp}.}
	\label{tab:BDW}
\end{table}	
To see the dependence of the difference on model parameters, let us define the ratios of decay widths as follows:
\begin{align}
	R_{\psi_i \psi_i^c}^\text{QFT} \coloneqq \frac{\sum_{j, \widehat{\alpha}} \Gamma^\text{scat}_{\chi_j \xi^c \to \widehat{\Phi}_{\widehat{\alpha}} \to \psi_2 \psi_2^c}}{\sum_{j, \widehat{\alpha}} \Gamma^\text{scat}_{\chi_j \xi^c \to \widehat{\Phi}_{\widehat{\alpha}} \to \psi_1 \psi_1^c}}, \qquad
	R_{\psi_i \psi_i^c}^\text{tot} \coloneqq \frac{\sum_{\widehat{\alpha}} \Gamma^\text{scat}_{\widehat{\Phi}_{\widehat{\alpha}} \to \psi_2 \psi_2^c}}{\sum_{\widehat{\alpha}} \Gamma^\text{scat}_{\widehat{\Phi}_{\widehat{\alpha}} \to \psi_1 \psi_1^c}}, \qquad
	R_{\psi_i \psi_i^c}^\text{QM} \coloneqq \frac{\Gamma^\text{QM}_{\Phi \to \psi_2 \psi_2^c}}{\Gamma^\text{QM}_{\Phi \to \psi_1 \psi_1^c}}.
\end{align}
Note that $R_{\psi_i \psi_i^c}^\text{tot}$ is an inclusive quantity incorporating all possible initial multiparticle states, while the others are quantities with initial $\Phi$. Varying the phase of $f_2$ such that $\minus \pi / 2 < \text{Arg} [f_2 / f_1] \leq \pi / 2$, we obtain the variation of those ratios as shown in figure \ref{fig:DWRatio}, in which the differences are clearly presented.
The variation of the CP asymmetry, $\mathcal{A}_\text{CP}$, in $\Phi^* \to \chi_1 \xi^c$ and $\Phi \to \chi_1^c \xi$ is also shown in figure \ref{fig:CPAsym}. Note that $\mathcal{A}_\text{CP} = 0$ when $\text{Arg} [f_2 / f_1] = n \pi / 2$ where $n$ is an integer.

\subsection{$K^0$-$\overline{K^0}$}	\label{sec:Kaon}
Now we consider an example in which the toy model imitates the $K^0$-$\overline{K^0}$ system. In $K^0$-$\overline{K^0}$ mixing, the total decay widths of physical particles are well-separated, and the analysis in terms of the integration over time discussed in section \ref{sec:DWtdep} is therefore useful. The model parameters for $\widehat{\Phi}_L \to \chi_i \xi^c$ $(\widehat{\Phi}_L \coloneqq \widehat{\Phi}_2)$ to mimic $K^0_L \to \pi^- \ell^+ \nu_\ell$ are given in tables \ref{tab:KInp} and \ref{tab:KOutp}. The mixing matrix is found to be
\begin{align}
	C_\Phi = \left( \begin{array}{cc} 0.705943 & 0.708265 + i \, 0.00214715 \\ 0.705943 & \minus 0.708265 - i \, 0.00214715 \end{array} \right)
	\eqqcolon \left( \begin{array}{cc} p & q \\ p & \minus q \end{array} \right),
\end{align}
which is a non-unitary matrix
\begin{align}
	C_\Phi^\dag C_\Phi = \left( \begin{array}{cc} 2 |p|^2 & 0 \\ 0 & 2 |q|^2 \end{array} \right)
	= \left( \begin{array}{cc} 0.996712  & 0 \\ 0 & 1.00329 \end{array} \right),
\end{align}
and its components satisfy $|p|^2 + |q|^2 = 1$. Defining the partial decay width of $\widehat{\Phi}_L$ as
\begin{align}
	\Gamma_{\widehat{\Phi}_L \to Y} \coloneqq \sum_{X, \widehat{\alpha}} \Gamma^{t > t_1}_{X \to \widehat{\Phi}_{\widehat{\alpha}} \to Y}
	= \sum_{j, \widehat{\alpha}} \Gamma^{t > t_1}_{\chi_j \xi^c \to \widehat{\Phi}_{\widehat{\alpha}} \to Y}
		+ \sum_{j, \widehat{\alpha}} \Gamma^{t > t_1}_{\chi_j^c \xi \to \widehat{\Phi}_{\widehat{\alpha}} \to Y}
		+ \sum_{j, \widehat{\alpha}} \Gamma^{t > t_1}_{\psi_j \psi_j^c \to \widehat{\Phi}_{\widehat{\alpha}} \to Y},
\end{align}
we write the asymmetry in the decay of $\widehat{\Phi}_L$ as
\begin{align}
	\mathcal{A}_L
	\coloneqq \frac{\Gamma_{\widehat{\Phi}_L \to \chi_i^c \xi} - \Gamma_{\widehat{\Phi}_L \to \chi_i \xi^c}}{\Gamma_{\widehat{\Phi}_L \to \chi_i^c \xi} + \Gamma_{\widehat{\Phi}_L \to \chi_i \xi^c}}.
\end{align}

The transition rates corresponding to the input parameters are given in table \ref{tab:KDW}. Above all, note the difference between $\sum_{\widehat{\alpha}} \Gamma^t_{X \to \widehat{\Phi}_{\widehat{\alpha}} \to Y}$ and $\sum_{\widehat{\alpha}} \Gamma^{t > t_1}_{X \to \widehat{\Phi}_{\widehat{\alpha}} \to Y}$, \textit{e.g.}, $\sum_{j, \widehat{\alpha}} \Gamma^t_{\chi_j \xi^c \to \widehat{\Phi}_{\widehat{\alpha}} \to \chi_1 \xi^c}$ and $\sum_{j, \widehat{\alpha}} \Gamma^{t > t_1}_{\chi_j \xi^c \to \widehat{\Phi}_{\widehat{\alpha}} \to \chi_1 \xi^c}$. To $\sum_{\widehat{\alpha}} \Gamma^{t > t_1}_{X \to \widehat{\Phi}_{\widehat{\alpha}} \to Y}$, only $\widehat{\Phi}_L$ contributes, since $\widehat{\Phi}_S$ and their interference have already decayed completely at $t = t_1$ up to working precision. The magnitude of the interference for $t_0 < t < t_1$ is, however, comparable to the contribution only of $\widehat{\Phi}_L$, and thus there exists a difference between $\sum_{\widehat{\alpha}} \Gamma^t_{X \to \widehat{\Phi}_{\widehat{\alpha}} \to Y}$ and $\sum_{\widehat{\alpha}} \Gamma^{t > t_1}_{X \to \widehat{\Phi}_{\widehat{\alpha}} \to Y}$. In other words, not all the contributions of $\widehat{\Phi}_L$ are taken into account in $\sum_{\widehat{\alpha}} \Gamma^{t > t_1}_{X \to \widehat{\Phi}_{\widehat{\alpha}} \to Y}$. The physical particles are dynamically generated by the various interacting fields, and it is not possible to define a decay width $\Gamma_{\widehat{\Phi}_{\widehat{\alpha}} \to Y}$ such that it truly considers everything related to each of physical particle. Note also that $\Gamma^t_{X \to \widehat{\Phi}_{\widehat{\alpha}} \to Y}$ is sometimes a complex number as shown in table \ref{tab:KDW}, and it cannot be interpreted as a decay width.

The big difference between $\sum_{\widehat{\alpha}} \Gamma^t_{X \to \widehat{\Phi}_{\widehat{\alpha}} \to \psi_1 \psi_1^c}$ and $\sum_{\widehat{\alpha}} \Gamma^{t > t_1}_{X \to \widehat{\Phi}_{\widehat{\alpha}} \to \psi_1 \psi_1^c}$ is particularly noteworthy. This happens because the transition $\widehat{\Phi}_L \to \psi_1 \psi_1^c$, which defines the latter, gets hugely suppressed although the final states $\psi_1 \psi_1^c$ almost evenly couple to $\Phi$ and $\Phi^*$. Note that, in the real $K^0$-$\overline{K^0}$ system, $K^0_L \to \pi^0 \pi^0, \pi^+ \pi^-$ get suppression while $K^0_S \to \pi^0 \pi^0, \pi^+ \pi^-$ are the dominant decay channels of $K^0_S$. Such an effect is a general characteristic of particle-antiparticle or flavor mixing when the coupling is almost universal: $|f_1| \sim |f_2| > 10 \, |h_j|$ is assumed for this example. In the perspective of CP violation, this occurs because $\widehat{\Phi}_L$ is almost CP-odd while $\psi_1 \psi_1^c$ is CP-even, which is the conventional explanation of this effect. In the perspective of dynamics, there exists a close cancellation among the various self-energy contributions to $\widehat{\Phi}_L$ when the couplings are almost universal, and thus $\widehat{\Phi}_L \to \psi_1 \psi_1^c$ gets suppressed so that $\widehat{\Phi}_L$ lives longer than $\widehat{\Phi}_S$.

\begin{table}[h]
	\center
	\begin{tabular}{|c|c|}
		\hline
		Input parameter for $K^0$-$\overline{K^0}$ & Value \\ \Xhline{2\arrayrulewidth}
		$m_\Phi$ & 497.611 MeV \cite{PDG} \\ \hline
		$m_{\chi_1}$ & 100 MeV \\ \hline
		$m_{\chi_2}$ & 300 MeV \\ \hline
		$f_1$ & $10^{-6.366} \, e^{i 0.00105 \pi}$ \\ \hline
		$f_2$ & $10^{-6.366} \, e^{-i 0.00105 \pi}$ \\ \hline
		$h_i$ & $10^{-7.579}$ \\ \hline
		$\mu_1$ & $m_\Phi$ \\ \hline
		$\mu_2$ & $10^{1.258} \, m_\Phi$ \\ \hline
	\end{tabular}
	\caption{Model parameters for $\widehat{\Phi}_L \to \chi_i \xi^c$ to mimic $K^0_L \to \pi^- \ell^+ \nu_\ell$. The masses $m_{\chi_i}$ are arbitrarily chosen.}
	\label{tab:KInp}
\end{table}
\begin{table}[h]
	\small
	\center
	\begin{tabular}{|c|c|c|}
		\hline
		Output parameter for $K^0$-$\overline{K^0}$ & Value & Observed value \\ \Xhline{2\arrayrulewidth}
		$p_{\widehat{\Phi}_S} = p_{\widehat{\Phi}_1}$ & $497.611 - i \, 3.67620 \cdot 10^{-12}$ MeV & $\cdot$ \\ \hline
		$p_{\widehat{\Phi}_L} = p_{\widehat{\Phi}_2}$ & $497.611 - i \, 6.44910 \cdot 10^{-15}$ MeV & $\cdot$ \\ \hline
		$\Gamma_{\widehat{\Phi}_S} = \Gamma_{\widehat{\Phi}_1}$ & $7.35240 \cdot 10^{-12}$ MeV & $(7.351 \pm 0.003) \cdot 10^{-12}$ MeV \cite{PDGOld} \\ \hline
		$\Gamma_{\widehat{\Phi}_L} = \Gamma_{\widehat{\Phi}_2}$ & $1.28984 \cdot 10^{-14}$ MeV & $(1.287 \pm 0.005) \cdot 10^{-14}$ MeV \cite{PDGOld} \\ \hline
		$\sum_{\widehat{\alpha}} \Gamma_{\widehat{\Phi}_{\widehat{\alpha}}}$ & $7.36529 \cdot 10^{-12}$ MeV & $\cdot$ \\ \hline
		$\Delta m_{\widehat{\Phi}} = m_{\widehat{\Phi}_L} - m_{\widehat{\Phi}_S}$ & $3.38370 \cdot 10^{-12}$ MeV & $(3.384 \pm 0.006) \cdot 10^{-12}$ MeV \cite{PDGOld} \\ \hline
		$|\epsilon|$ & $2.23609 \cdot 10^{-3}$ & $(2.228 \pm 0.011) \cdot 10^{-3}$ \cite{PDG} \\ \hline
		$\mathcal{A}_L$ & $3.28785  \cdot 10^{-3}$ & $(3.32 \pm 0.06) \cdot 10^{-3}$ \cite{PDG} \\ \hline
	\end{tabular}
	\caption{Output parameters corresponding to the input parameters given in table \ref{tab:BInp}. The observed values of lifetime directly available in the references are $\hbar / \Gamma_{\widehat{\Phi}_S} = \hbar / \Gamma_{K_S} = (0.8954 \pm 0.0004) \cdot 10^{-10}$ s and $\hbar / \Gamma_{\widehat{\Phi}_L} = \hbar / \Gamma_{K_L} = (5.116 \pm 0.021) \cdot 10^{-8}$ s.}
	\label{tab:KOutp}
	\small
	\center
	\begin{tabular}{|c|c|}
		\hline
		Output parameter for $K^0$-$\overline{K^0}$ & Value \\ \Xhline{2\arrayrulewidth}
		$\sum_j \Gamma^t_{\chi_j \xi^c \to \widehat{\Phi}_1 \to \chi_1 \xi^c}$ & $6.18889 \cdot 10^{-18} + i \, 2.95597 \cdot 10^{-18}$ MeV \\ \hline
		$\sum_j \Gamma^t_{\chi_j \xi^c \to \widehat{\Phi}_2 \to \chi_1 \xi^c}$ & $1.69720 \cdot 10^{-15} - i \, 2.95597 \cdot 10^{-18}$ MeV \\ \hline
		$\sum_{j, \widehat{\alpha}} \Gamma^t_{\chi_j \xi^c \to \widehat{\Phi}_{\widehat{\alpha}} \to \chi_1 \xi^c} = \sum_{j, \widehat{\alpha}} \Gamma^\text{scat}_{\chi_j \xi^c \to \widehat{\Phi}_{\widehat{\alpha}} \to \chi_1 \xi^c}$ & $1.70339 \cdot 10^{-15}$ MeV \\ \hline
		$\sum_{j, \widehat{\alpha}} \Gamma^t_{\chi_j^c \xi \to \widehat{\Phi}_{\widehat{\alpha}} \to \chi_1 \xi^c} = \sum_{j, \widehat{\alpha}} \Gamma^\text{scat}_{\chi_j^c \xi \to \widehat{\Phi}_{\widehat{\alpha}} \to \chi_1 \xi^c}$ & $1.70167 \cdot 10^{-15}$ MeV \\ \hline
		$\sum_{j, \widehat{\alpha}} \Gamma^t_{\psi_j \psi_j^c \to \widehat{\Phi}_{\widehat{\alpha}} \to \chi_1 \xi^c} = \sum_{j, \widehat{\alpha}} \Gamma^\text{scat}_{\psi_j \psi_j^c \to \widehat{\Phi}_{\widehat{\alpha}} \to \chi_1 \xi^c}$ & $3.43266 \cdot 10^{-15}$ MeV \\ \hline
		$\sum_{j, \widehat{\alpha}} \Gamma^{t > t_1}_{\chi_j \xi^c \to \widehat{\Phi}_{\widehat{\alpha}} \to \chi_1 \xi^c}$ & $1.51860 \cdot 10^{-15}$ MeV \\ \hline
		$\sum_{j, \widehat{\alpha}} \Gamma^{t > t_1}_{\chi_j^c \xi \to \widehat{\Phi}_{\widehat{\alpha}} \to \chi_1 \xi^c}$ & $1.52861 \cdot 10^{-15}$ MeV \\ \hline
		$\sum_{j, \widehat{\alpha}} \Gamma^{t > t_1}_{\psi_j \psi_j^c \to \widehat{\Phi}_{\widehat{\alpha}} \to \chi_1 \xi^c}$ & $2.77889 \cdot 10^{-17}$ MeV \\ \Xhline{2\arrayrulewidth}
		$\sum_j \Gamma^t_{\chi_j \xi^c \to \widehat{\Phi}_1 \to \psi_1 \psi_1^c}$ & $3.20030 \cdot 10^{-15} + i \, 1.13720 \cdot 10^{-17}$ MeV \\ \hline
		$\sum_j \Gamma^t_{\chi_j \xi^c \to \widehat{\Phi}_2 \to \psi_1 \psi_1^c}$ & $6.77910 \cdot 10^{-17} - i \, 1.13720 \cdot 10^{-17}$ MeV \\ \hline
		$\sum_{j, \widehat{\alpha}} \Gamma^t_{\chi_j \xi^c \to \widehat{\Phi}_{\widehat{\alpha}} \to \psi_1 \psi_1^c} = \sum_{j, \widehat{\alpha}} \Gamma^\text{scat}_{\chi_j \xi^c \to \widehat{\Phi}_{\widehat{\alpha}} \to \psi_1 \psi_1^c}$ & $3.26809 \cdot 10^{-15}$ MeV \\ \hline
		$\sum_{j, \widehat{\alpha}} \Gamma^t_{\chi_j^c \xi \to \widehat{\Phi}_{\widehat{\alpha}} \to \psi_1 \psi_1^c} = \sum_{j, \widehat{\alpha}} \Gamma^\text{scat}_{\chi_j^c \xi \to \widehat{\Phi}_{\widehat{\alpha}} \to \psi_1 \psi_1^c}$ & $3.20556 \cdot 10^{-15}$ MeV \\ \hline
		$\sum_{j, \widehat{\alpha}} \Gamma^t_{\psi_j \psi_j^c \to \widehat{\Phi}_{\widehat{\alpha}} \to \psi_1 \psi_1^c} = \sum_{j, \widehat{\alpha}} \Gamma^\text{scat}_{\psi_j \psi_j^c \to \widehat{\Phi}_{\widehat{\alpha}} \to \psi_1 \psi_1^c}$ & $3.66339 \cdot 10^{-12}$ MeV \\ \hline
		$\sum_{j, \widehat{\alpha}} \Gamma^{t > t_1}_{\chi_j \xi^c \to \widehat{\Phi}_{\widehat{\alpha}} \to \psi_1 \psi_1^c}$ & $4.20503 \cdot 10^{-17}$ MeV \\ \hline
		$\sum_{j, \widehat{\alpha}} \Gamma^{t > t_1}_{\chi_j^c \xi \to \widehat{\Phi}_{\widehat{\alpha}} \to \psi_1 \psi_1^c}$ & $4.23277 \cdot 10^{-17}$ MeV \\ \hline
		$\sum_{j, \widehat{\alpha}} \Gamma^{t > t_1}_{\psi_j \psi_j^c \to \widehat{\Phi}_{\widehat{\alpha}} \to \psi_1 \psi_1^c}$ & $7.69482 \cdot 10^{-19}$ MeV \\ \Xhline{2\arrayrulewidth}
		$\frac{\Gamma_{\Phi \to \chi_2 \xi^c}}{\Gamma_{\Phi \to \chi_1 \xi^c}} = \frac{\sum_{j, \widehat{\alpha}} \Gamma^\text{scat}_{\chi_j \xi^c \to \widehat{\Phi}_{\widehat{\alpha}} \to \chi_2 \xi^c}}{\sum_{j, \widehat{\alpha}} \Gamma^\text{scat}_{\chi_j \xi^c \to \widehat{\Phi}_{\widehat{\alpha}} \to \chi_1 \xi^c}} = \frac{\Gamma^\text{QM}_{\Phi \to \chi_2 \xi^c}}{\Gamma^\text{QM}_{\Phi \to \chi_1 \xi^c}}$ & $0.869311$ \\ \hline
		$\frac{\Gamma_{\Phi \to \psi_2 \psi_2^c}}{\Gamma_{\Phi \to \psi_1 \psi_1^c}} = \frac{\sum_{j, \widehat{\alpha}} \Gamma^\text{scat}_{\chi_j \xi^c \to \widehat{\Phi}_{\widehat{\alpha}} \to \psi_2 \psi_2^c}}{\sum_{j, \widehat{\alpha}} \Gamma^\text{scat}_{\chi_j \xi^c \to \widehat{\Phi}_{\widehat{\alpha}} \to \psi_1 \psi_1^c}}$ & $0.976160$ \\ \hline
		$\frac{\Gamma^\text{QM}_{\Phi \to \psi_2 \psi_2^c}}{\Gamma^\text{QM}_{\Phi \to \psi_1 \psi_1^c}}$ & $0.998262$ \\ \hline
	\end{tabular}
	\caption{Transition rates corresponding to the input parameters given in table \ref{tab:KInp}. The values of $\Gamma^t_{X \to \widehat{\Phi}_{\widehat{\alpha}} \to Y}$ are complex in general, and thus each mass eigenstates cannot be separately considered. Note the difference between $\sum_{\widehat{\alpha}} \Gamma^t_{X \to \widehat{\Phi}_{\widehat{\alpha}} \to Y}$ and $\sum_{\widehat{\alpha}} \Gamma^{t > t_1}_{X \to \widehat{\Phi}_{\widehat{\alpha}} \to Y}$. In addition, $t_1 = 8.47360 \cdot 10^{12}$ MeV$^{-1} = 5.57743 \cdot 10^{-9}$ s, whose definition is given by equation \ref{eq:T1}.}
	\label{tab:KDW}
\end{table}	

\section{Beyond the leading order in perturbation}		\label{sec:BeyLO}
In this paper, we have calculated the decay widths of unstable particles only up to the leading order in perturbation. In some cases, however, we must go beyond the leading order. The CP violation in the direct decay of mesons, for example, requires consideration of the loop corrections to the vertices, and the calculation of the decay widths should be performed up to the next-leading order in perturbation. Here, we briefly discuss what kind of changes is needed to go beyond the leading order. In that case, not all of $\chi_i$ and $\psi_i$ can be regarded as stable particles because of decays such as $\chi_i \to \chi_j \xi \xi^c$, and thus they cannot be regarded as external states. In addition to that, the \textit{next-leading order} calculation requires consideration of the followings:
\begin{enumerate}
	\item Two-loop corrections in the self-energy.
	\item One-loop corrections in the vertices.
	\item $R_{\widehat{\Phi}_{\widehat{\alpha}}} \neq 1$.
	\item Distinction between $\Sigma_\Phi (p_{\widehat{\Phi}_1}^2)$ and $\Sigma_\Phi (p_{\widehat{\Phi}_2}^2)$.
	\item Interference between $s$- and non-$s$-channel scattering processes.
\end{enumerate}
Note that we have been using $\Sigma_\Phi (p_{\widehat{\Phi}_1}^2) = \Sigma_\Phi (p_{\widehat{\Phi}_2}^2) = \Sigma_\Phi (m_\Phi^2)$ which is correct only up the leading order. The interference between $s$- and non-$s$-channel scattering processes is generated by cutting the vertex loops, and it also contributes to a resonance since one of them is in the $s$ channel. The detailed discussion is beyond the scope of this paper.

\section{Conclusion}
We have discussed the mixing of an unstable particle and its antiparticle in the quantum field theory using a simple toy model that imitates a neutral meson system. Renormalizing the theory step-by-step in a non-physical scheme and exactly diagonalizing the propagator up to the infinite order in perturbation, we have shown that the physical particles in the plane-wave limit, \textit{i.e.}, the particles that propagate like free particles until they decay, should be interpreted as quasiparticles which cannot be regarded as external states of physical processes. Moreover, we have shown that each of physical particles as a quasiparticle cannot be separately observed because of their interferences, even when there exist two clearly distinguishable resonance peaks in scattering mediated by those unstable particles, or when only one of physical particles appear to eventually survive due to a hierarchy in their lifetimes. Several methods to calculate the decay widths of physical unstable particles from scattering have been discussed in the quantum field theory, and they have been found to be mutually consistent with each other. Specific examples have been provided to verify all those results and to explain their meanings more clearly.

Those physical particles cannot be associated with one-particle states in the Hilbert space, and thus the understanding of decays as phenomena in which one-particle states evolve in time to decay into stable particles is wrong. In consequence, the dynamics of unstable particles with particle-antiparticle mixing cannot be described in quantum mechanics, where a particle in the plane-wave limit is always described by a one-particle state that evolves in time as a solution of the Schr\"odinger equation while keeping its identity as a specific linear combination of basis states of the Hilbert space. In other words, quantum mechanics is not a proper non-relativistic limit of the quantum field theory in the presence of particle-antiparticle mixing. As a result, the neutral meson system cannot be studied in the frameworks of quantum mechanics. Using an example, we have explicitly shown that there indeed exists a discrepancy between the results from the quantum field theory and quantum mechanics, and the difference is sometimes very large while it is sometimes negligible. In conclusion, the physics of the neutral meson mixing and CP violation should be reanalyzed using the method developed in this paper, all the more because we are observing anomalies in meson physics.

\section*{Acknowledgement}
This work was supported by the National Center for Theoretical Sciences, Hsinchu.

\appendix

\section{Calculation of the self-energy}		\label{sec:SECal}
For the loops of $\chi_i (k) \xi^c (p - k)$ and $\chi_i^c (k) \xi (p - k)$, we can write
\begin{align}
	i \Sigma_{0 \Phi}^{\chi_i \xi} (p^2)
	&\coloneqq \minus \frac{1}{2} \mu^{4 - d} \int \frac{d^d k}{(2 \pi)^d} (\minus i h_i^*) \, \text{Tr} \bigg[ \frac{i}{\slashed{k} - m_{\chi_i}} \frac{i}{\slashed{k} - \slashed{p}} \bigg] (\minus i h_i)
	= \mu^{4 - d} \frac{1}{2} |h_i|^2 \int \frac{d^d k}{(2 \pi)^d} \frac{\text{Tr} [(\slashed{k} + m_{\chi_i}) (\slashed{p} - \slashed{k})]}{(k^2 - m_{\chi_i}^2) (p - k)^2} \nonumber \\
	&= \mu^{4 - d} \frac{d}{2} |h_i|^2 \int \frac{d^d k}{(2 \pi)^d} \int_0^1 dx \frac{p \cdot k - k^2}{[(1 - x) (k^2 - m_{\chi_i}^2) + x (p - k)^2 ]^2},
\end{align}
where $d = 4 - \epsilon$ is the spacetime dimension and $\text{Tr} [\gamma^\mu \gamma^\nu] = d \eta^{\mu \nu}$ have been used. The factor 1/2 is the coefficient of $\big[ \overline{\xi} (x_1) \chi_i (x_1) \Phi^\dag (x_1) \big] \big[ \overline{\chi_i} (x_2) \xi (x_2) \Phi (x_2) \big]$ in the Taylor expansion of $e^{i \int dx \, \mathcal{L}_\text{int}}$ which appears in $\langle \Omega | T \{\Phi (y) \Phi^\dag (x)\} | \Omega \rangle$. No symmetric combination of contractions cancels this factor in this theory. Manipulating the Feynman parameter and the momentum integration in the standard way, we rewrite
\begin{align}
	\int \frac{d^d k}{(2 \pi)^d} \frac{p \cdot k - k^2}{[(1 - x) (k^2 - m_{\chi_i}^2) + x (p - k)^2]^2}
	= \int \frac{d^d \ell}{(2 \pi)^d} \frac{x (1 - x) p^2 - \ell^2}{[\ell^2 - (1 - x) (m_{\chi_i}^2 - x p^2)]^2},
\end{align}
where $\ell = k - x p$. Hence,
\begin{align}
	\mu^{4 - d} \frac{1}{2} &\int \frac{d^d k}{(2 \pi)^d} \frac{\text{Tr} [(\slashed{k} + m_{\chi_i}) (\slashed{p} - \slashed{k})]}{(k^2 - m_{\chi_i}^2) (p - k)^2}
	= \mu^{4 - d} \frac{d}{2} \int_0^1 dx \int \frac{d^d \ell}{(2 \pi)^d} \frac{x (1 - x) p^2 - \ell^2}{[\ell^2 - (1 - x) (m_{\chi_i}^2 - x p^2)]^2} \nonumber \\
	&= i \mu^{4 - d} \frac{d}{2} \bigg\{ p^2 \int_0^1 dx \, x (1 - x) \frac{\Gamma (2 - \frac{d}{2})}{(4 \pi)^{d/2}} [(1 - x) (m_{\chi_i}^2 - x p^2)]^{-(2 - d/2)} \nonumber \\
		&\qquad + \int_0^1 dx \, \frac{d}{2} \frac{\Gamma (1 - \frac{d}{2})}{(4 \pi)^{d/2}} [(1 - x) (m_{\chi_i}^2 - x p^2)]^{-(1 - d/2)} \bigg\} \nonumber \\
	&= i \frac{p^2}{16 \pi^2} \bigg[ \bigg( 1 - \frac{2 m_{\chi_i}^2}{p^2} \bigg) \frac{2}{\epsilon} + F (p^2) \bigg] + \mathcal{O} (\epsilon),
\end{align}
where
\begin{align}
	F (p^2) &\coloneqq \minus \frac{1}{6} - 2 \int_0^1 dx \, (1 - x) \bigg( 3 x - \frac{2 m_{\chi_i}^2}{p^2} \bigg) \bigg[ \log{(1 - x)} + \log{\frac{m_{\chi_i}^2 - x p^2}{\widetilde{\mu}^2}} \bigg] \nonumber \\
	&= \bigg( \frac{3}{2} - \frac{2 m_{\chi_i}^2}{p^2} \bigg)
			- \frac{m_{\chi_i}^2}{p^2} \bigg( 3 - \frac{6 m_{\chi_i}^2}{p^2} + \frac{2 m_{\chi_i}^4}{p^4} \bigg) \log{\frac{m_{\chi_i}^2}{\widetilde{\mu}^2}}
			- \bigg( 1 - \frac{5 m_{\chi_i}^2}{p^2} + \frac{6 m_{\chi_i}^4}{p^4} - \frac{2 m_{\chi_i}^6}{p^6} \bigg) \log{\frac{m_{\chi_i}^2 - p^2}{\widetilde{\mu}^2}} \nonumber \\
	&= \bigg[ \frac{3}{2} - \log{\frac{m_{\chi_i}^2}{\widetilde{\mu}^2}} - \bigg( 1 - \frac{5 m_{\chi_i}^2}{p^2} + \frac{6 m_{\chi_i}^4}{p^4} - \frac{2 m_{\chi_i}^6}{p^6} \bigg) \log{\frac{m_{\chi_i}^2 - p^2}{m_{\chi_i}^2}} \bigg]
		- \frac{2 m_{\chi_i}^2}{p^2} \bigg( 1 - \log{\frac{m_{\chi_i}^2}{\widetilde{\mu}^2}} \bigg).
\end{align}
We therefore obtain
\begin{align}
	\Sigma_{0 \Phi}^{\chi_i \xi} (p^2) &= p^2 \Sigma_{0 \Phi}'^{\chi_i \xi} (p^2) + \delta \Sigma_\Phi,
\end{align}
where
\begin{align}	
	\Sigma_{0 \Phi}'^{\chi_i \xi} (p^2) &= \sum_i \frac{|h_i|^2}{16 \pi^2} \bigg[ \frac{2}{\epsilon} + \frac{3}{2} - \log{\frac{m_{\chi_i}^2}{\widetilde{\mu}^2}} - \bigg( 1 - \frac{5 m_{\chi_i}^2}{p^2} + \frac{6 m_{\chi_i}^4}{p^4} - \frac{2 m_{\chi_i}^6}{p^6} \bigg) \log{\frac{m_{\chi_i}^2 - p^2}{m_{\chi_i}^2}} \bigg] + \mathcal{O} (\epsilon), \\
	\delta \Sigma_\Phi &= \minus \sum_i \frac{|h_i|^2}{8 \pi^2} m_{\chi_i}^2 \bigg( \frac{2}{\epsilon} + 1 - \log{\frac{m_{\chi_i}^2}{\widetilde{\mu}^2}} \bigg) + \mathcal{O} (\epsilon).
\end{align}
The contribution from the loop of $\psi_i \psi_i^c$ is calculated in the same way, and it is given by
\begin{align}
	\Sigma_{0 \Phi}^{\psi_i} (p^2) = p^2 \Sigma_{0 \Phi}'^{\psi_i} (p^2)
	= \frac{p^2}{16 \pi^2} \left( \begin{array}{cc} |f_i|^2 & f_i^{* 2} \\ f_i^2 & |f_i|^2 \end{array} \right) \bigg( \frac{2}{\epsilon} + \frac{3}{2} - \log{\frac{p^2}{\widetilde{\mu}^2}} + i \pi \bigg) + \mathcal{O} (\epsilon),
\end{align}
where we have used $\log{(\minus 1)} = i \pi$. The non-renormalized self-energy matrix of $\Phi$ is written as
\begin{align}
	(\Sigma_{0 \Phi})_{\beta \alpha} (p^2) = \sum_i  \bigg[ (\Sigma_{0 \Phi}^{\psi_i})_{\beta \alpha} (p^2) + \delta_{\beta \alpha} \Sigma_{0 \Phi}^{\chi_i \xi} (p^2) \bigg].
\end{align}


\end{document}